%% file: Large-GEM-TPC-PANDA-Prototype.tex
\documentclass[final,12pt]{elsarticle}

\usepackage{graphicx}
\usepackage{multirow}
\usepackage{subcaption}

\usepackage{bookmark}

\usepackage{tikz}
\usetikzlibrary{arrows,shapes}
\definecolor{TikzTorquise}{RGB}{128,255,230}
\definecolor{TikzBlue}{RGB}{0,0,128}
\definecolor{TikzViolet}{RGB}{212,42,255}
\input{myTikzSet}
\usepackage[separate-uncertainty=true,multi-part-units=single]{siunitx}

\usepackage{upgreek}

\usepackage{lineno}

\usepackage{sty/myunits} 
\usepackage{sty/mymath} 
\usepackage{framed, color}
\graphicspath{	{./figures/intro/}
				{./figures/design/}
				{./figures/fieldcage/}
				{./figures/calibration/}
				{./figures/mediaflange/}
				{./figures/cooling/}
				{./figures/fee_daq/}
			}

\input{newcom}

\journal{Nuclear Instruments and Methods in Physics Research Section A}

\newlength{\twosubht}
\newsavebox{\twosubbox}
\DeclareSIUnit[]\liter{\text{\ensuremath{\ell}}}

\begin{document}

\begin{frontmatter}

  \title{A Large Ungated TPC with GEM Amplification}

  \author[tum,ucl]      {M. Berger}
  \author[tum,hiskp]	{M. Ball}
  \author[tum,ucl]      {L. Fabbietti}
  \author[tum,hiskp]	{B. Ketzer\corref{cor1}}
  \ead{Bernhard.Ketzer@cern.ch}
  \author[gsi]      {R. Arora}
  \author[hiskp]    {R. Beck}         
  \author[tum]      {F. B\"ohmer}
  \author[tum,ucl]  {J.-C. Chen}
  \author[tum,ucl]  {F. Cusanno\fnref{fn1}}
  \author[tum]      {S. D{\o}rheim}
  \author[gsi]      {J. Hehner}
  \author[uniH]     {N. Herrmann}     
  \author[tum]      {C. H\"oppner}
  \author[hiskp]    {D. Kaiser}     
  \author[gsi]      {M. Ki\u s}     
  \author[gsi]      {V. Kleipa}
  \author[tum]      {I. Konorov}
  \author[gsi]      {J. Kunkel}
  \author[gsi]      {N. Kurz}     
  \author[gsi]      {Y. Leifels}     
  \author[smi]      {P. M\"ullner}
  \author[tum,ucl]  {R. M\"unzer}    
  \author[tum]      {S. Neubert}
  \author[tum]      {J. Rauch}
  \author[gsi]      {C.J. Schmidt}
  \author[hiskp]    {R. Schmitz}     
  \author[gsi]      {D. Soyk}
  \author[tum]      {M. Vandenbroucke}
  \author[gsi]      {B. Voss}         
  \author[hiskp]    {D. Walther}
  \author[smi]      {J. Zmeskal}
%
%
\cortext[cor1]{Corresponding author}
\fntext[fn1]{Deceased}
  \address[tum]{Technische Universit\"at M\"unchen, Physik Department, 85748 Garching, Germany} 
  \address[hiskp]{Universit\"at Bonn, Helmholtz-Institut f\"ur
    Strahlen- und Kernphysik, 53115 Bonn, Germany}
  \address[gsi]{GSI Helmholtzzentrum f\"ur Schwerionenforschung GmbH, 64291 Darmstadt, Germany} 
  \address[smi]{Stefan Meyer Institut f\"ur Subatomare Physik, Wien, Austria}
  \address[uniH]{Universit\"at Heidelberg, Heidelberg, Germany}
  \address[ucl]{Excellence Cluster ’Origin and Structure of the Universe’, 85748 Garching, Germany}
\begin{abstract}
A Time Projection Chamber (TPC) is an ideal device for the detection
of charged particle tracks in a large volume covering a solid angle of
almost $4\pi$.  
The high density of hits on a given particle track facilitates the
task of pattern recognition in a high-occupancy environment and in
addition provides particle identification by measuring the specific
energy loss for each track. 
For these
reasons, TPCs with Multiwire Proportional Chamber (MWPC)
amplification have been and are widely used in experiments
recording heavy-ion collisions. A significant drawback, however,
is the large dead time of the order of \SI{1}{\milli\second} per
event generated by the use of a gating grid, which is mandatory to
prevent ions created in the amplification region from drifting
back into the drift volume, where they would severely distort the
drift path of subsequent tracks.  For experiments with higher
event rates this concept of a conventional TPC operating with a
triggered gating grid can therefore not be applied without a
significant loss of data.  A continuous readout of the signals is
the more appropriate way of operation.  This, however, constitutes
a change of paradigm with considerable challenges to be met
concerning the amplification region, the design and bandwidth of
the readout electronics, and the data handling.  A mandatory
prerequisite for such an operation is a sufficiently good
suppression of the ion backflow from the avalanche region, which
otherwise limits the tracking
and particle identification capabilities of such a detector.  Gas
Electron Multipliers (GEM) are a promising candidate to combine
excellent spatial resolution with an intrinsic suppression of
ions.  In this paper we describe the design, construction and the
commissioning of a large TPC with GEM amplification and without
gating grid (GEM-TPC).  The design requirements have driven
innovations in the construction of a light-weight field-cage, a
supporting media flange, the GEM amplification and the readout
system, which are presented in this paper.  We further describe
the support infrastructure such as gas, cooling and slow control. 
Finally, we report on the operation
of the \gt in the FOPI experiment, and describe the calibration
procedures which are applied to achieve the design performance of the device. 
\end{abstract}


\end{frontmatter}
\input{intro/intro.tex}
\input{design/design.tex}
\input{fieldcage/fieldcage.tex}
\input{mediaflange/mediaflange.tex}
\input{readout/readout.tex}
\input{fee_daq/fee_daq_new.tex}
\input{cooling/cooling.tex}
\input{gas/gas.tex}
\input{Control/Control.tex}
\input{commissioning/commissioning.tex}
%
%
\input{calibration/calibration.tex}
\input{conclusions/conclusions.tex}
%
\input{ack/ack.tex}

\newpage

\bibliographystyle{elsarticle-num}

\bibliography{bibliography/hadron,bibliography/compass,bibliography/panda,bibliography/detectors,bibliography/mpgd}

\end{document}

%% file: myTikzSet.tex
\tikzset
{
task/.style = {draw=TikzBlue, line width=0.6mm, rectangle,rounded corners=5mm, fill=TikzTorquise,text centered,node distance=2.5cm,text width=7cm, minimum height=2cm,align=center},
simtask/.style = {draw=TikzBlue, line width=0.6mm, rectangle,rounded corners=5mm, fill=TikzViolet,text centered,node distance=2.5cm,text width=7cm, minimum height=2cm,align=center},
arrow/.style = {draw=TikzBlue,line width=0.6mm,->},
darrow/.style= {draw=black,line width=0.6mm,->,dashed},
object/.style={draw=black,line width=0.6mm, rectangle, align=flush left, node distance=8cm,text width=4.5cm,dashed},
myline/.style={draw=black, line width=0.6mm,dashed}
}

%% file: newcom.tex
\newcommand{\figref}[1]{Fig.~\ref{#1}}
\newcommand{\Figref}[1]{Figure~\ref{#1}}
\newcommand{\tabref}[1]{Table~\ref{#1}}
\newcommand{\Tabref}[1]{Table~\ref{#1}}
\newcommand{\secref}[1]{section~\ref{#1}}
\newcommand{\Secref}[1]{Section~\ref{#1}}

\newcommand{\gt}{GEM-TPC\ }

\newcommand{\Xminus}{\relax\ifmmode\mathrm{X^-}%
                \else$\mathrm{X^-}$\fi}%
\newcommand{\X}{\relax\ifmmode\mathrm{X}%
                \else$\mathrm{X}$\fi}%

\DeclareSIUnit{\sqrtcm}{\ensuremath{\sqrt{\text{\centi\meter}}}}

%% file: intro/intro.tex
\section{Introduction}
\label{intro}
\noindent
A gas-filled Time Projection Chamber (TPC) \cite{Nygren:1978rx} with
its low material budget, large solid-angle coverage and excellent
pattern recognition capabilities is an ideal detector for
three-dimensional tracking and identification of charged particles.
TPCs have been already used in many experiments such as PEP-4
\cite{Madaras:1982cj}, ALEPH \cite{Atwood:1991bp}, DELPHI
\cite{Brand:1989qv}, NA49 \cite{Wenig:1998vv,Afanasev:1999iu}, and are
currently being employed e.g.\ in STAR
\cite{Ackermann:1999kc,Ackermann:2002yx} and ALICE \cite{Alme:2010ke}.
In a conventional TPC, electron-ion pairs created by an ionizing
particle within the drift volume are separated by an electric field.
Electrons drift towards the readout anode, where they produce a charge
avalanche in Multiwire Proportional Chambers (MWPCs)
\cite{Charpak:1968} which induces a detectable signal in the cathode
pads.  In order to prevent the ions produced in the avalanche from
drifting back into the drift volume of the TPC, where they would
severely distort the drift field, a switchable electrostatic gate must
be employed.
This, however, inevitably leads to dead times and limits the TPC readout rate to values of the order of few \si{\kilo\hertz}. 

\begin{figure}[!ht]
  \begin{center}
    \centering
    \includegraphics[width=0.8\textwidth]{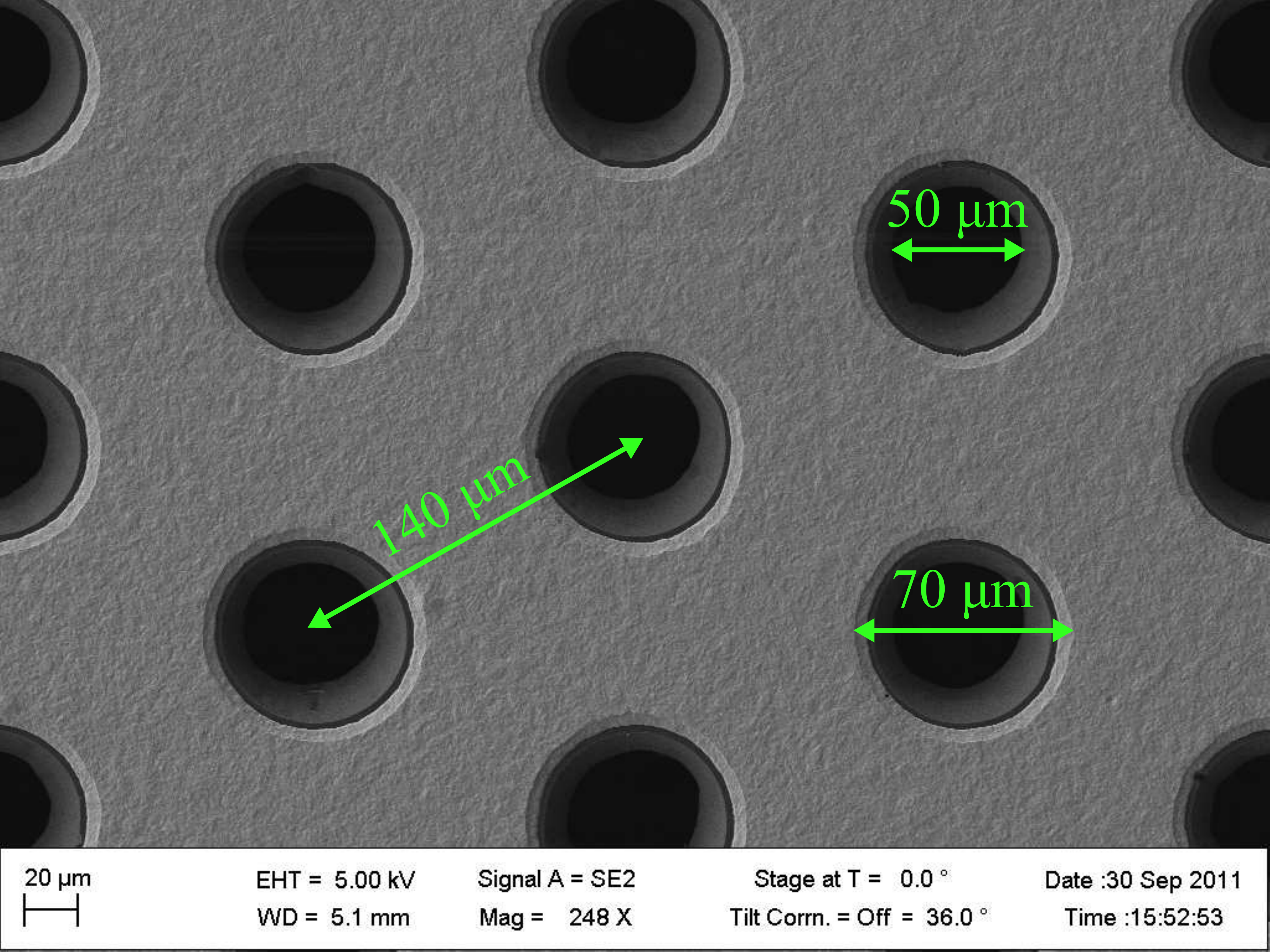}
    \caption {\label{fig:gem_photo} Electron microscope image of
      a GEM foil. The scale is given on the bottom left side.}
  \end{center}
\end{figure}

In order to efficiently employ TPCs in experiments with high interaction rates, new techniques of ion-backflow suppression have to be found which do not require the use of a gating grid. 
Gas Electron Multiplier (GEM) foils \cite{Sauli:97} represent an
optimal solution since they offer an intrinsic suppression of the ion
backflow \cite{Sauli:02a}.  A GEM consists of a \SI{50}{\micro\meter}
thin insulating polyimide foil with copper-coated surfaces, typically
\SI{5}{\micro\meter} thick.  The foil is perforated by
photolithography and etching processes, forming a dense, regular
hexagonal pattern of holes.  In the standard geometry the holes have a
double-conical shape with an inner diameter of \SI{\approx
  50}{\micro\meter}, an outer diameter of \SI{\approx
  70}{\micro\meter}, and a pitch of \SI{140}{\micro\meter}.
\Figref{fig:gem_photo} shows an electron microscope image of a
standard GEM foil.
The small dimensions of the amplification structures lead to very
large electric fields inside the holes of the GEM foil of the order of
\SI{50}{\kilo\volt\per\cm} if a moderate voltage difference of
\SIrange{200}{400}{\volt} (depending on the gas) is applied between
the metal layers.
Such fields are sufficient to create an avalanche inside the hole of
the GEM foil. 
Avalanche electrons are extracted from the hole by the electric field
applied below the GEM foil and can be further multiplied in subsequent
GEM foils or collected at the anode. Detectors based on a triple-GEM
amplification have been pioneered by the COMPASS  
experiment at CERN
\cite{Altunbas:02a,Ketzer:04a,Abbon:2007pq,Ketzer:07a}, and are  
now routinely 
used in several particle physics experiments like LHCb
\cite{Bencivenni:2002jr}, PHENIX
\cite{Fraenkel:2005wx}, and 
TOTEM \cite{Bagliesi:2010zz}.
New applications
include the use of GEM-based detectors in KLOE-2
\cite{Bencivenni:2007zz} and CMS
\cite{Abbaneo:2010dg}.  

\begin{figure}[!ht]
  \begin{center}
    \begin{subfigure}[b]{.49\textwidth}
      \centering
      \includegraphics[width=\textwidth]{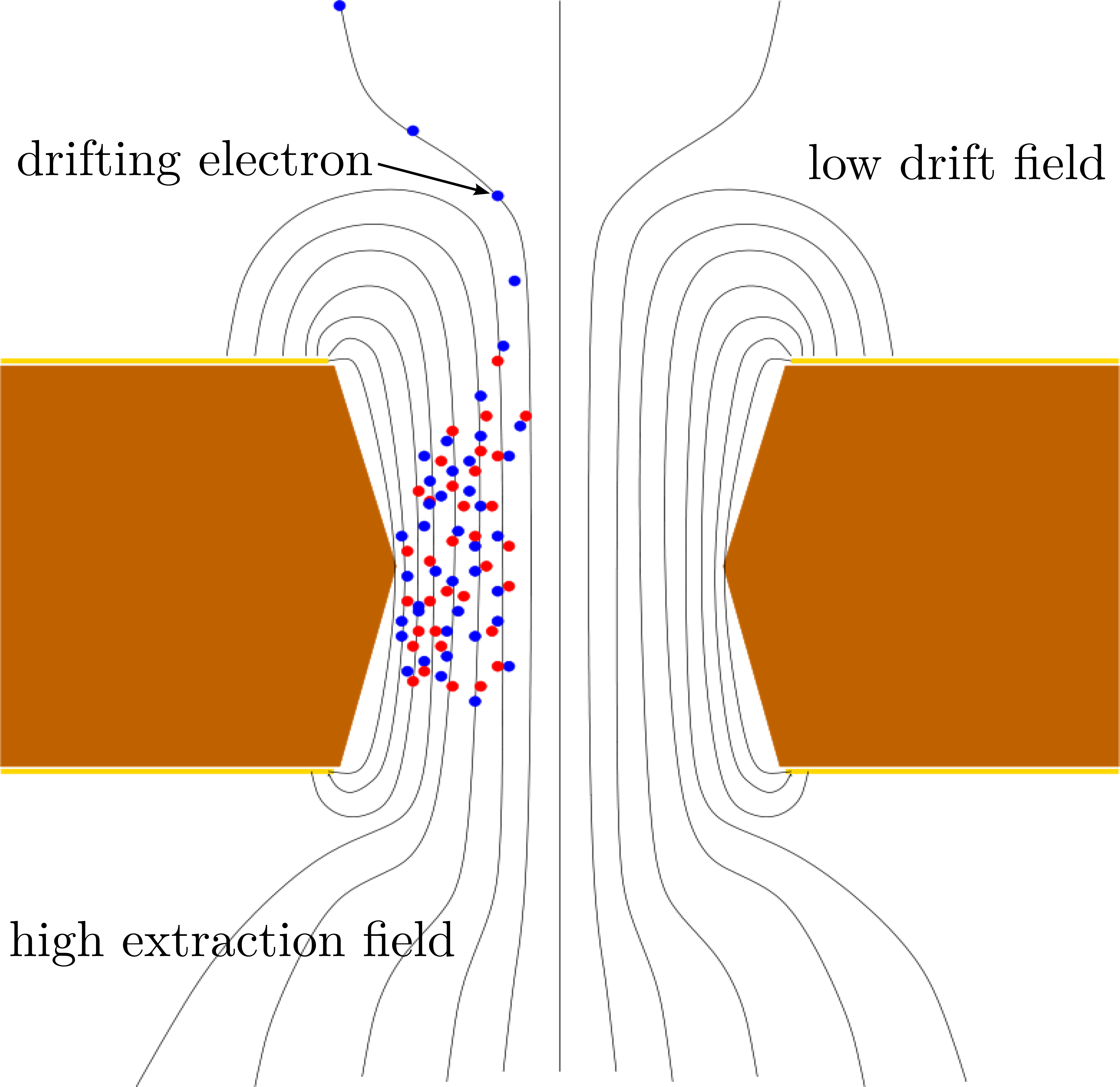}
      \caption{}
      \label{label1}
    \end{subfigure}
    \begin{subfigure}[b]{.49\textwidth}
      \centering
      \includegraphics[width=\textwidth]{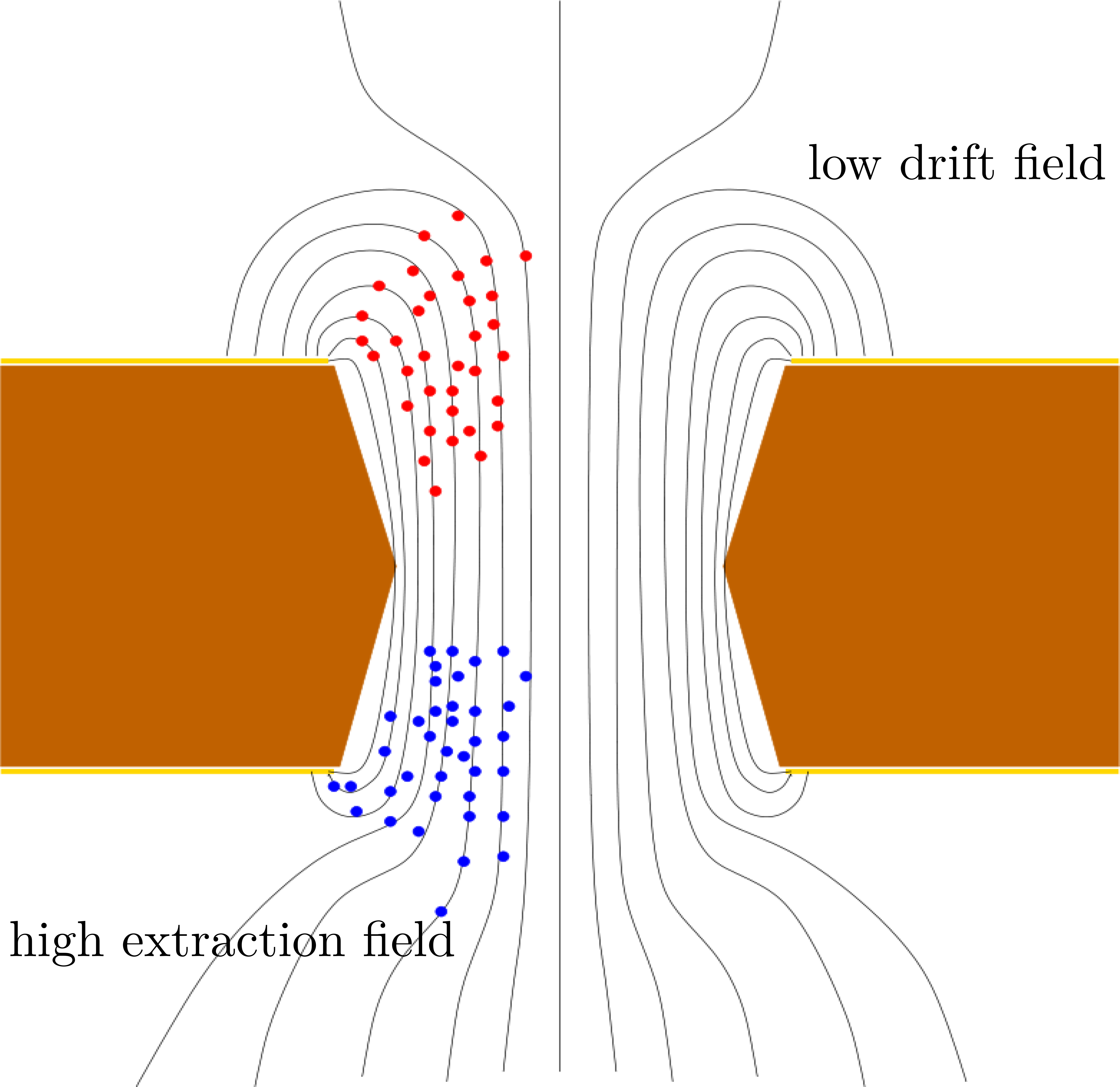}
      \caption{}
      \label{label2}
    \end{subfigure}
    \caption{Suppression of ion backflow in a GEM: (a) electrons are
      guided into the holes by the low drift field, where avalanches
      of electron-ion pairs are generated.  (b) The asymmetric field
      configuration of low drift field and higher extraction field
      together with the small ion mobility lead to efficient backflow
      suppression.}
    \label{fig:gem_fields}
  \end{center}
\end{figure}

Owing to their small diffusion, the ions created in the avalanche
inside a GEM hole 
closely follow the electric field lines leading to the top side of the
GEM foil and are collected there.  Depending on the electric field
settings, only a fraction of the ions created in the avalanche 
drifts back into the drift
volume.  This mechanism of ion backflow suppression is illustrated in
\figref{fig:gem_fields}.  

Although the ion suppression in a multi-GEM stack is of  
the order of $\EE*{-2}$ to a few $\EE*{-3}$ 
and hence does not reach the same level as 
a gating grid (see e.g. \cite{Ball:2014qaa,CERN-LHCC-2015-002} for a
summary of the present status of ion backflow measurements), 
the resulting track distortions due to ion backflow  
are limited to a few \si{\cm}, and can thus be corrected using
advanced calibration techniques.  This provides the possibility to
operate a TPC without the gating grid in a continuous, trigger-less
mode, allowing an increase in event rate
by two orders of magnitude, as foreseen e.g.\ in ALICE at CERN
\cite{Ketzer:2013laa,CERN-LHCC-2013-020}. 

In order to prove the concept of an ungated TPC with GEM amplification
the detector presented in
this paper was built.  The GEM-TPC was commissioned with cosmic
rays and with particle beams at the FOPI experiment at GSI in
Darmstadt (Germany), and was employed for an experimental
campaign with a secondary pion beam.
This paper describes the design and construction of the GEM-TPC 
including the auxiliary systems which were custom-made for this
detector. 
\Secref{sec:design} gives a general overview of the design of the
GEM-TPC. In 
\secref{sec:FC} the 
electro-mechanical  
design of the field-cage
is presented, followed in \secref{sec:mediaflange} by   
a detailed description of the so-called media flange, which 
hosts all supply connections and is the central support element 
of the detector. 
The GEM amplification stage is described in 
\secref{sec:ro:gem}, the pad-plane with hexagonal-shaped signal pads 
in \secref{sec:prototype.design.readout}. 
The
front-end electronics and the data acquisition system 
are discussed in 
\secref{sec:FEE_DAQ}. \Secref{sec:cooling} describes the cooling
system for the front-end electronics, while the gas system
and chosen gas mixture are presented in \secref{sec:gas}.  The slow
control system used to operate the GEM-TPC detector is treated in
\secref{sec:slowcontrol}.
Finally, in \secref{sec:Com} the results from the commissioning in the
FOPI experiment are presented followed in \secref{sec:CalSys} by a
discussion of the calibration procedures applied in order to reach the
design performance of the GEM-TPC. 

%% file: design/design.tex
\section{General Design}
\label{sec:design}
\noindent 
The GEM-TPC presented in this paper has a cylindrical shape with the
two end-planes serving as cathode and anode, respectively, and an
inner bore for the beam. 
The key parameters of the GEM-TPC are listed in
\tabref{tab:tpc.properties}. 
\begin{table}[tbp]
  \centering
  \begin{tabular}{llrr} \hline\hline
    Parameter 			&
    & Value & Unit			\\ \hline
    \multicolumn{2}{l}{Outer Dimensions:}		& {} &			\\
    & Inner Diameter 	& 104 				& $\mm$			\\
    & Outer Diameter 	& 309 				& $\mm$			\\
    & Total Length		& 843 			& $\mm$ \\
    \multicolumn{2}{l}{Active Volume:} 			& {} &			\\
    & Inner Diameter 	& 110.2				&$\mm$			\\
    & Outer Diameter 	& 299.0				&$\mm$			\\
    & Drift Length 		& 727.8				&$\mm$			\\
    & Gas Volume 		& 48
    &$\mathrm{dm}^3$			\\
    \multicolumn{2}{l}{GEM amplification stage:} & & \\
    & Effective gain & $1\EE{3} - 2\EE{3}$ & \\
    & GEM thickness & $50$ & $\upmu\m$ \\
    & GEM pitch     & $140$ & $\upmu\m$ \\
    & GEM hole \O & $70$ & $\upmu\m$ \\
    & Transfer gaps & 2 & $\mm$ \\
    & Induction gap & 4 & $\mm$ \\
    \multicolumn{2}{l}{Electronics} & &  \\
    & No. of readout pads & 10,254 & \\
    & ASIC & AFTER (T2K) & \\
    & Sampling clock & $15.55$ & $\MHz$ \\
    & Buffer depth & 511 & samples \\
    & Dynamic range & 120 & $\fC$ \\
    & No. of front-end cards & 42 & \\
    & ADC resolution & 11 & bit \\
    & Sensitivity & 393 & $e^-/\mathrm{ADC\ ch.}$ \\
    & Average ENC & 720 & $e^-$ \\
    \hline \hline
  \end{tabular}
  \caption{Design parameters of the GEM-TPC.}
  \label{tab:tpc.properties}
\end{table}

The outer diameter of
the TPC was chosen to fit inside the Central Drift Chamber (CDC)
of FOPI (see \secref{sec:Com}) while the inner diameter was chosen 
to host different target dimensions. The dimensions also match the
requirements for the Crystal Barrel experiment at ELSA
\cite{Lang:2010zza}, so that a similar detector could be used as a
tracking system in this experiment in the future.

The design of the detector takes into 
account the fixed-target nature of the FOPI experiment, 
with the ionization electrons drifting in one
direction only, different from the symmetric setup with a central
cathode and two anodes as used in collider experiments.
The cathode is
oriented in the forward direction with respect to the beam,
while the readout anode and the associated 
readout electronics are situated in the backward region in order to
minimize the material exposed to particles emitted in interactions
with the target, which is mounted in the inner bore. 

The detection gas 
is enclosed in a
vessel formed by two concentric 
cylinders and the two end planes. In addition to containing the
detector gas, this vessel also serves as field cage to realize the
required uniform electric field parallel to the cylinder axis.  
In order to keep the multiple scattering and interactions of traversing
particles as low 
as possible a very thin but rigid composite material is used for the
field-cage. On the cathode side the field-cage
vessel is closed by an end plate of similar sandwich structure as the
field cage.  

\begin{figure}[tbp]
  \centering
  \includegraphics[width=\textwidth]{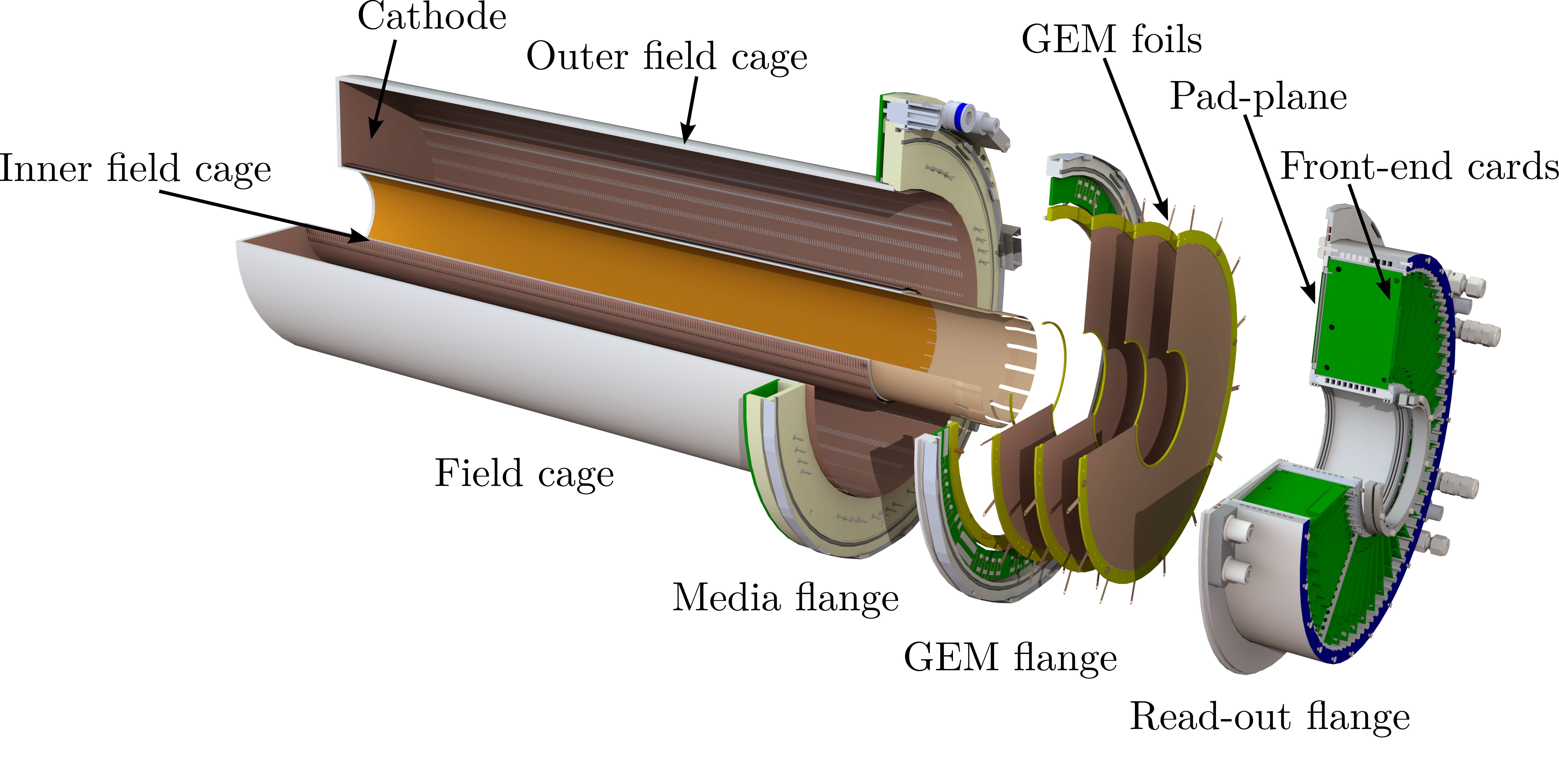}
  \caption{Cross-section of the large GEM-TPC.}
  \label{fig:prototype.exptpc}
\end{figure}

\begin{figure}[tbp]
  \centering
  \includegraphics[width=0.9\textwidth]{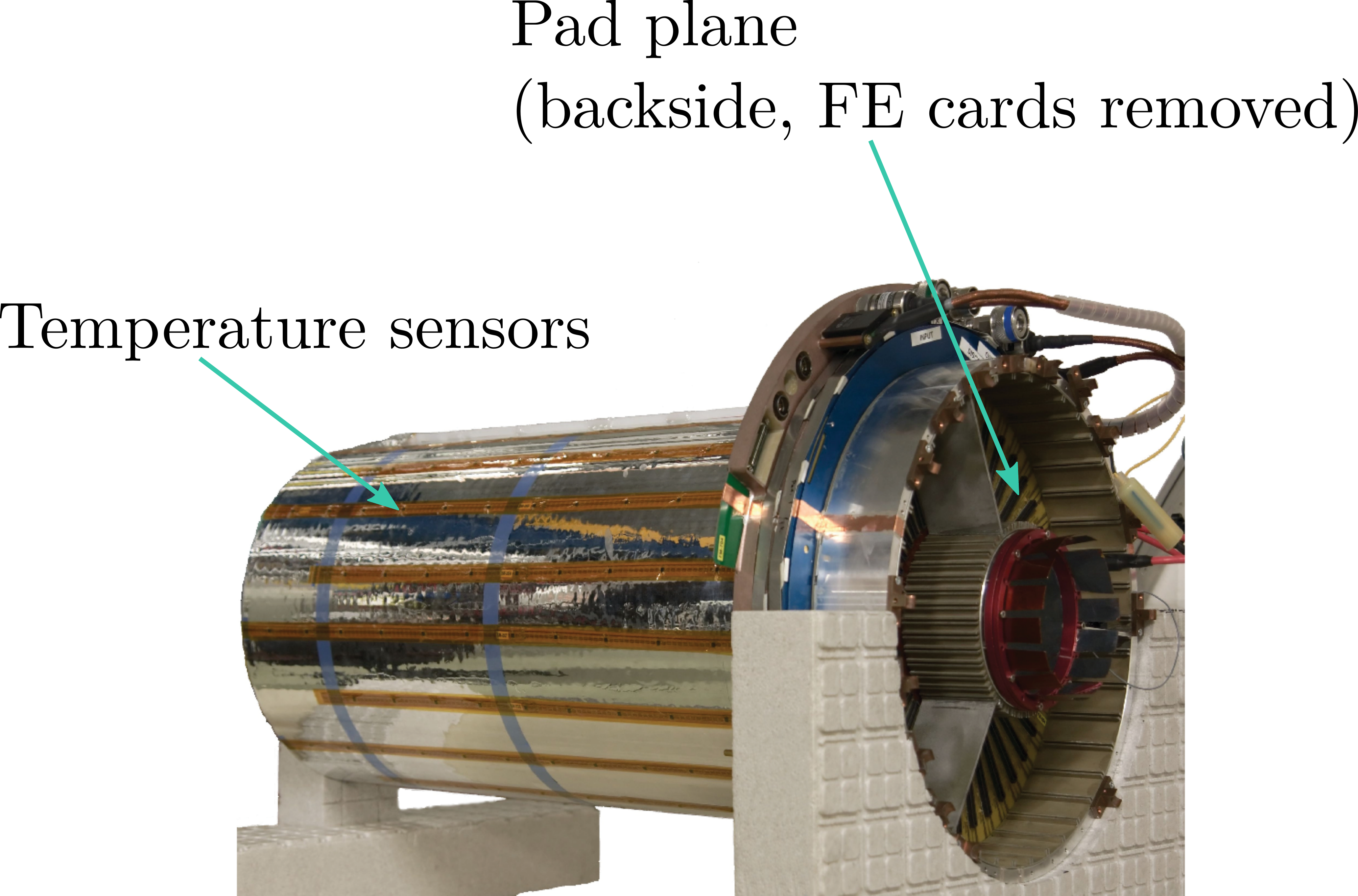}
  \caption{Photo of the large GEM-TPC.}
  \label{fig:prototype.realtpc}
\end{figure}

On the anode side a rigid flange called media flange is
glued to the field-cage.  The media flange is used to fix the detector
to external structures and houses all the connections for gas supply,
high voltage, low voltage and sensor read-out.  
Two further flanges
are mounted to the upstream side of the media flange: the GEM flange
housing the GEM stack, and the read-out flange holding the
pad-plane, which is 
patterned with $10,254$ hexagonal pads for a two-dimensional localization of 
the incoming charge. The printed circuit boards holding the front-end
(FE) electronics are directly connected to the back side of the
pad-plane in a fan-shaped arrangement and are supported by two rigid
rings made of Aluminum. 
These flanges 
are fixed to the media flange by external metal clamp bands and are sealed
by rubber O-rings. 
\Figref{fig:prototype.exptpc} and \figref{fig:prototype.realtpc} show
a cross-section and a photo of the final large GEM-TPC, respectively. 

%


The whole apparatus is therefore composed of three mechanically
separable parts: the field-cage cylindrical vessel with the media
flange, the GEM flange and the read-out flange.  
This modular setup
allows an easy maintenance of the detector system and to 
supply and read out the GEM-TPC exclusively from the anode side.
The GEM-TPC was mounted inside the existing FOPI spectrometer at GSI
\cite{FOPI1} and employed in experimental campaigns with hadron and
secondary pion beams at intermediate energies of few
\si{\giga\electronvolt}.
%
One major innovation in this TPC is the construction of a light-weight
field-cage (\secref{sec:FC}) with a material budget below
\SI{1}{\percent} of a radiation length.  This keeps the multiple
scattering in front of the other sub-detectors small to prevent a
deterioration of their tracking performance.  The target position is
chosen such that most of the produced particles are emitted in the
forward direction, where the material budget of the TPC is lowest,
while all the rigid support structures as well as the all
infrastructure connections are located upstream of the target
position.  Since the overall center of gravity of the TPC is close to
the detector backward end-cap, only few fixation points at the media
flange are necessary to hold the \gt in place.  Another major
achievement within this project is the adaption of the T2K AFTER chip
(\secref{sec:FEE_DAQ}) for the specific needs of the GEM-TPC.

%% file: fieldcage/fieldcage.tex
\section{Field-cage}
\label{sec:FC}
\subsection{General Concept}
\noindent
Owing to physical and mechanical constraints, the cylindrical vessel
of the TPC has to fulfill several requirements: it has to serve as gas
container and field-cage at the same time, provide shielding of the HV
to the outside while keeping the amount of material in forward and
radial directions at the minimum.  The main purpose of the integrated
gas vessel and the field-cage is to define an electrostatic field
with a relative homogeneity better than \num{e-4} inside the gas
volume to prevent that field inhomogeneities effect the detector
resolution \cite{Alme:2010ke}.  These requirements can be fulfilled
employing a layered composite material.  Because of its dual purpose
realisation, the field-cage walls must be gas-tight and mechanically
stable 
against changes in gas pressure and temperature.  A temperature
stability within \SI{1}{\degree} should be achieved during the system
operation.  The vessel should furthermore withstand a potential
difference of about \SI{30}{\kilo\volt} between cathode and anode and
must be electrically insulated.
\subsection{Technical Realisation}
\subsubsection{Composite Material}
\begin{figure}[tbp] \centering
  \includegraphics[width=0.9\textwidth]{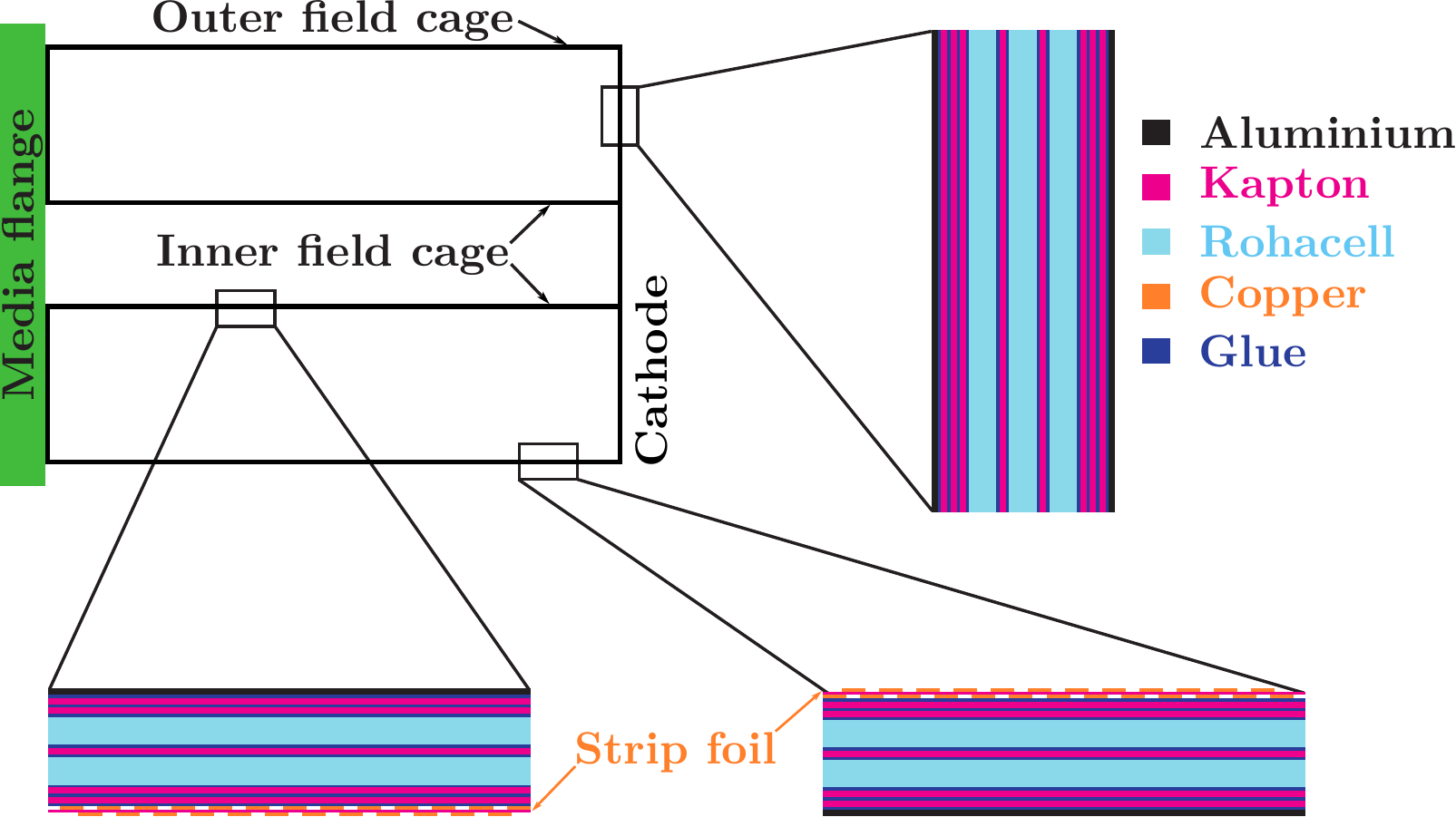}
  \caption{Cross-sections of the layout of the barrel and cathode
    end-cap walls, respectively. The shown layer stacking is not to
    scale.}
  \label{fig:fieldcage.sand}
\end{figure}

\begin{figure}[tbp] \centering
  \includegraphics[width=0.4\textwidth]{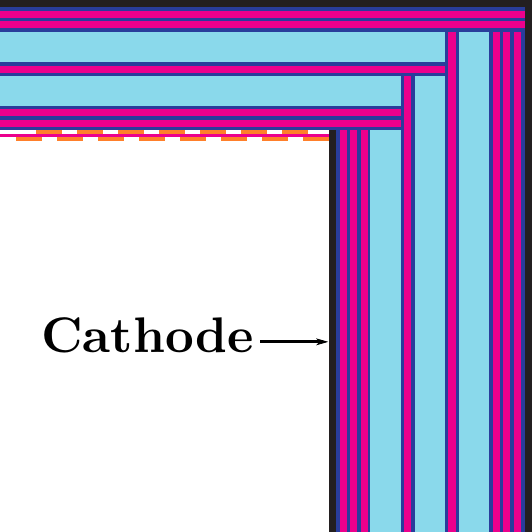}
  \caption{Cross-section of the interface between the field-cage barrel
    and the cathode
    end plane. The color code is the same as in
    \figref{fig:fieldcage.sand}.  
  }
  \label{fig:fieldcage.catcorn}
\end{figure}

\begin{figure}[tbp]
  \centering
  \includegraphics[width=0.75\textwidth]{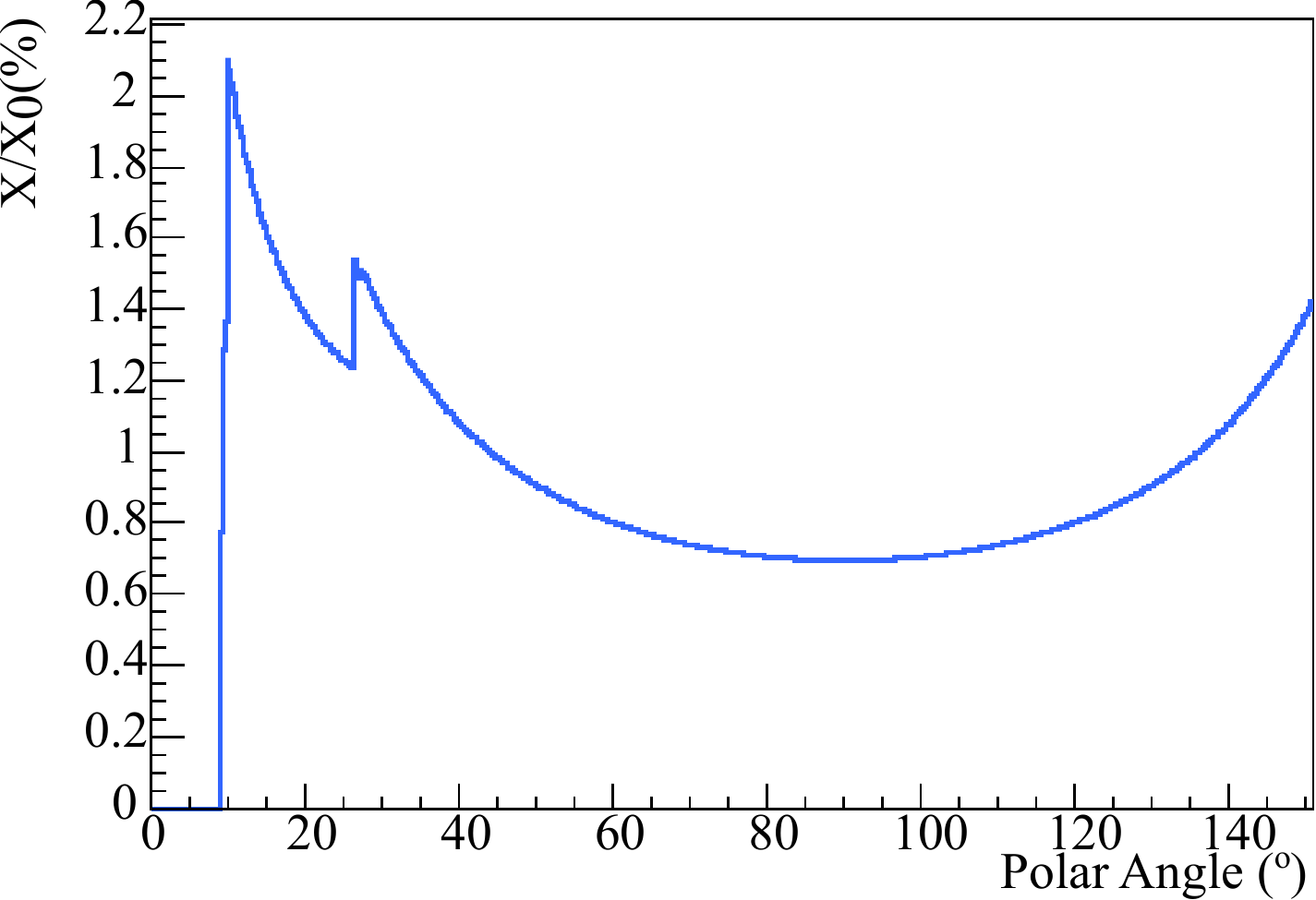}
  \caption{Total thickness of the field-cage vessel filled with
    Ar/CO$_2$ gas in units of a
    radiation length $X_0$ plotted against the polar angle of 
    particles emitted from the target position (see
    \figref{fig:prototype.scheme} for a sketch of the geometry).} 
  \label{pic:proto.radlen}
\end{figure}

\noindent The materials, the thicknesses of the different layers that
compose the barrel walls and the cathode end-cap expressed in
\si{\micro\meter} and in units of a radiation length are listed in the
tables \tabref{tab:matBar} and \tabref{tab:matEC}, respectively.  The
layer stacking is sketched schematically in
\figref{fig:fieldcage.sand}.  One can see that the layout of the two
concentric cylinders building the barrel walls and the cathode end-cap
differ slightly.  The cathode end-cap is composed of three
Rohacell\textsuperscript{\textregistered} 
layers to stabilize the structure and a further aluminized
Kapton\textsuperscript{\textregistered}  foil
is added to the side facing the gas volume to realize the cathode
plane.
The barrel and the cathode end-cap are joined together by gluing the extremities as shown in \figref{fig:fieldcage.catcorn}.\\
\begin{table}[tbp]
  \begin{subtable}[t]{.49\textwidth}
    \centering \vspace{0pt}
    \begin{tabular}{l | S[table-format=4.1] S[table-format=1.4]}
      \hline\hline
      & \multicolumn{2}{c}{Thickness}						\\ 
      Material	& {(\si{\micro\meter})} & {(\si{\percent}X$_0$)}	\\ 
      \hline
      Copper			& 5		& 0.035									\\
      Kapton			&25 	& 0.009 								\\
      Copper			& 5 	& 0.035 								\\
      Kapton			&25		& 0.009 								\\ 
      Kapton			&125 	& 0.045 								\\ 
      Rohacell		&2000	& 0.026   								\\ 
      Kapton			&125 	& 0.045   								\\ 
      Rohacell		&2000 	& 0.026   								\\ 
      Kapton			&125 	& 0.045  								\\ 
      Kapton			&25 	& 0.009 								\\ 
      Aluminium		&0.2 	& 0.0002  								\\ 
      Epoxy glue tot.	&140 	& 0.028     							\\ 
      &		&										\\
      Barrel Walls	&4600  	& 0.31		  							\\  
      \hline\hline
    \end{tabular}
    \vfill
    \caption{}
    \label{tab:matBar}
  \end{subtable}
  \begin{subtable}[t]{.49\textwidth}
    \centering \vspace{0pt}
    \begin{tabular}{l| S[table-format=4.1] S[table-format=1.4]}
      \hline\hline
      & \multicolumn{2}{c}{Thickness} 				\\ 
      Material 		& {(\si{\micro\meter})} & {(\si{\percent}X$_0$)}\\ 
      \hline
      Aluminium		&0.2	& 0.0002								\\
      Kapton			&25 	& 0.009									\\
      Kapton			&25		& 0.009 								\\
      Kapton			&125 	& 0.045									\\
      Rohacell		&2000	& 0.026									\\
      Kapton			&125 	& 0.045									\\
      Rohacell		&2000 	& 0.026									\\
      Kapton			&125 	& 0.045									\\
      Rohacell		&2000	& 0.026									\\
      Kapton			&125 	& 0.045									\\
      Kapton			&25 	& 0.009									\\
      Kapton			&25 	& 0.009									\\
      Aluminium		&0.2 	& 0.0002								\\
      Epoxy glue tot.	&140 	& 0.028    								\\
      \\
      Cathode End-Cap	&8900  	&0.33 									\\
      \hline\hline
    \end{tabular}
    \caption{}
    \label{tab:matEC}
  \end{subtable}
  \caption{Materials, corresponding thickness of the layers that compose the barrel walls (a) and the cathode end-cap (b). The items are ordered from the inner to the outer surface of the field-cage.}
\end{table}
Special moulds were developed and constructed for the assembly of the 
cylindrical sandwich structures. 
The moulds are built by panelling an Aluminium cylinder with vacuum
grooves with Metapor\textsuperscript{\textregistered} slabs which are
then milled into a cylindrical 
form.  The outer strip foil or the inner strip foil are rolled around
the respective mould and fixed by aspirating air through the porous
Metapor material.  The different
materials are then glued successively 
on top of the previous layer.  Vacuum bags are used to apply pressure
and to flatten the layers during the gluing procedure of the sandwich
structure.  The vessel is electrically insulated by an additional
Aluminium-coated Kapton layer of \SI{30}{\micro\meter} placed on its
outside surface. This layer is connected to the detector-ground
potential providing a shielded Faraday cage.  Two gas supply lines and
one HV line are attached to the outer field-cage parallel to the
cylinder axis.  The two gas lines are connected to the gas inlet
realized with a Vespel\textsuperscript{\textregistered} block hosting
two inner channels.  The block 
is inserted inside a cutout of the interface between the cathode and
the field-cage to allow the gas flow through the channels into the
cylinder.
The high voltage supply cable is pierced through the field-cage and
soldered onto a copper strip which in turn is glued with conductive
adhesive onto the cathode.\\ 
The thickness of the field-cage vessel in units of a radiation length
$X_0$ was evaluated within the Root analysis framework
\cite{Brun199781}.  A detailed geometry of the large GEM-TPC was
created and the thickness was evaluated along tracks with different
emission angles.  The emission point for the considered particle
tracks was placed on the central axis of the GEM-TPC at a distance of
\SI{30}{\cm} upstream of the cathode
plane (see \figref{fig:prototype.scheme}).  The obtained polar-angle
dependence of the thickness in units 
of a radiation length $X_0$ is shown in \figref{pic:proto.radlen}.
At polar angles of \SI{\approx 9}{\degree} 
particles start to hit the inner field-cage wall in forward
direction, causing a sharp increase of the material budget. Particles
cross the cathode end plane of the field-cage up to an angle of 
\SI{\approx26}{\degree}, traversing less material of the inner barrel
wall with increasing polar angle. At \SI{\approx26}{\degree} they
start traversing both the inner 
and the outer barrel walls, which have a shorter radiation length than
the cathode (see \tabref{tab:matEC}), causing an increase in the
material budget. The remaining smooth structure with a minimum at
\SI{90}{\degree} reflects the thickness of the walls crossed by particles. 
Above a polar
angle of \SI{160}{\degree} the thickness of the field-cage increases
significantly (not shown in Figure \ref{pic:proto.radlen}) due to the
large material budget of the media, GEM and readout flanges.  For fixed
target experiments such as FOPI where most of the tracks go in forward
direction this high material budget in the backward direction has no
impact on the overall detector performance.
\subsubsection{Strip Foil and Voltage Divider}
\label{Divi}
\begin{figure}[tbp]
  \begin{center}
    \includegraphics[width=.7\textwidth]{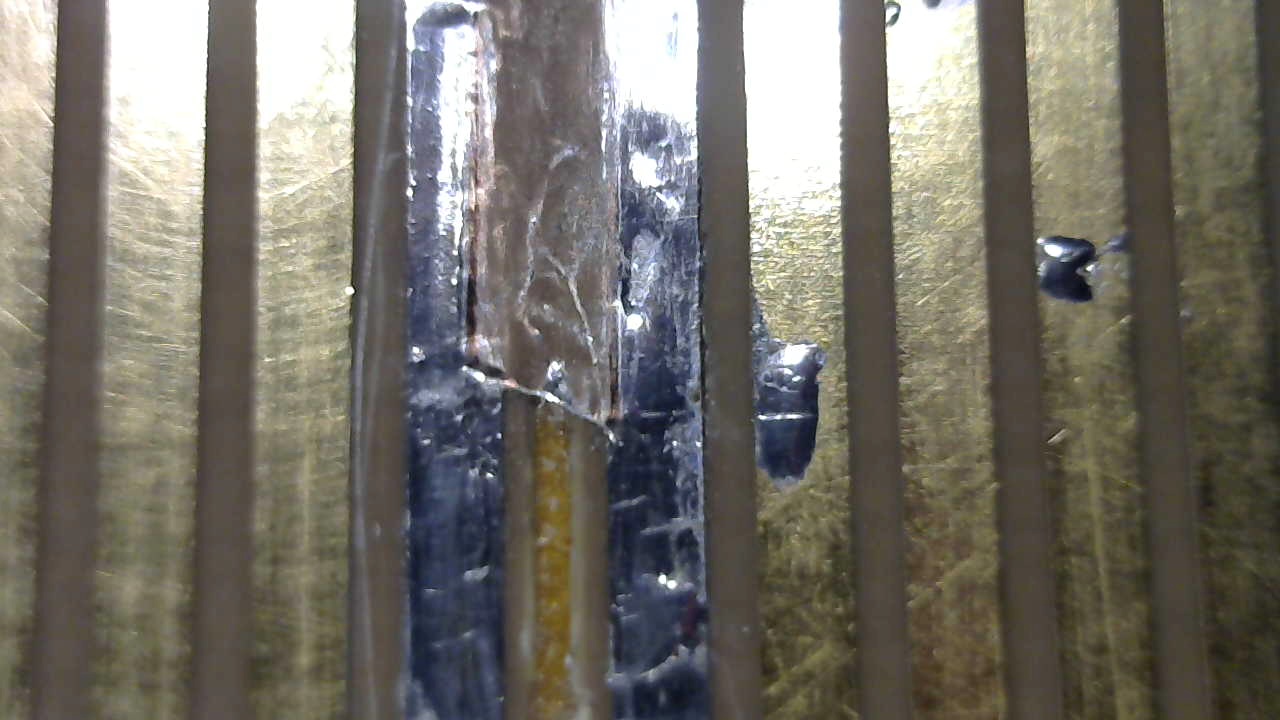}
    \caption{Copper bridge soldered onto the field-cage strips at the
      border of the two spliced strip-foils.}
    \label{fig:fieldcage.copperbridge}
  \end{center}
\end{figure}

\begin{figure}[tbp]
  \begin{center}
    \begin{subfigure}[b]{.49\textwidth}
      \centering
      \includegraphics[width=\textwidth]{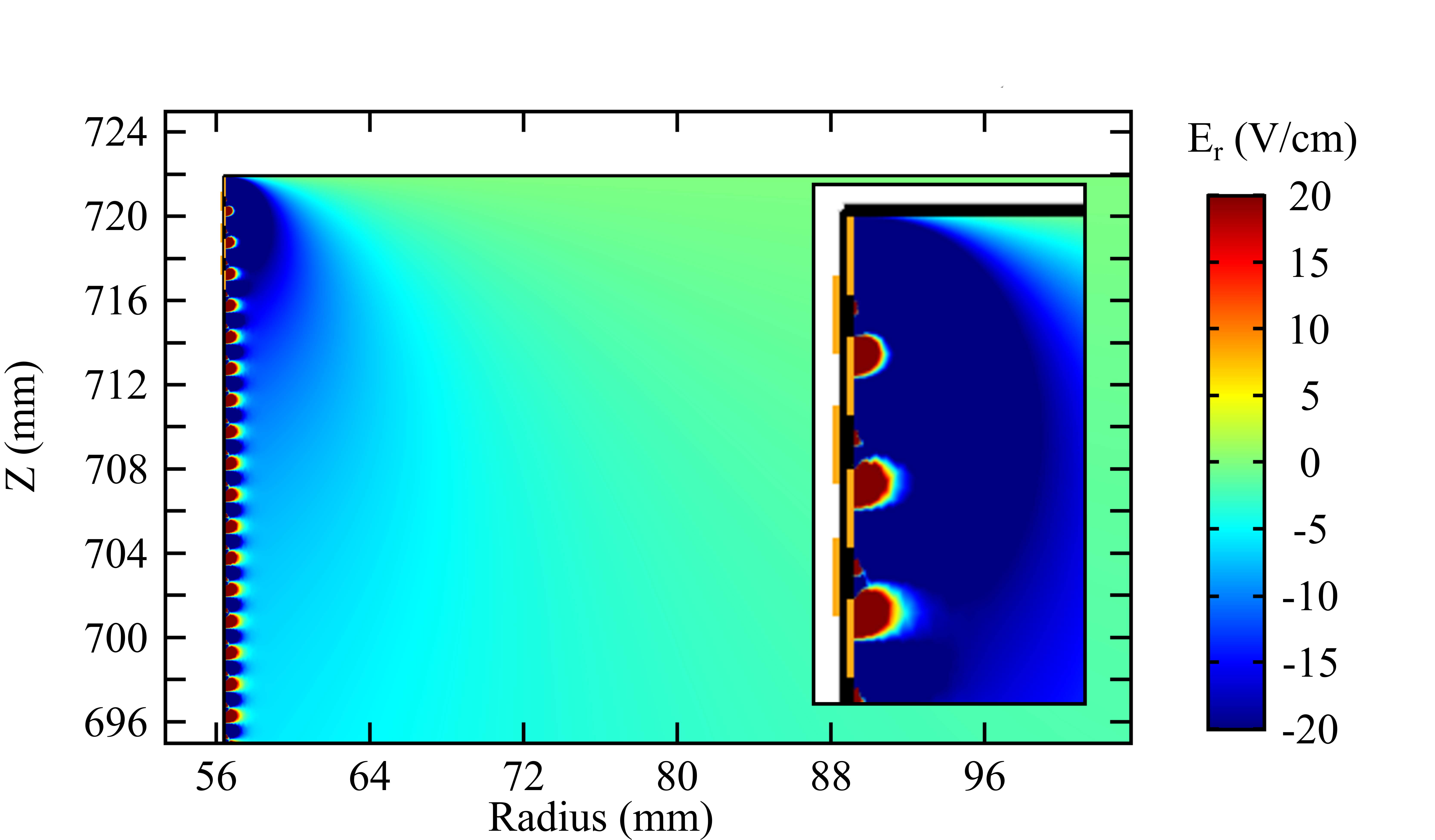}
      \caption{}
      \label{label1}
    \end{subfigure}
    \begin{subfigure}[b]{.49\textwidth}
      \centering
      \includegraphics[width=\textwidth]{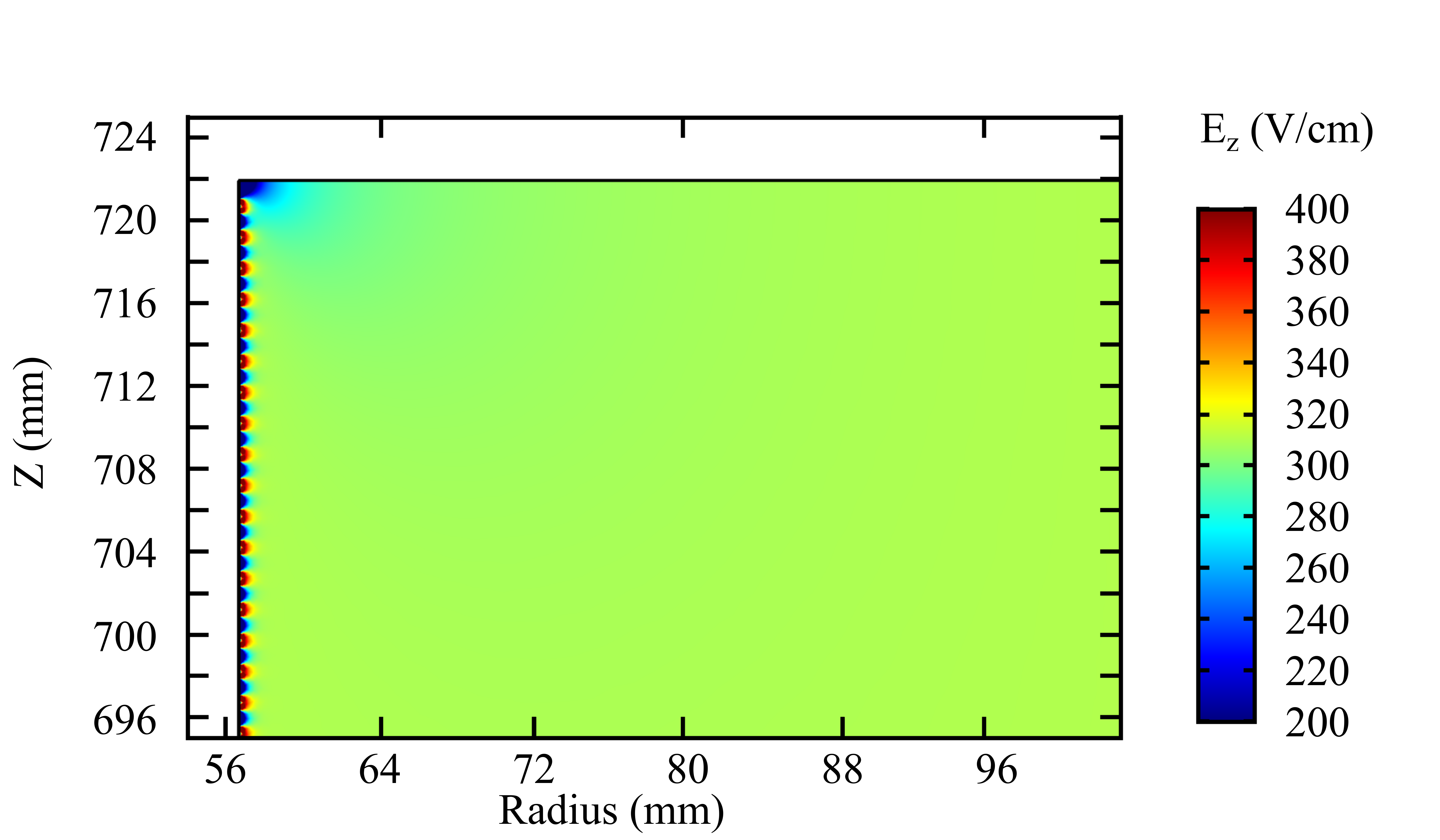}
      \caption{}
      \label{label2}
    \end{subfigure}
    \caption[shortCaption]{Results of a finite-element simulation of
      the 2-D map of the electric field in the radial (left panel) and
      $z$ (right panel) direction in the region close to the cathode for
      an ideal set-up. See text for details.}
    \label{fig:fieldcage.cathode1}
  \end{center}
\end{figure}

\begin{figure}[tbp]
  \begin{center}
    \begin{subfigure}[b]{.49\textwidth}
      \centering
      \includegraphics[width=\textwidth]{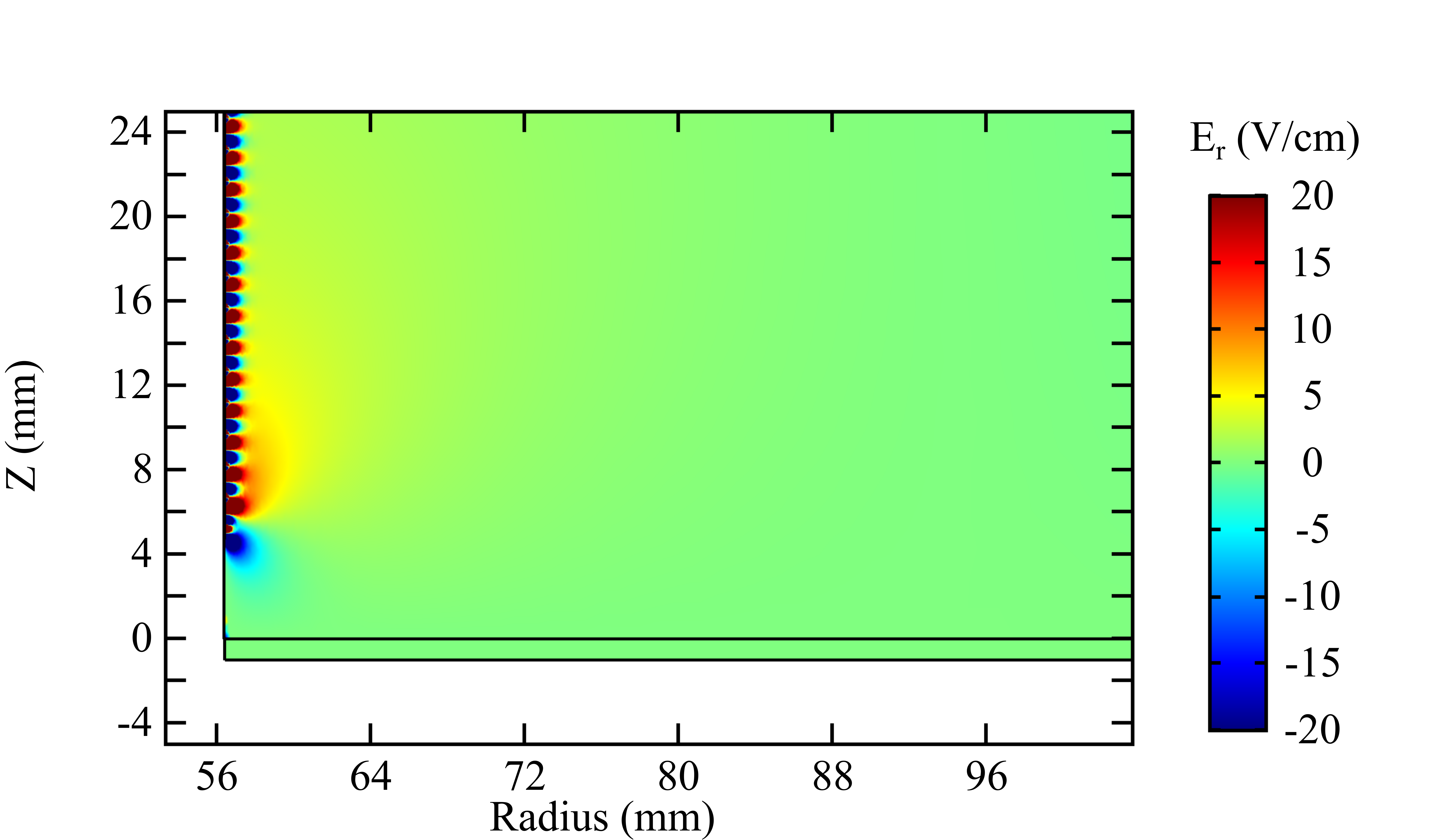}
      \caption{}
      \label{label1}
    \end{subfigure}
    \begin{subfigure}[b]{.49\textwidth}
      \centering
      \includegraphics[width=\textwidth]{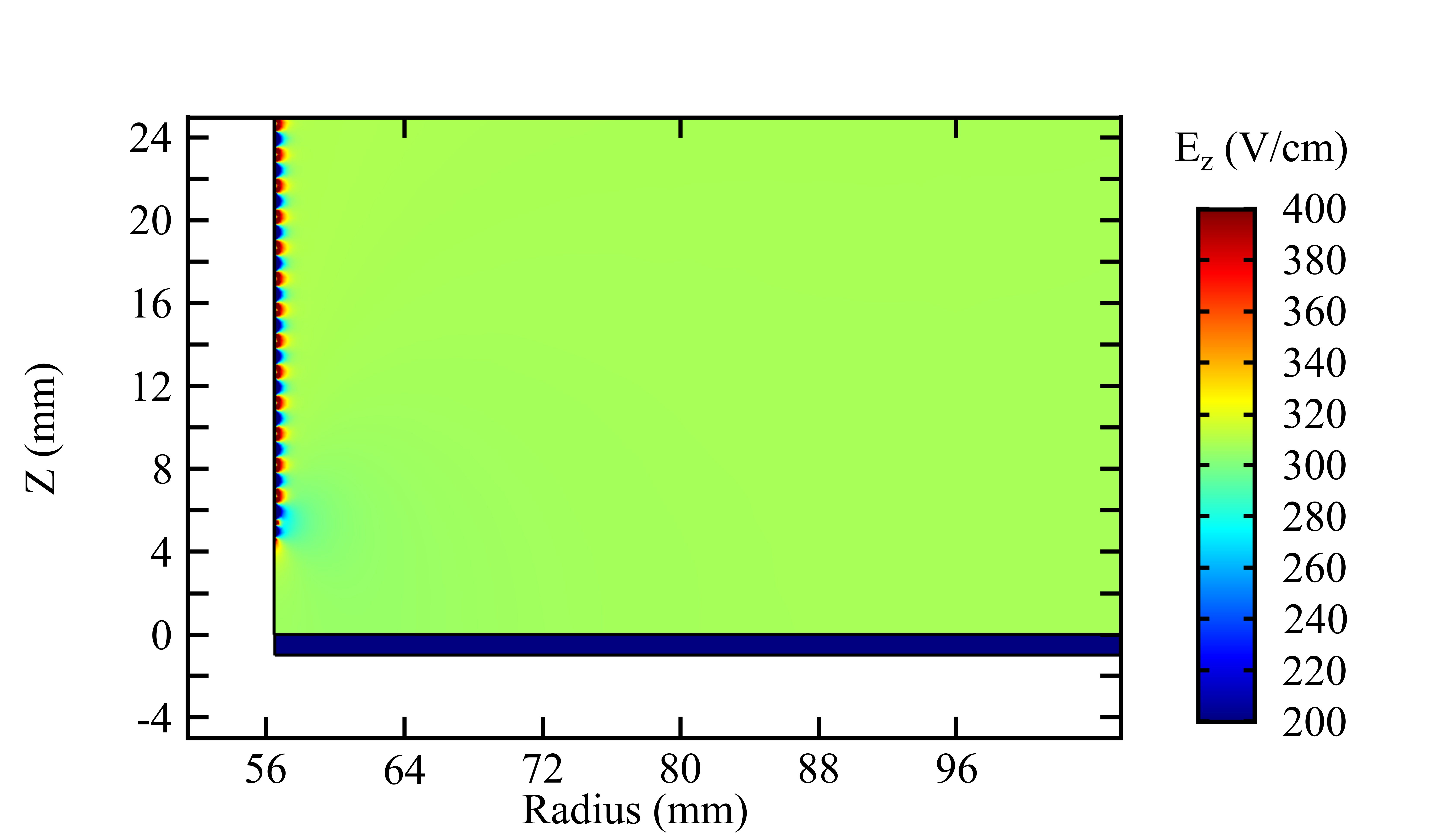}
      \caption{}
      \label{label2}
    \end{subfigure}
    \caption[shortCaption]{Results of a finite-element simulation of
      the 2-D map of the electric field in the radial (left panel) and
      $z$ (right panel) direction in the region close to the last strip
      for an ideal set-up. See text for details.}
    \label{fig:fieldcage.ls1}
  \end{center}
\end{figure}

\begin{figure}[tbp]
  \begin{center}
    \begin{subfigure}[b]{.49\textwidth}
      \centering
      \includegraphics[width=\textwidth]{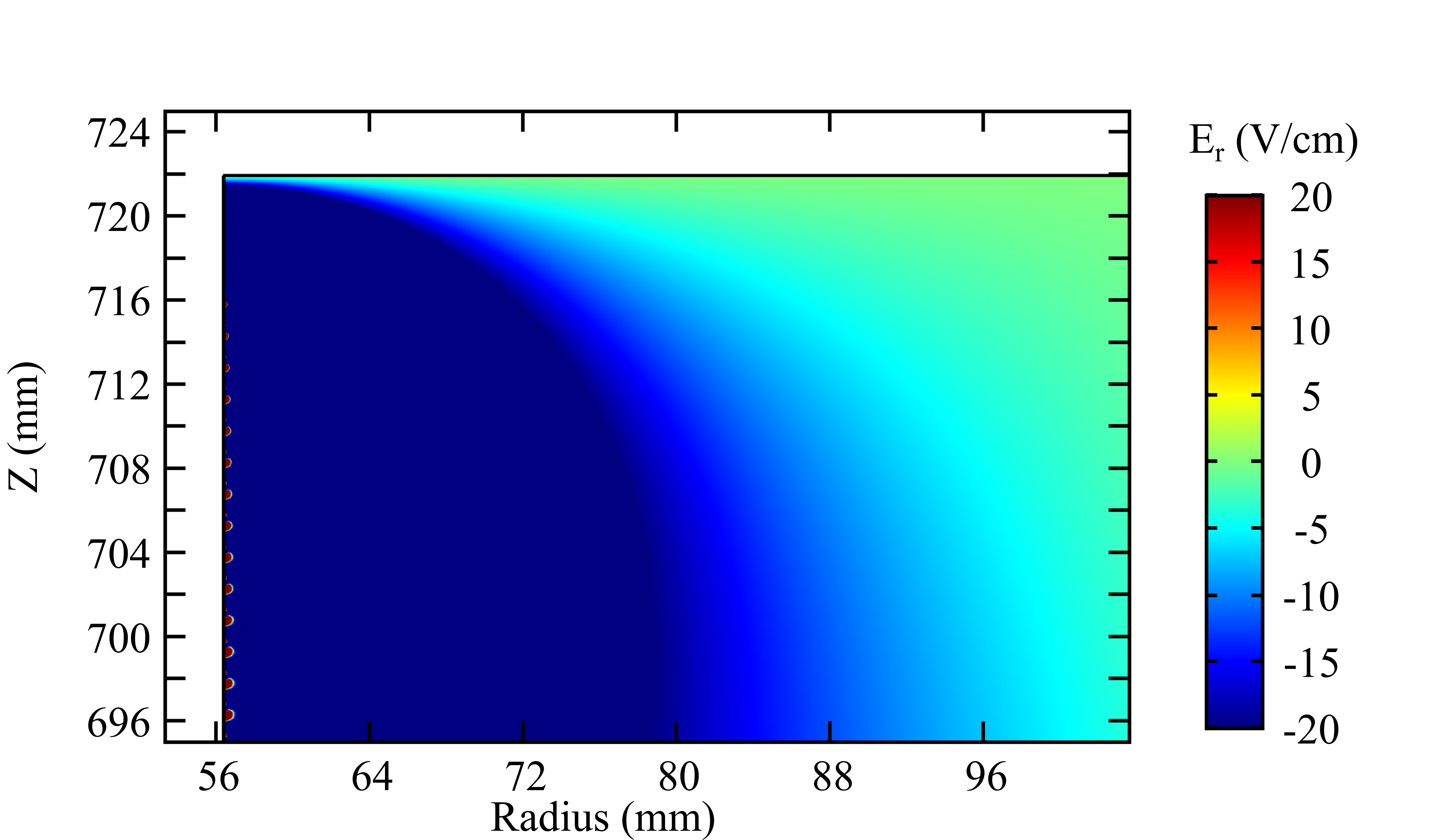}
      \caption{}
      \label{label1}
    \end{subfigure}
    \begin{subfigure}[b]{.49\textwidth}
      \centering
      \includegraphics[width=\textwidth]{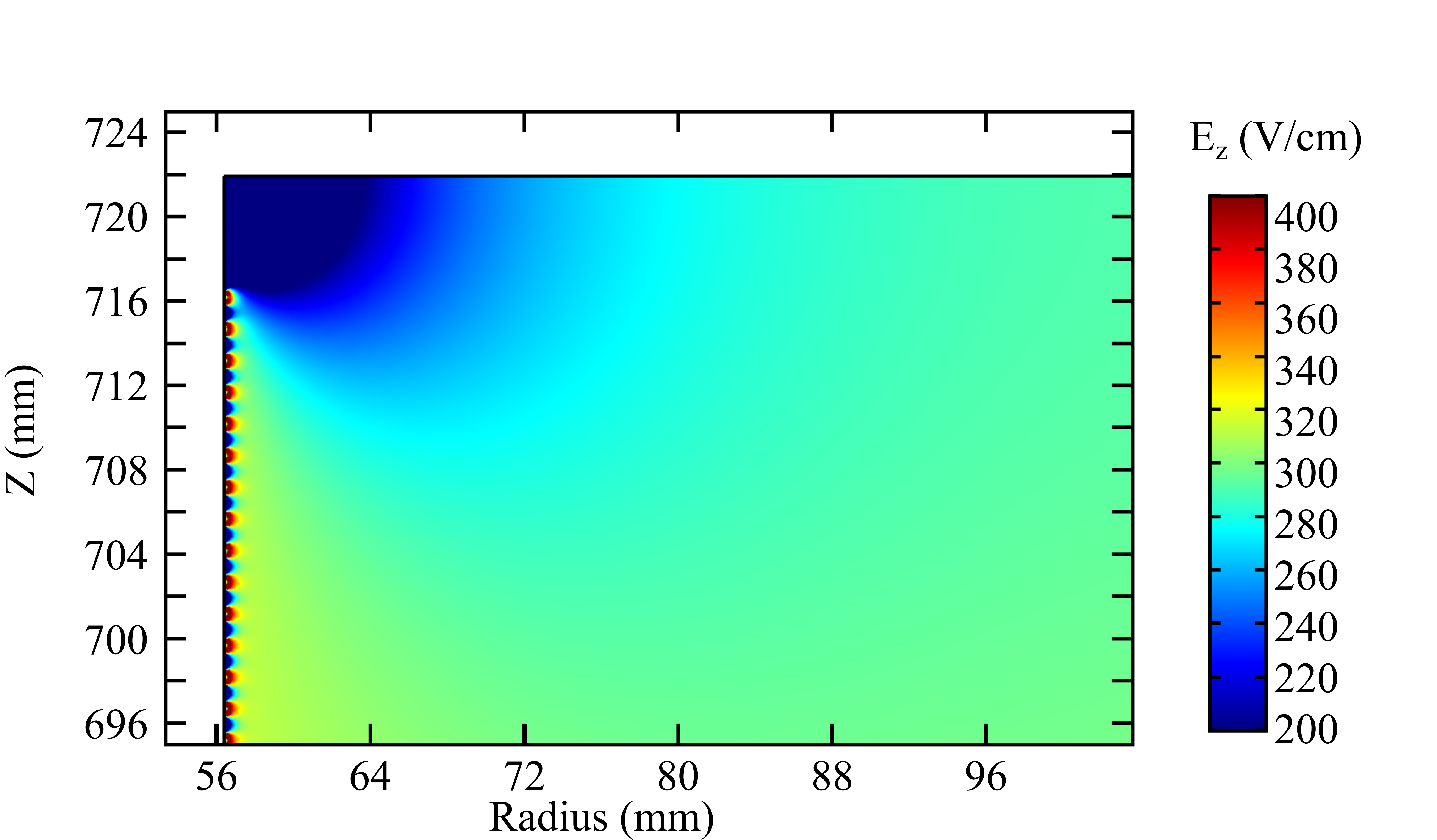}
      \caption{}
      \label{label2}
    \end{subfigure}
    \caption[shortCaption]{Results of a finite-element simulation of
      the 2-D map of the electric field in the radial (left panel) and
      $z$ (right panel) direction in the region close to the cathode for
      the real set-up. See text for details.}
    \label{fig:fieldcage.cathode2}
  \end{center}
\end{figure}

\begin{figure}[tbp]
  \begin{center}
    \begin{subfigure}[b]{.49\textwidth}
      \centering
      \includegraphics[width=\textwidth]{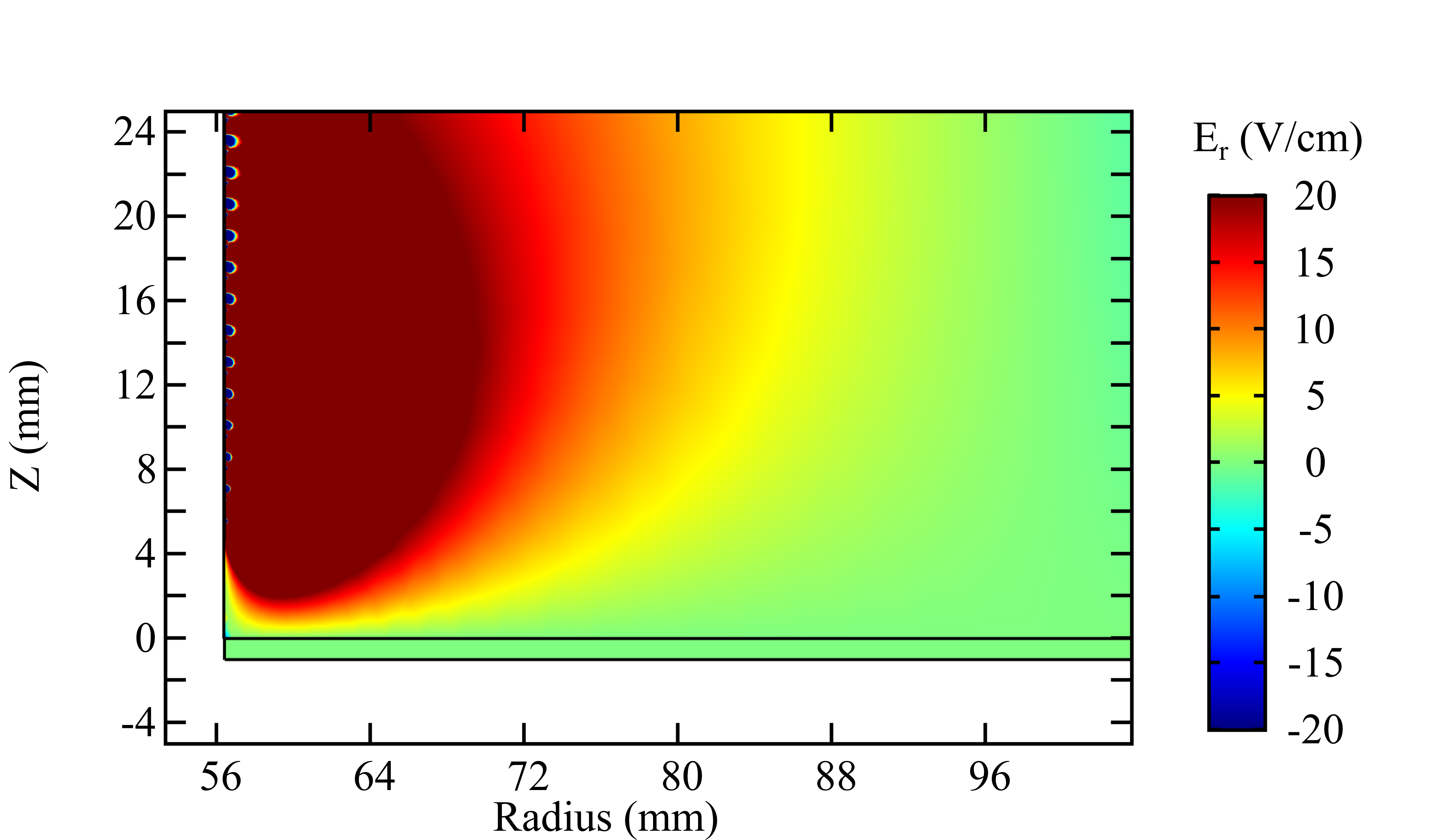}
      \caption{}
      \label{label1}
    \end{subfigure}
    \begin{subfigure}[b]{.49\textwidth}
      \centering
      \includegraphics[width=\textwidth]{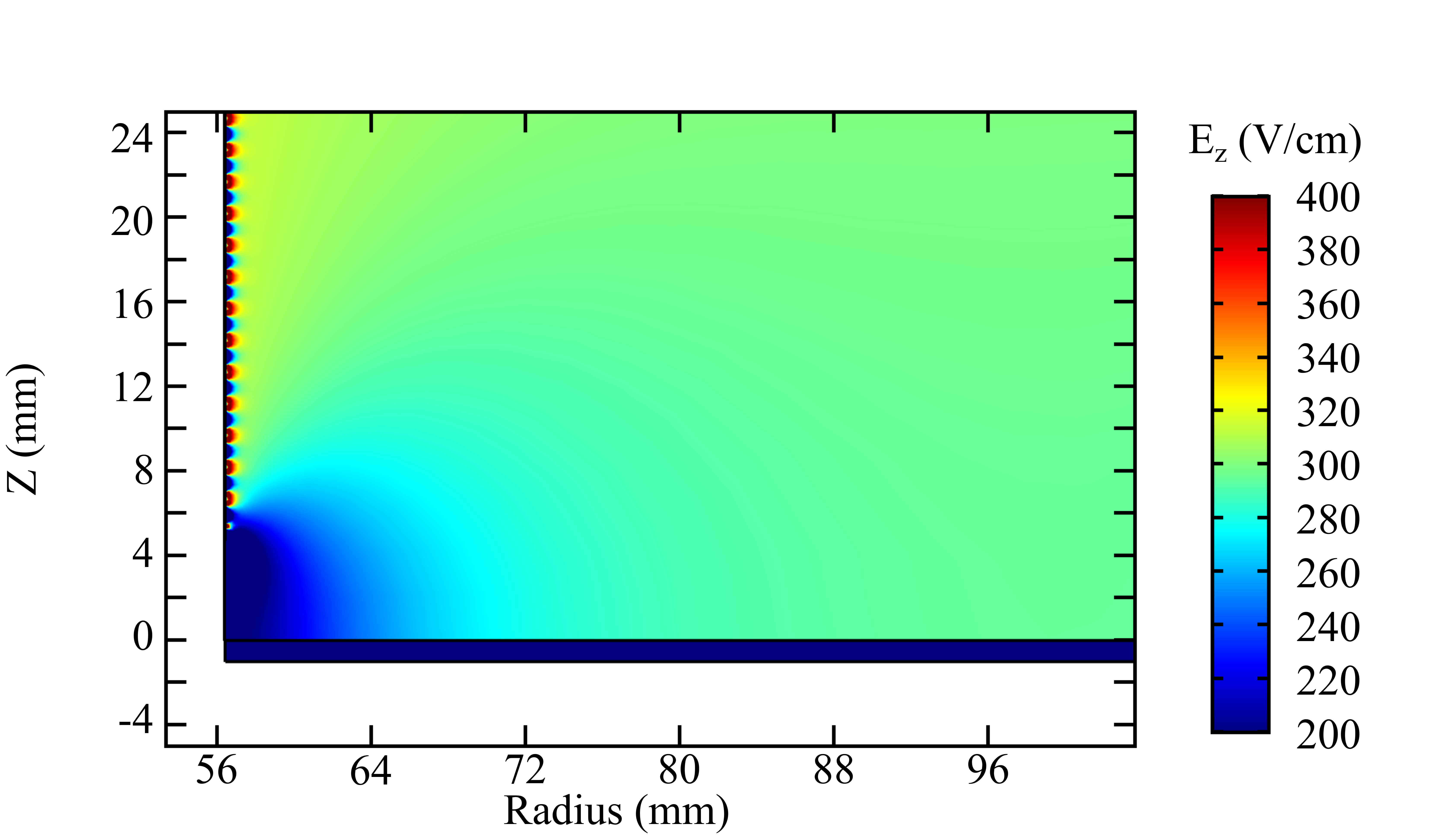}
      \caption{}
      \label{label2}
    \end{subfigure}
    \caption[shortCaption]{ Results of a finite-element simulation of
      the 2-D map of the electric field in the radial (left panel) and
      $z$ (right panel) direction in the region close to the last strip
      for the real set-up. See text for details.}
    \label{fig:fieldcage.ls2}
  \end{center}
\end{figure}

\noindent
The layers of the inner and outer field-cage cylinders facing the gas
volume are composed of \SI{25}{\micro\meter} thick foils of
Kapton and 
FR4, respectively.
Both foils are coated with copper on both sides.\\
Both copper surfaces are patterned in sets of parallel strips aligned
transversally to the beam axis and staggered between the two sides in
order to improve the homogeneity of the field.  This way, 968
concentric cylindrical rings of copper with a width of \SI{1}{\mm} and
a pitch of \SI{1.5}{\mm} cover the walls of the drift volume along the
inner and outer radius, respectively.  By connecting the strips with
resistors a stepwise degrading potential is obtained, from the
cathode value being about \SI{-30}{\kilo\volt} to the potential of the
last strip, close to the first GEM foil, with about
\SI{-3}{\kilo\volt}.
\begin{figure}[tbp]
  \centering
  \includegraphics[width=.98\textwidth]{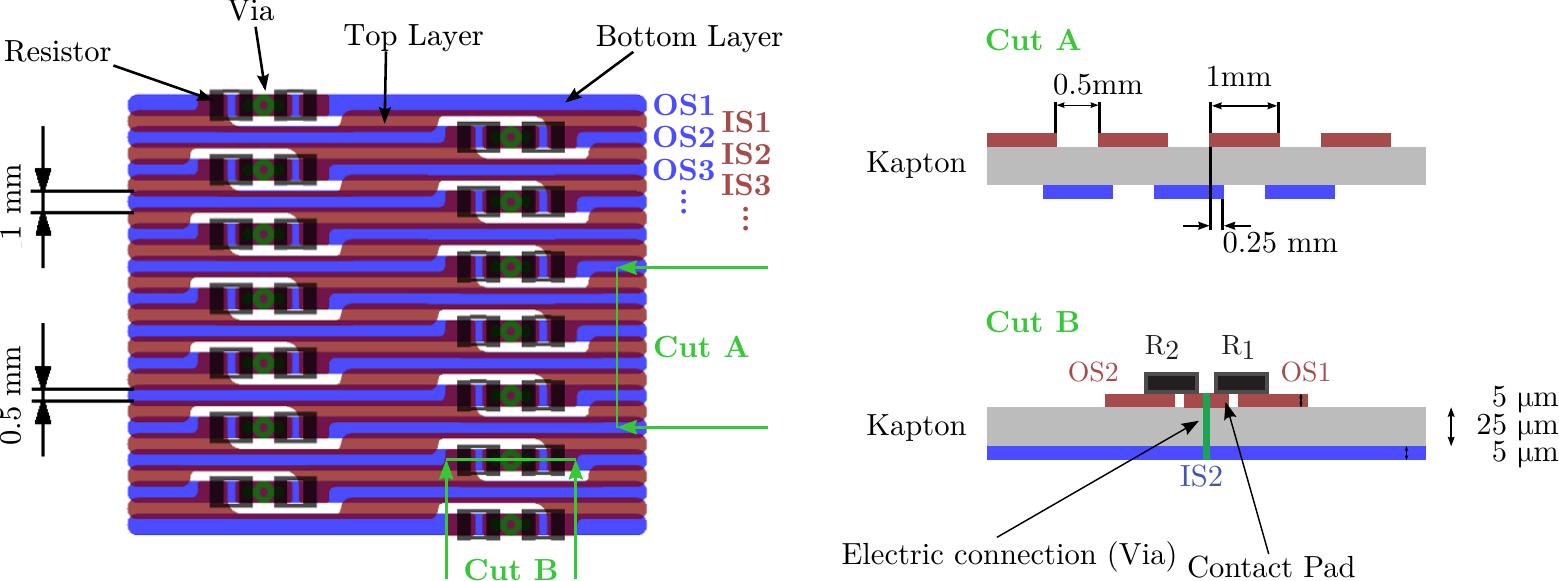}
  \caption{Left: Schematic view of two resistor lines composed by the
    outer (OS) and inner (IS) strips and the SMD 0805 resistors (R1, R2) that connect the strips. Right: Schematic view of the
    transversal cross section of the strip foil. See text for
    details.}
  \label{fig:fieldcage.Conn}
\end{figure}
\Figref{fig:fieldcage.Conn} shows a schematic view of the strip
arrangement.  The top view shown on the left panel of
\figref{fig:fieldcage.Conn} indicates in red the outer strips which
are facing outwards with respect to the gas volume and have special
soldering pads to allow for the connection between neighbouring strips
through resistors.
\begin{figure}[tbp] \centering
  \includegraphics[width=0.7\textwidth]{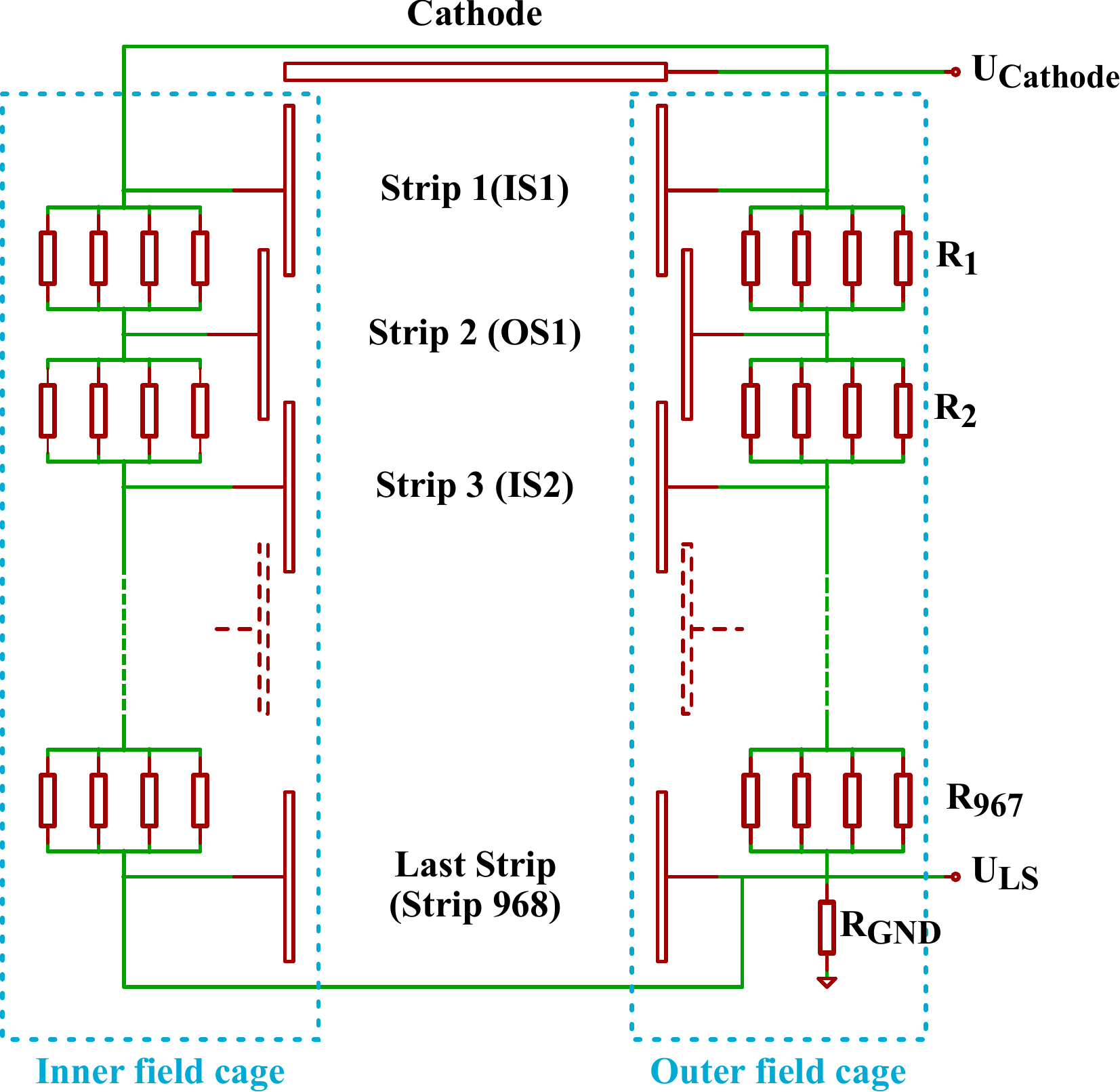}
  \caption{ Sketch of the arrangement of the resistor chain. Only one
    half of the axially symmetric cross-section of the cylindrical barrel
    is shown. See text for details.}
  \label{fig:fieldcage.ResChain}
\end{figure}
\begin{figure}[tbp] \centering
  \includegraphics[width=.5\textwidth]{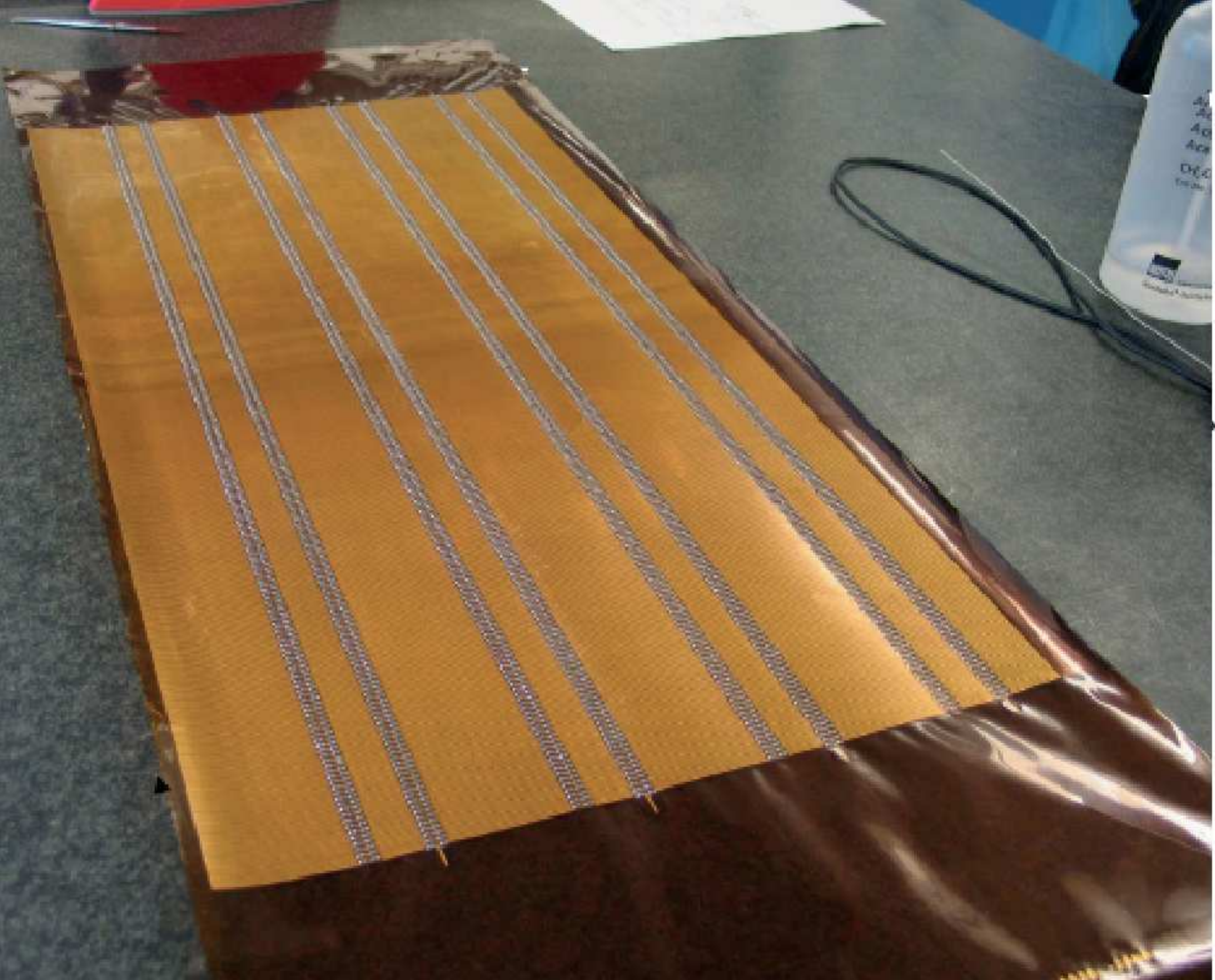}
  \caption{Photograph of the entire strip-line foil for the inner
    field-cage. The 8 rows with the SMD 0805 resistors that connect
    the strips are visible. See text for details.}
  \label{fig:fieldcage.pixFoilAll}
  \includegraphics[width=.9\textwidth]{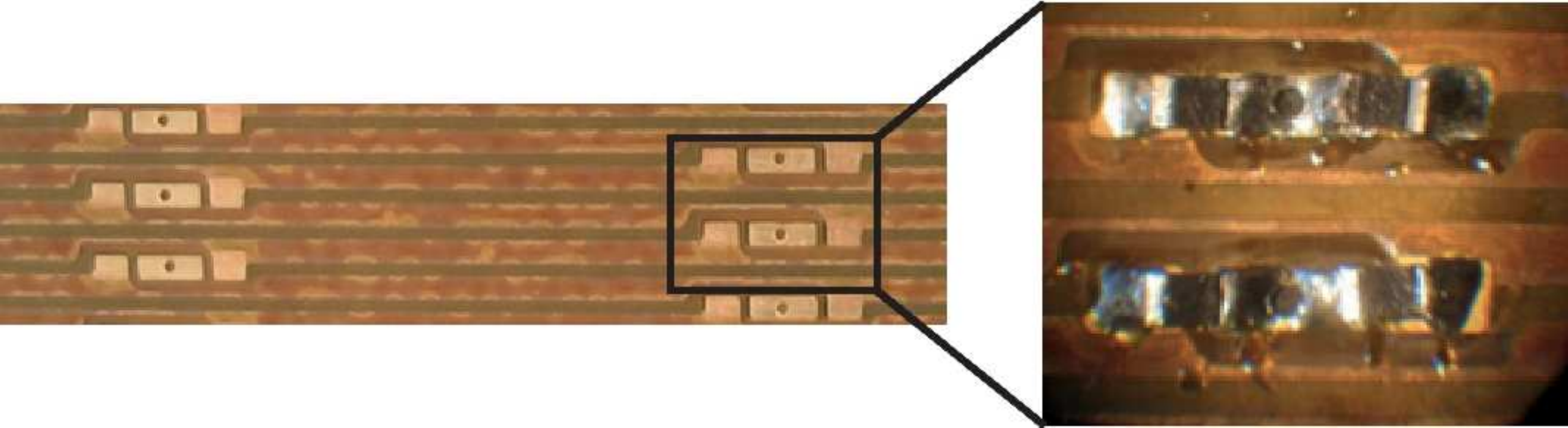}
  \caption{Photograph of the strip-line electrodes. The inlet shows a
    photograph of the enlarged view on the SMD 0805 resistors soldered
    on the respective pads.}
  \label{fig:fieldcage.pixFoil}
\end{figure}
The right panel of \figref{fig:fieldcage.Conn} shows schematic 
cross-sections of the strip foil.  In Cut B the top of the Kapton
foil with the two soldering pads of the neighbouring outer strips OS1
and OS2 are shown together with the contact pad placed between the two
outer strips.  The contact pad is directly connected to the inner
strip IS2.  The electric connection across the \SI{25}{\micro\meter}
thick Kapton foil is represented by the green line and green circle in
the right and left panel of \figref{fig:fieldcage.Conn}, respectively.
By placing the resistor $R_1$ between OS1 and the contact pad one connects OS1 to IS2 and by placing a second resistor $R_2$ one connects IS2 to OS2. \\
This way, all the 968 copper strips can be connected to build a
resistor chain.  Sequential connections are staggered to avoid contact
between the resistors.  \Figref{fig:fieldcage.ResChain} shows the
scheme of the resistor chain.  Only half of the symmetric
cross-section of the barrel is shown.  $U_\mathrm{Cathode}$ and
$U_\mathrm{LS}$ represent the cathode and the last strip
potential, respectively.  One can see that the end of the resistor
chain is connected to a high voltage power supply.  This allows not
only to decouple the field-cage high voltage from the high voltage of
the GEM amplification system but also to vary the potential of the
last strip.
This enables to adjust the electric field between the last strip and the first GEM foil in order to minimize the distortions at the junction between the field-cage and the first GEM foil.\\
Since the power supply used is not able to sink currents, a resistor
$R_\text{GND}$ of \SI{20}{\mega\ohm} connected to ground is placed
between the power supply output and the last strip.
This ensures a positive current output for the high voltage supply.\\
Each of the individual \SI{4.2}{\mega\ohm} 0805 SMD resistors has a
precision specification of \SI{1}{\percent}.  The resistors were
measured by hand and four elements were grouped in parallel such that
the precision of the total resistance per step could be increased to
the 
needed value of \SI{0.1}{\percent}.  The total resistance between two
adjacent strips is then \SI{1.05}{\mega\ohm}.
This results in a total resistance $R_\text{F}$ of
\SI{1014.3}{\mega\ohm} over the resistor chain for the inner and outer
cylinder, respectively. \\ 
Figure~\ref{fig:fieldcage.pixFoilAll} shows a photograph of the entire
strip-foil of the inner field-cage.  One can see the 8 rows of
resistors resulting from the parallel connection and the staggering
connecting the inner to the outer strips.
\Figref{fig:fieldcage.pixFoil} shows an enlarged version of the
previous image where the outer strips and the soldered resistors are
visible.  Due to the limitation of the maximal width of the FR4 base
material to \SI{600}{\mm}, the outer strip-foil is assembled by
splicing two FR4 foils together where the splicing runs in parallel to
the strips of the field-cage.  Since the two spliced foils have to be
electrically connected a copper bridge is soldered in between these
two foils.  This interconnection can be seen in
\Figref{fig:fieldcage.copperbridge}.
A finite-element simulation of the GEM-TPC field-cage was carried out
to investigate the homogeneity of the drift field.  A two dimensional
geometry has been used exploiting the rotational symmetry of the
GEM-TPC and different settings of the potentials have been considered
to emulate the ideal and the real setup. 
The resulting field distortions within the drift volume were 
evaluated.\\ 
Figures \ref{fig:fieldcage.cathode1}-\ref{fig:fieldcage.ls2} show the
simulated two-dimensional maps of the electric field for the ideal and
real setups, respectively.  The color code shows the values of the
radial and the 
longitudinal electric field components, respectively, for the
different settings. 
The ideal set-up corresponds to the mechanical specifications given
so-far and to the following voltage settings for the cathode, last
strip and top side of the first GEM foil:
$U_\mathrm{CATHODE}=\SI{-25802.4}{\volt}$,
$U_{\mathrm{LS}}=\SI{-3612.6}{\volt}$ and
$U_{\mathrm{GEM1}}=\SI{-3458.6}{\volt}$.  These settings lead to a
longitudinal electric field of \SI{309.6}{\volt\per\cm} in the center
of the drift volume.  \Figref{fig:fieldcage.cathode1} shows the radial
and longitudinal components of the electric field close to the cathode
for the ideal case; the left panel shows also a schematic
representation of the first three strips close to the cathode.
\Figref{fig:fieldcage.ls1} shows the two field components close to the
last strip for the ideal settings.
Despite these expected local variations of the field close to the
field-defining structures, the required field homogeneity of
$\EE*{-4}$ is reached at a maximal distance of $2.2\,\mm$ from the
field cage walls, corresponding to $95.6\%$ of the total gas volume,
provided that the mechanical specifications are met.    


\begin{figure}[tbp]
  \begin{center}
    \begin{subfigure}[b]{.49\textwidth}
      \centering
      \includegraphics[width=\textwidth]{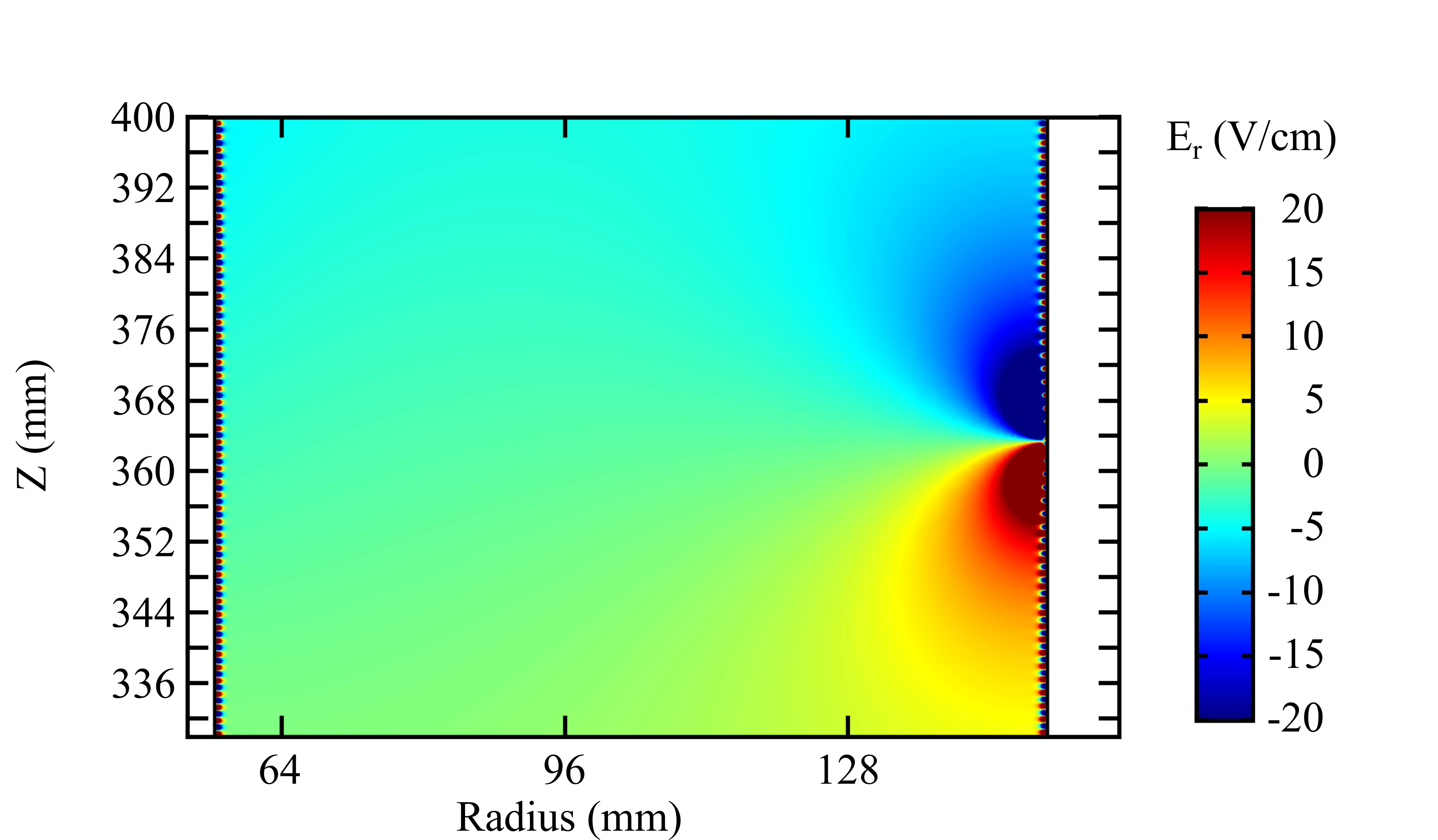}
      \caption{}
      \label{label1}
    \end{subfigure}
    \begin{subfigure}[b]{.49\textwidth}
      \centering
      \includegraphics[width=\textwidth]{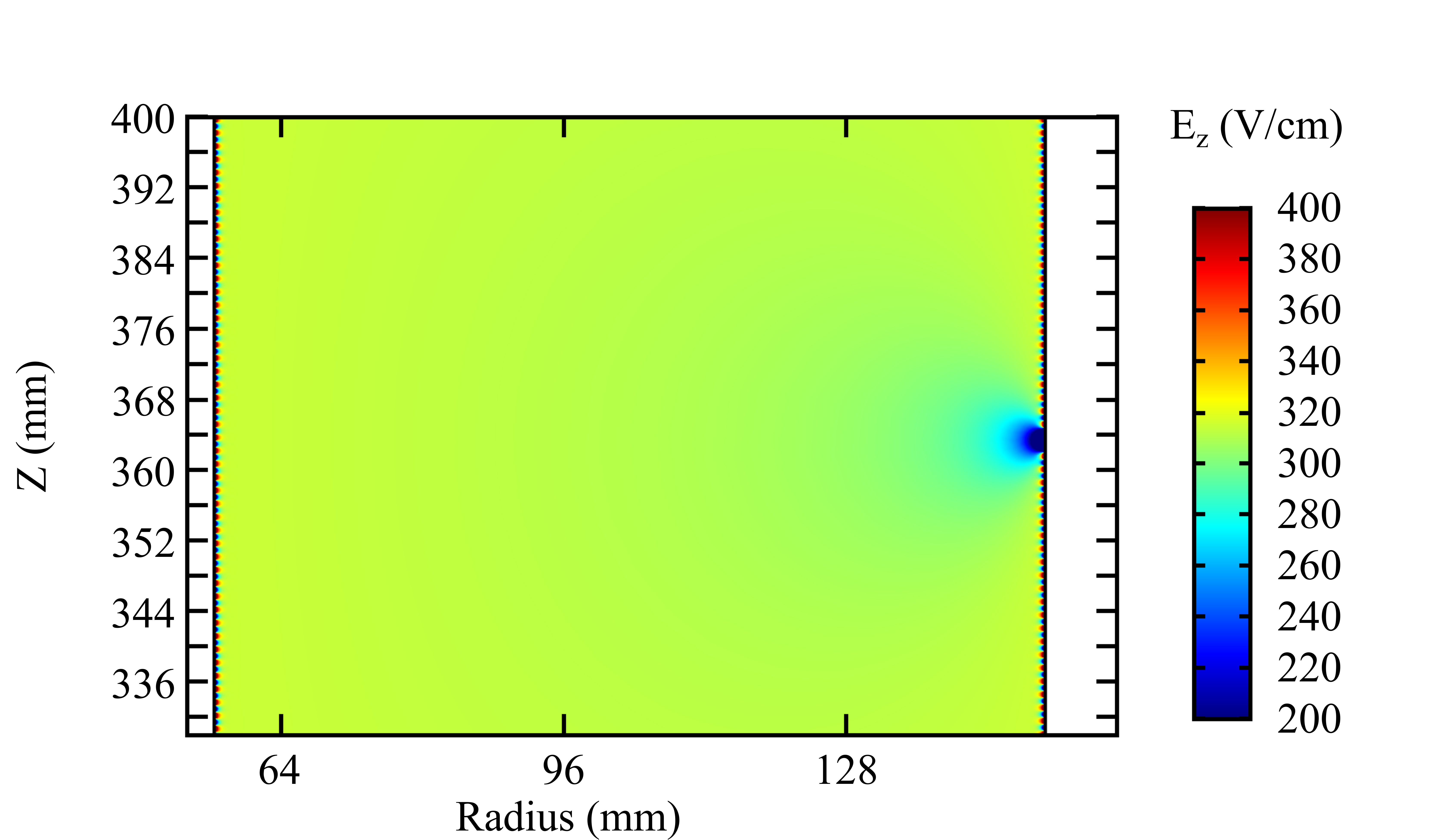}
      \caption{}
      \label{label2}
    \end{subfigure}
    \caption[shortCaption]{ Results of a finite-element simulation of
      the 2-D map of electric field in the radial (left panel) and $z$
      (right panel) direction in the region close to the strip foil
      splicing for the ideal set-up. See text for details.}
    \label{fig:fieldcage.splicing}
  \end{center}
\end{figure}

\begin{figure}[tbp]
  \begin{center}
    \subcaptionbox{}{\includegraphics[width=.8\textwidth]{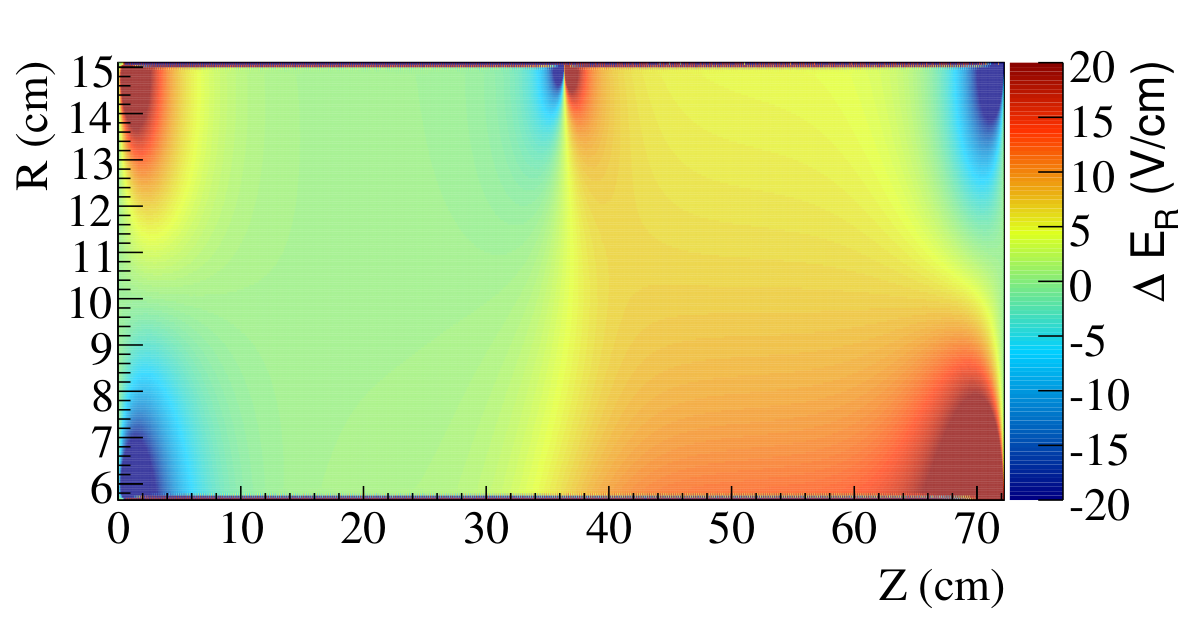}}\par\medskip
    \subcaptionbox{}{\includegraphics[width=.8\textwidth]{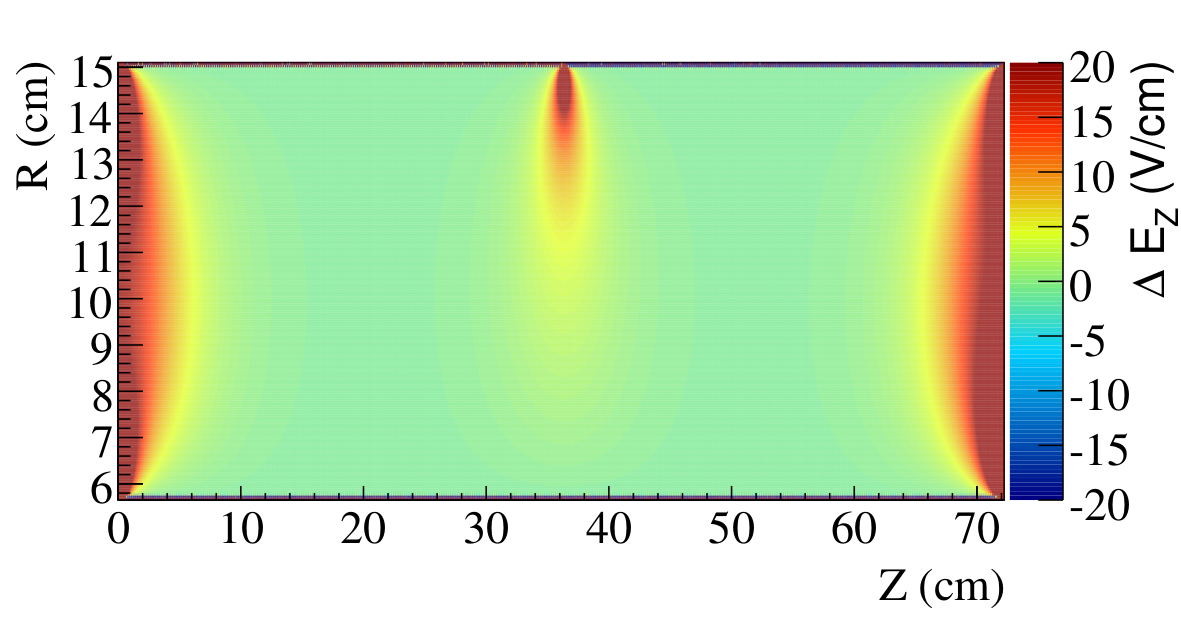}}
    \caption[shortCaption]{Deviation from the nominal field in radial
      (a) and longitudinal (b) direction as a function of the radius
      and $z$ position. The nominal radial field is
      \SI{0}{\volt\per\cm}, the nominal longitudinal field is
      \SI{309.6}{\volt\per\cm}.}
    \label{fig:fieldcage.dEz}
  \end{center}
\end{figure}

The connection of the first strip to the cathode is realised via
conductive glue; during the construction of the field-cage, however, the
first three strips on the outer and the first four strips on the inner
cage were short-circuited by accident and are therefore 
at the same potential as
the cathode.  This configuration modifies the field close
to the cathode surface as shown in \figref{fig:fieldcage.cathode2},
where the same quantities as in \figref{fig:fieldcage.cathode1} are
presented for the real setup.  Moreover, during the detector
commissioning too low a voltage was set to the last strip:
$U_{\mathrm{CATHODE}}=\SI{-25940}{\volt}$,
$U_{\mathrm{LS}}=\SI{-3528.6}{\volt}$ and
$U_{\mathrm{GEM1}}=\SI{-3458.6}{\volt}$.  These settings lead to a
longitudinal electric field of \SI{309.9}{\volt\per\cm} in the center
of the drift volume, which is very similar to the ideal setting, but
also 
to an additional radial component of the electric field close to the
last strip and the first GEM foil as one can see in
\figref{fig:fieldcage.ls2}.
The copper bridge used to connect the two spliced foils of the outer
strip foil introduces further field distortions as shown in
\Figref{fig:fieldcage.splicing}.  There panels a) and b) show the
introduced distortions in the radial component and along the
$z$-direction of the electric field, respectively.  These distortions
are calculated with the ideal setting mentioned above.  The resulting
distortions at the copper bridge for the real setting are not shown as
they do not differ from the ideal setting.
The deviation of the drift field obtained for the real setup are
summarized in \figref{fig:fieldcage.dEz}.  Panel a) of
\figref{fig:fieldcage.dEz} shows the deviation from zero of the radial
component of the electric field as a function of the radius and the $z$
position and panel b) presents the deviation of the $z$-component of the
electric field from the nominal value as a function of the same
spatial coordinates.

%% file: mediaflange/mediaflange.tex
\section{Media Flange}
\label{sec:mediaflange}
\begin{figure}[ht]
  \begin{center} \centering
    \includegraphics[width=0.65\textwidth]{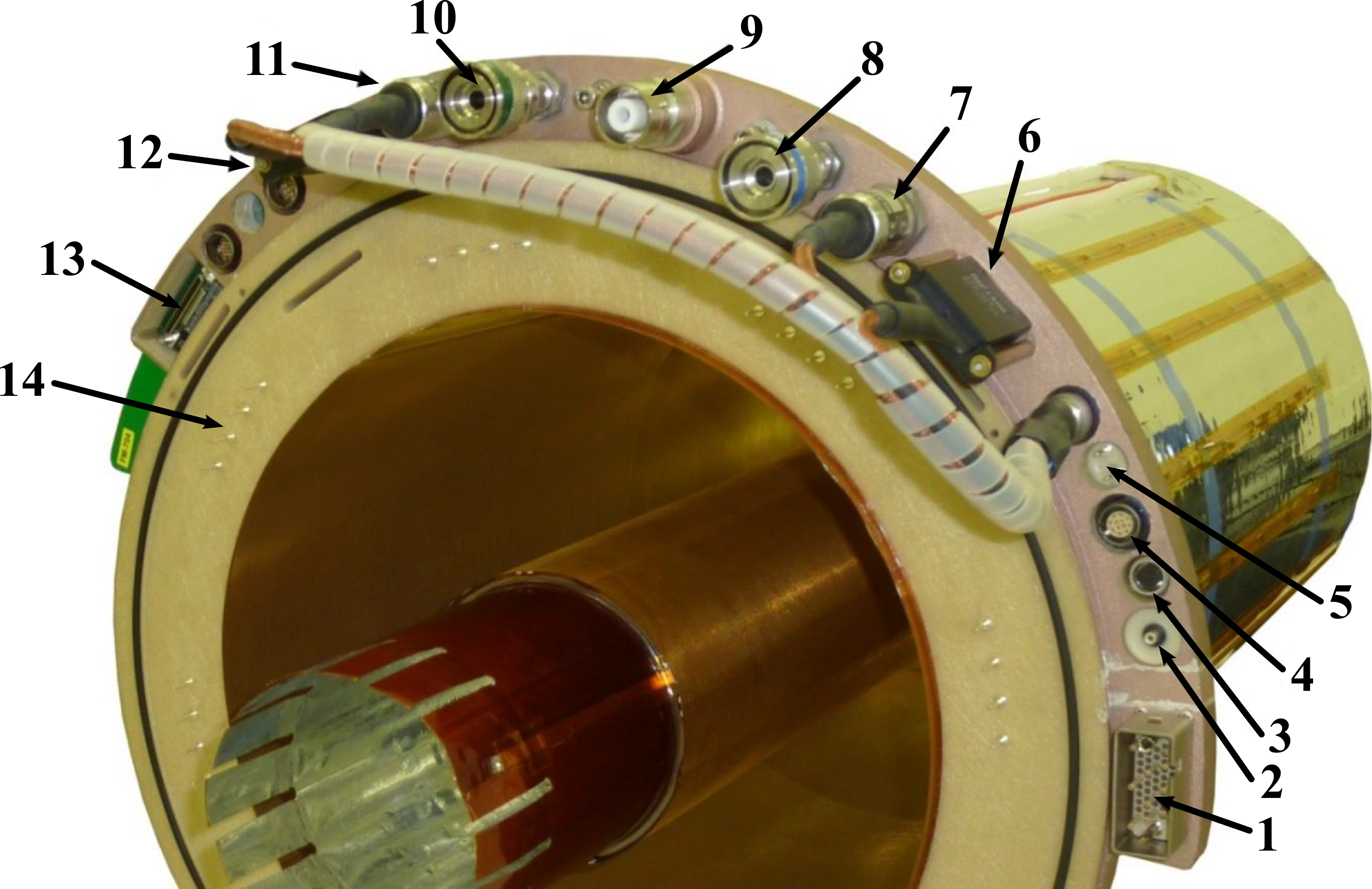}
    \caption {Photograph of the media flange glued to the field-cage,
      with (1) high-voltage connection for GEM foils, (2) pin for last
      strip high voltage, (3) connector for field-cage temperature
      sensors, (4) connector for gas flow and pressure sensors, (5)
      pins for ground connection, (6,12) mass flow meters, (7,11)
      pressure sensors, (8) gas outlet, (9) cathode high-voltage
      connection, (10) gas inlet, (13) connector for pad-plane
      temperature sensors, (14) pins for GEM-foil high voltage.}
    \label{fig:mediaflange.photo}
  \end{center}
\end{figure}

\begin{figure}[ht]
  \sbox\twosubbox{%
    \resizebox{\dimexpr.95\textwidth}{!}{%
      \includegraphics[height=3cm]{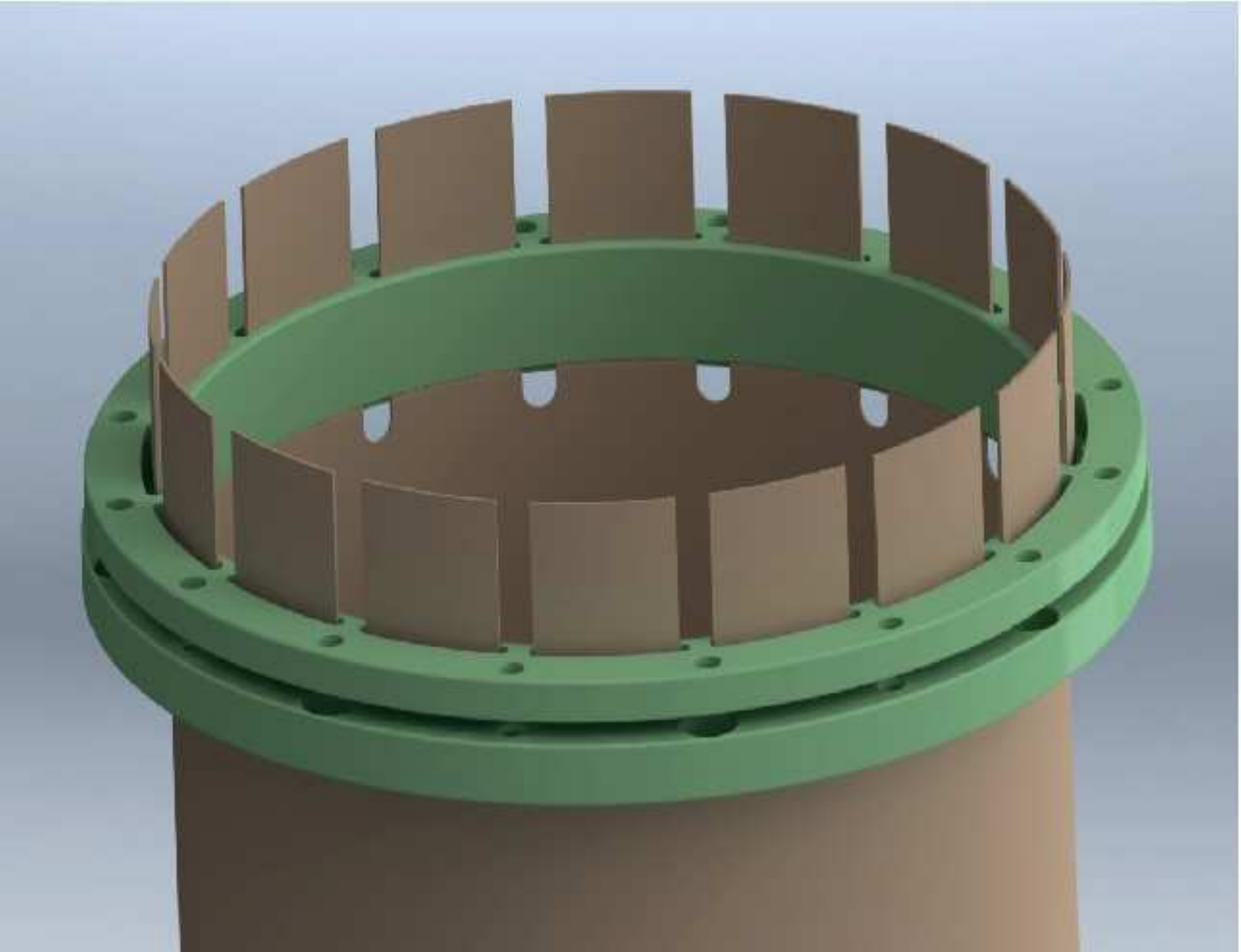}%
      \includegraphics[height=3cm]{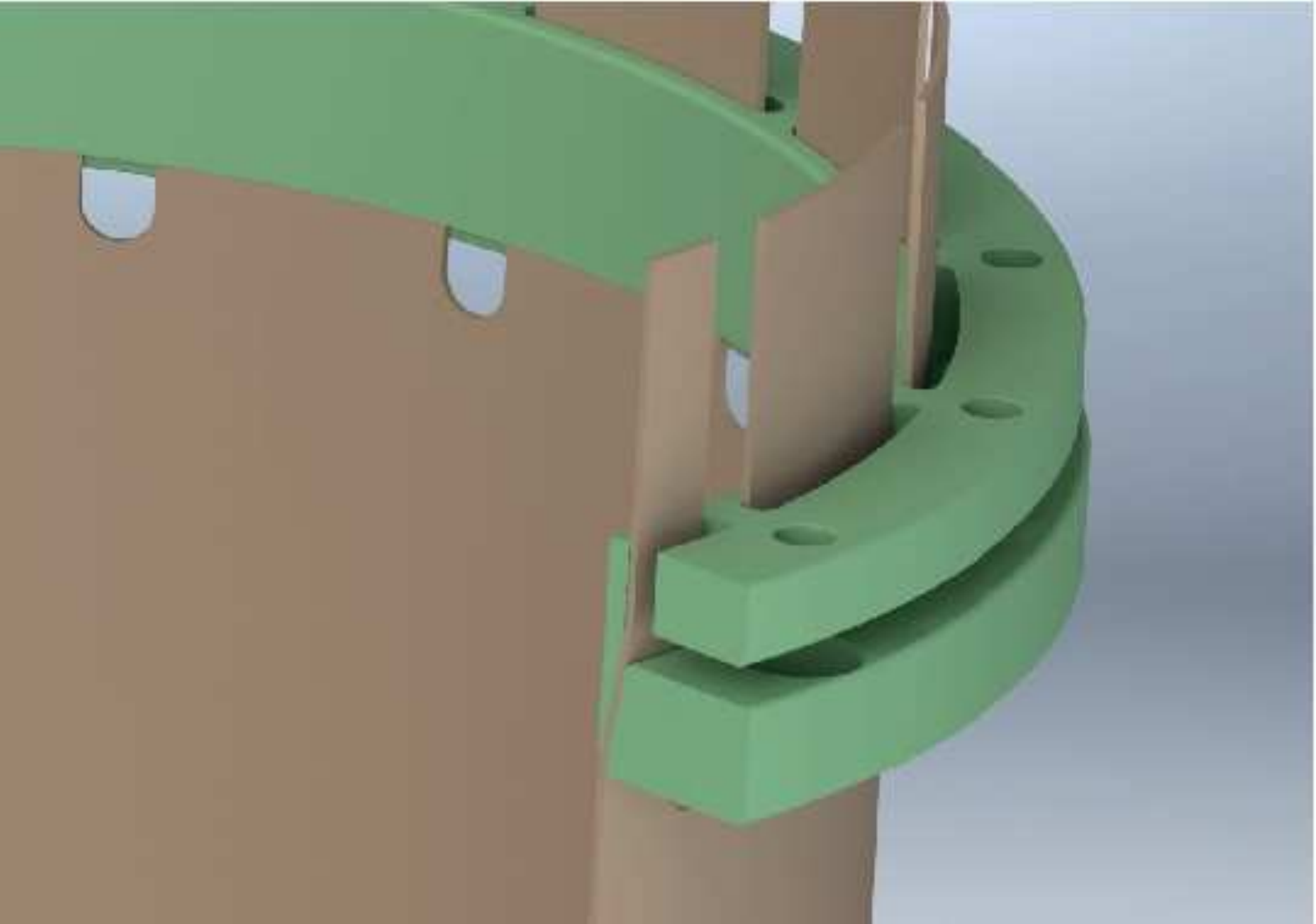}%
    }%
  }%
  \setlength{\twosubht}{\ht\twosubbox}
	
	\begin{center}
          \subcaptionbox{}{
            \includegraphics[height=\twosubht]{./figures/mediaflange/Clap1_1.pdf}}
          \subcaptionbox{}{
            \includegraphics[height=\twosubht]{./figures/mediaflange/Clap2_1.pdf}}
	\end{center}
	\caption{(a) Schematic view of the support ring used to
          fix the inner field-cage flaps to the flange. (b)
          Details of how the flaps are inserted in the slotted holes
          of the support ring.}
	\label{fig:fieldcage.flaps}
\end{figure}

\noindent The media flange is a rigid support flange which is glued to
the field-cage vessel.  It provides mechanical stability and serves as
a mounting structure for the whole detector, is the main support for
the GEM and readout flanges and provides the interfaces for all
external supplies like gas, high voltage and sensors for measuring gas
flow, temperature and pressure.  \Figref{fig:mediaflange.photo}
displays a photo of the media flange glued onto the field-cage vessel.
It is made of two pieces glued together in order to create internal
channels which are used for the distribution of the gas. 
The main material used in the fabrication of the media flange is
Stesalit 1011.  The gas in- and outlets are realized by quick coupling
connectors, type Swagelok Instrumentation Quick-Connect QC4.  The high
voltage for the drift cathode is supplied via a Radiall socket, type
R341018.  The high voltages for the GEM electrodes (6 channels) and
the last strip of the field-cage are supplied via a multi-pin REDEL
socket, type SLG.H51.  The GEM high voltage channels are distributed
through the media flange via pins to the GEM flange (see
\secref{sec:ro:gem}).  The last strip channel is connected within the
media flange to the outer field-cage and via an external cable to the
inner field-cage.
An additional cable connects the inner field-cage via the resistor
($R_\text{GND}$ in \figref{fig:fieldcage.ResChain}) to ground.\\ 
To measure the pressure at the gas inlet and outlet, two compact
pressure transducers, type Measurement Specialties M5141, are mounted
on the media flange.  They are suited for measuring the pressure in
the range from \SIrange{0}{3.5}{\bar}.  Additionally, there are two
Sensirion ASF 1400 mass flow meters, installed in series with the
inlet and outlet gas channel.
All sensor signals are read out by the GEM-TPC Slow Control system, described in \secref{sec:slowcontrol}.\\
In order to fix the inner field-cage to the readout flange, it ends on
the anode side with Kapton flaps that are used to connect it to a
two-piece support ring.  \Figref{fig:fieldcage.flaps} shows how the
flaps are inserted into the slotted holes of the upper support ring.
The two parts of the support ring are screwed together clamping the
flaps in between them.  Finally, the whole structure is screwed to the
mounting system for the FE cards.  By adjusting the longitudinal
position of the ring the tension and therefore the rigidity of the
inner field-cage can be controlled.

%% file: readout/readout.tex
\section{The GEM Amplification Stage}
\label{sec:ro:gem}
\subsection{GEM Design, Framing and Implementation}
\begin{figure}[tbp]
  \sbox\twosubbox{%
    \resizebox{\dimexpr.95\textwidth}{!}{%
      \includegraphics[height=3cm]{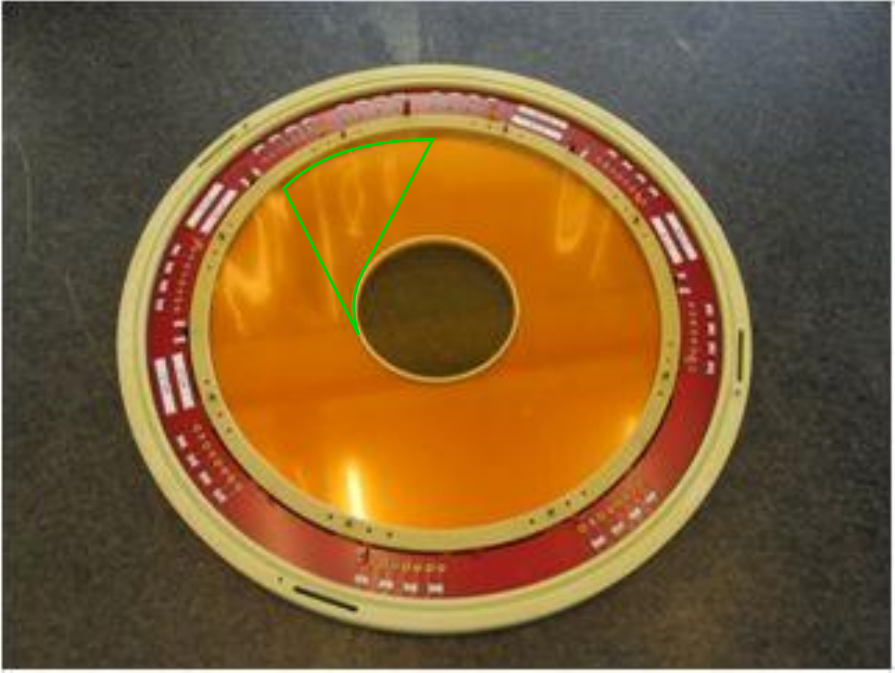}%
      \includegraphics[height=3cm]{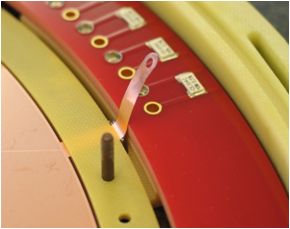}
    }%
  }%
  \setlength{\twosubht}{\ht\twosubbox}
  \begin{center}
    \subcaptionbox{}{\includegraphics[height=\twosubht]{./figures/readout/gem-flange.pdf}}
    \subcaptionbox{}{\includegraphics[height=\twosubht]{./figures/readout/gem-flange-flap.jpg}}
  \end{center}
  \caption{(a) Photograph of a framed GEM in the GEM flange. The green
    contour shows one of the eight sectors. (b) HV flap that connects
    the GEM foil to the HV distribution board (red PCB in the
    figure).}
  \label{fig:ro:gem_flange_flap}
\end{figure}

\noindent The GEMs are designed to fit to the dimensions of the
GEM-TPC, so that the inner and outer radii of the GEM active area
match the inner and outer 
radius of the TPC field cage, respectively. This results in an active
area of 
the GEM foil of 
around \SI{630}{cm^2}.  One side of each GEM foil is divided into eight
sectors to decrease the capacitance of each sector reducing
significantly the discharge probability as well as the amount of
stored charge which has to be dissipated in the case of a discharge
\cite{Bachmann:01e}.  This segmentation furthermore allows an
operation of the TPC even in case of several non-operational sectors
per foil. The eight sectors are shaped like aperture blades (see
\figref{fig:ro:gem_flange_flap} a) and are separated by
\SI{400}{\micro\meter} thick insulating borders where the copper layer
was etched away.  Each sector has a flap on the outer radius to
connect it to the high voltage distribution board located inside the
GEM flange (see \figref{fig:ro:gem_flange_flap} b).  The high
voltage distribution board also contains \SI{10}{\mega\ohm}
bias resistors for each sector.
\begin{figure}[tbp]
  \begin{center}
    \includegraphics[width=.6\textwidth]{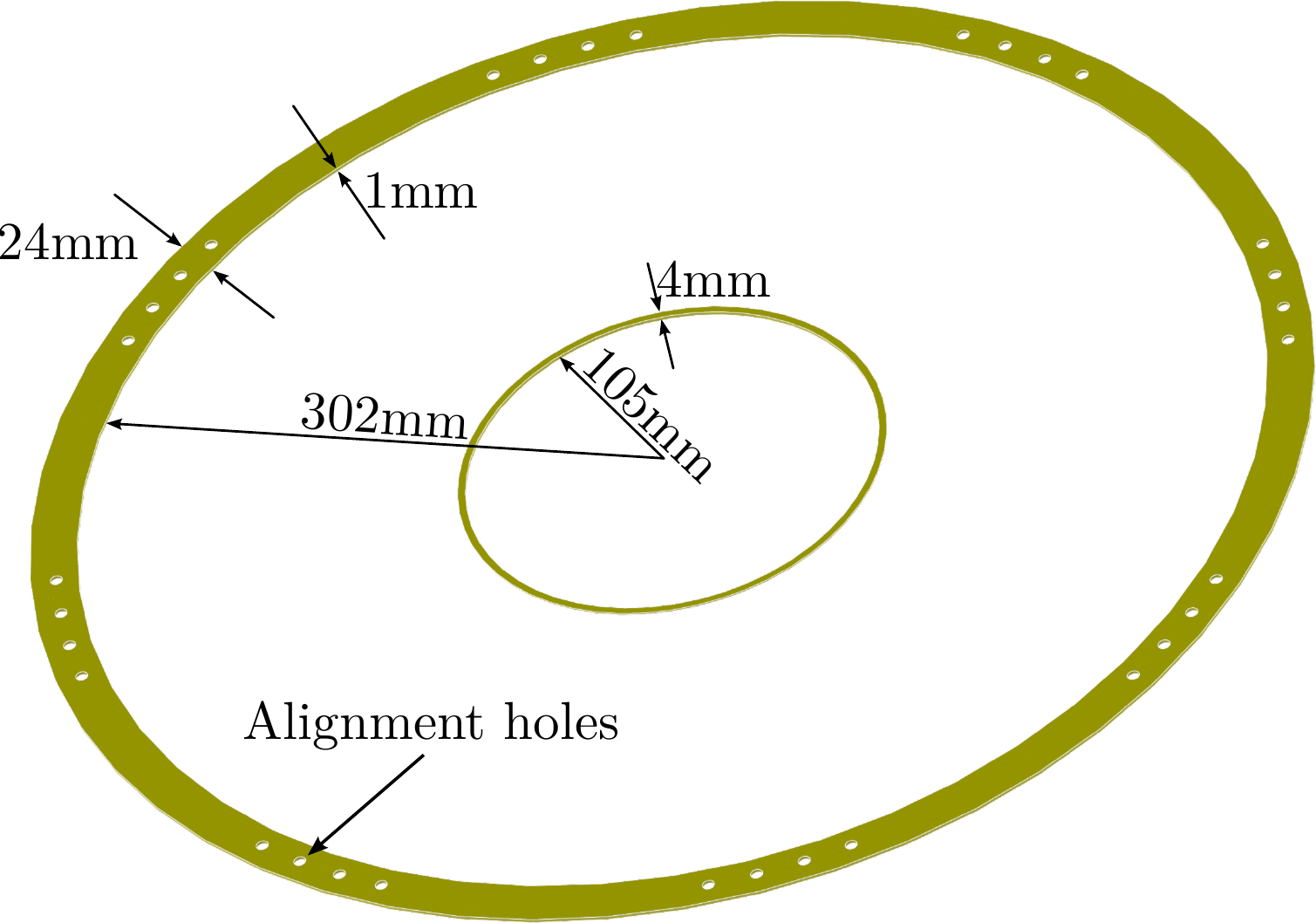}
    \caption{The inner and outer frame for the GEM foils with their
      dimensions and the alignment holes for installation inside the
      GEM flange.}
    \label{pic:gemframes}
  \end{center}
\end{figure}
\begin{figure}[tbp]
  \begin{center}
    \includegraphics[width=.8\textwidth]{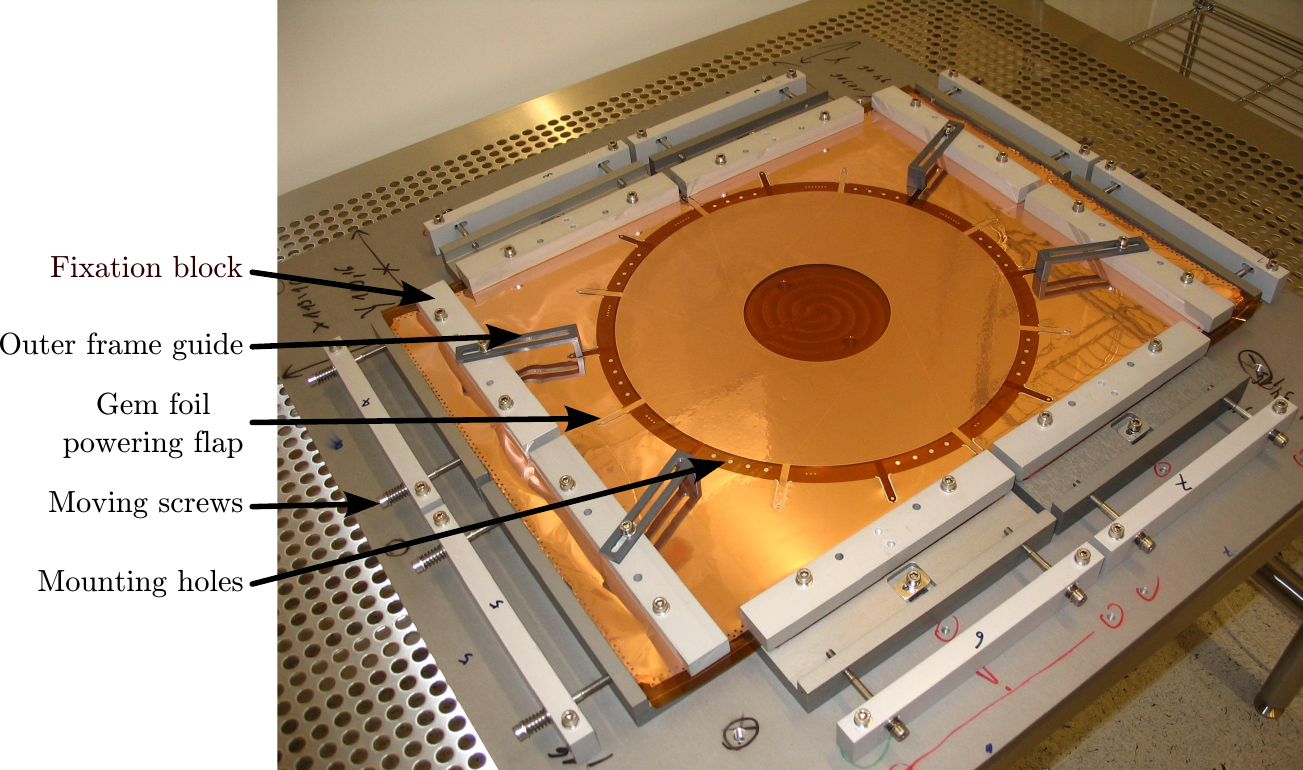}
    \caption{A GEM foil clamped inside the stretching tool during
      framing. One of the eight sectors is marked by a green line.}
    \label{fig:ro:framed_gem}
  \end{center}
\end{figure}
As the distance between two GEM foils or a GEM foil and the read-out
anode determines the field strength, a well defined and constant
spacing throughout the whole foil has to be maintained.  This is
achieved by gluing \SI{1}{\mm} thick FR4 frames onto each side of a
foil while applying tension to the foil.
The frames and their dimensions are depicted in \figref{pic:gemframes}.\\
The foil is stretched by applying tension on the four sides of the
uncut rectangular foil. To this end, the foil edges are clamped
between two fixation blocks 
where the lower one is movable and has a silicone rubber inlay to
provide the necessary friction. The upper block fixes the edge to the
lower block. By
moving the fixation blocks with the help of screws, the foil can be
stretched until no wrinkles are left.   \Figref{fig:ro:framed_gem}
shows an uncut foil mounted 
inside the stretching tool and stretched via the fixation blocks.

When the GEM foil is completely 
flat the inner and outer FR4 rings are glued on the upper side of the
GEM foil with Araldite
AW106 (hardener: 953U).
Alignment holes are distributed along the outer circumference of the
GEM foil and on the outer FR4 frame.  The alignment holes on the outer
GEM frame are visible in \figref{pic:gemframes}.  The outer ring is
aligned by placing pins through the alignment holes of the frames and
the GEM foils into holes
on the metal plate underneath the foil.
A special tool employing the same alignment pins as for the outer ring
is used for the deployment of the inner ring.  This tool is basically
a metal plate with concentric groves for the two frames which holds
the inner frame with the help of a small air pump.
This metal plate additionally serves as a weight to apply pressure on
both rings while curing the glue.  
The glue is cured for about ten hours at a temperature of around
\SI{60}{\degree}. 

In the next step the foil is flipped and the whole procedure is
repeated for the other side of the GEM foil.  After the stretching and
framing procedure is completed the unnecessary material around the 
outer frame is removed.  The framed GEM foils are then permanently
under tension and flat and can be assembled into a stack fixed to the
GEM flange.  Before, during and after the gluing high voltage tests of
the 
GEM foils are performed to ensure the functionality and high voltage
stability.  To perform these tests, the foil is set under nitrogen
atmosphere and voltages up to \SI{550}{\volt} are applied to each
sector via a \SI{10}{\mega\ohm} bias resistor, while all remaing
sectors as well as the un-sectorized side are grounded.
The ramping scheme and the dwell time for each HV step are listed in
\tabref{tab:ro:hvtest}. 
\begin{table}[tbp]
  \begin{center}
    \begin{tabular}{S|c}
      \hline\hline
      {Voltage (\si{\volt})}	& Dwell time			\\
      \hline
      100					& Until currents are stable	\\ 
      300					& Until currents are stable	\\
      400					& Until currents are stable	\\ 
      450					& \SI{1}{\minute} 			\\
      500					& \SI{3}{\minute} 			\\
      550					& \SI{3}{\minute} 			\\
      0					& Until currents are stable	\\ 
      550					& \SI{3}{\minute}			\\
      \hline
      \hline
    \end{tabular}
    \caption[shortcaption]{High voltage test ramping scheme.}
    \label{tab:ro:hvtest}
  \end{center}
\end{table}
In \figref{fig:ro:gem_flange_flap} one can see a framed GEM mounted in
the GEM flange.

\begin{figure}[tbp]
  \centering
  \includegraphics[width=0.8\textwidth]{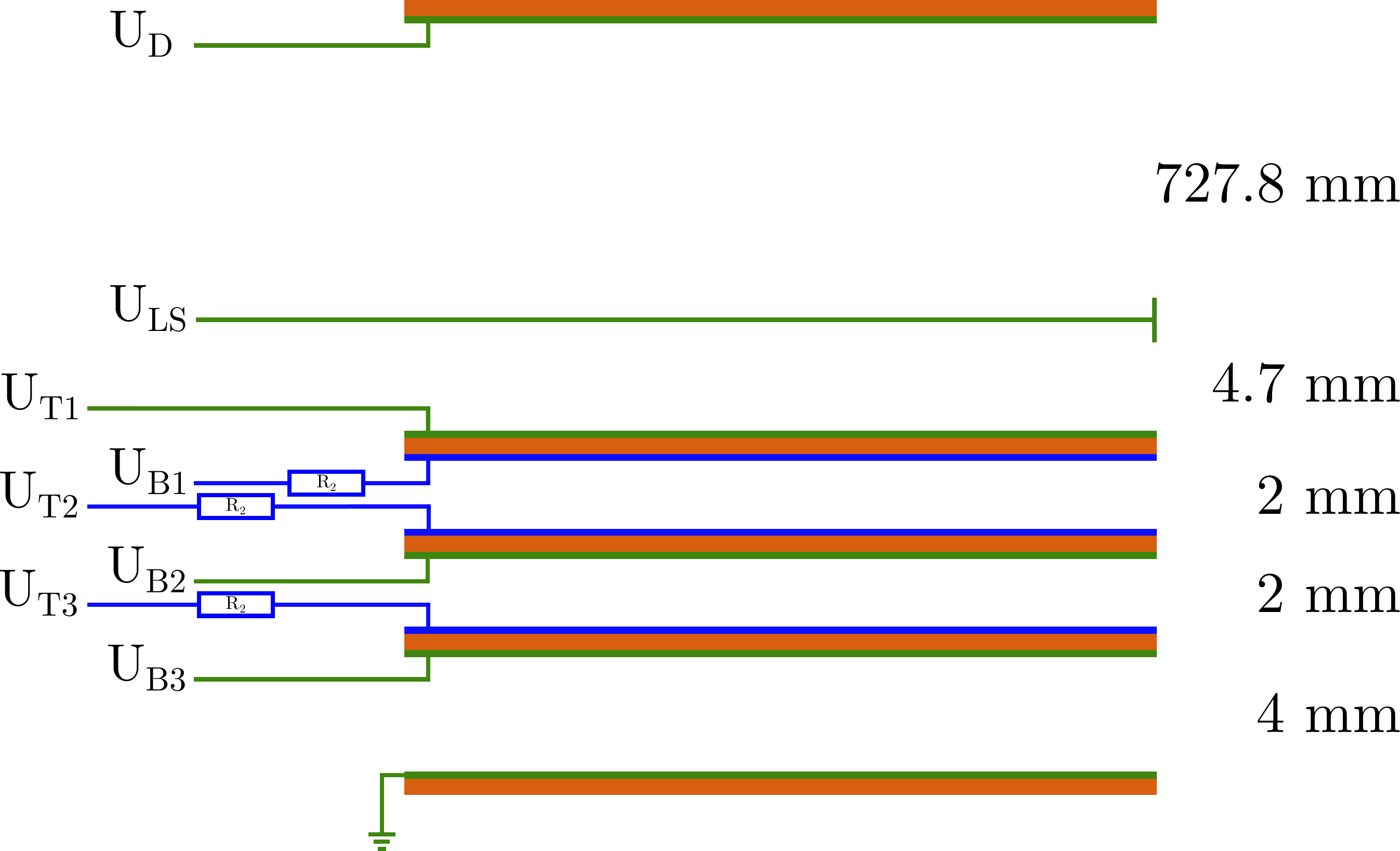}
  \caption{Nominal HV scheme of the GEM-TPC
    where $U_\mathrm{D}$ is the drift voltage, $U_{\mathrm{LS}}$ is
    the voltage of the last strip before the GEM foils and
    $U_{\mathrm{T}i}$ and $U_{\mathrm{B}i}$ are the top and bottom voltages
    of GEM foil $i$, respectively. The segmented sides of the foils are
    depicted in blue, the unsegmented in green. The spacings between
    the foils are indicated as well.}
  \label{pic:gemstack}
\end{figure}

The GEM flange can hold up to four GEM foils which are rotated with
respect to each other to allow for a proper mounting of the flaps to
the HV distribution.  The positioning of the framed GEM foils is
realized with alignment pins.  Finally, the flaps of each sector are
connected to their corresponding high voltage pin on the distribution
board, visible in red in \figref{fig:ro:gem_flange_flap}, which is
also mounted inside the GEM flange.  The unsegmented side of a GEM
foil is connected with just one flap to the high voltage distribution
board.  Pins connect the high voltage distribution board to the media
flange where a Redel connector (\figref{fig:mediaflange.photo}(1)) is
placed to connect the \gt to the high voltage module.  The frames with
their thickness of \SI{1}{\mm} define the spacing in-between two GEM
foils as shown in \figref{pic:gemstack}.

\subsection{GEM HV scheme}
\label{sec:ro:gem-HV-scheme}
\noindent
For the sake of flexibility, we chose an HV supply for the GEM stack
which includes a separate HV channel for each side of a GEM foil.
On the one hand, operation with a resistor chain is known to be more
reliable and 
stable because it avoids by design that large potential differences
build up between electrodes in case of a trip of a single sector or in
case of different time constants of discharging the foils.
On the other hand, it is not possible in this configuration to test
different voltage settings, which was an important asset for the
present detector. 
 
\begin{figure}[tbp]
  \centering
  \includegraphics[width=0.75\textwidth]{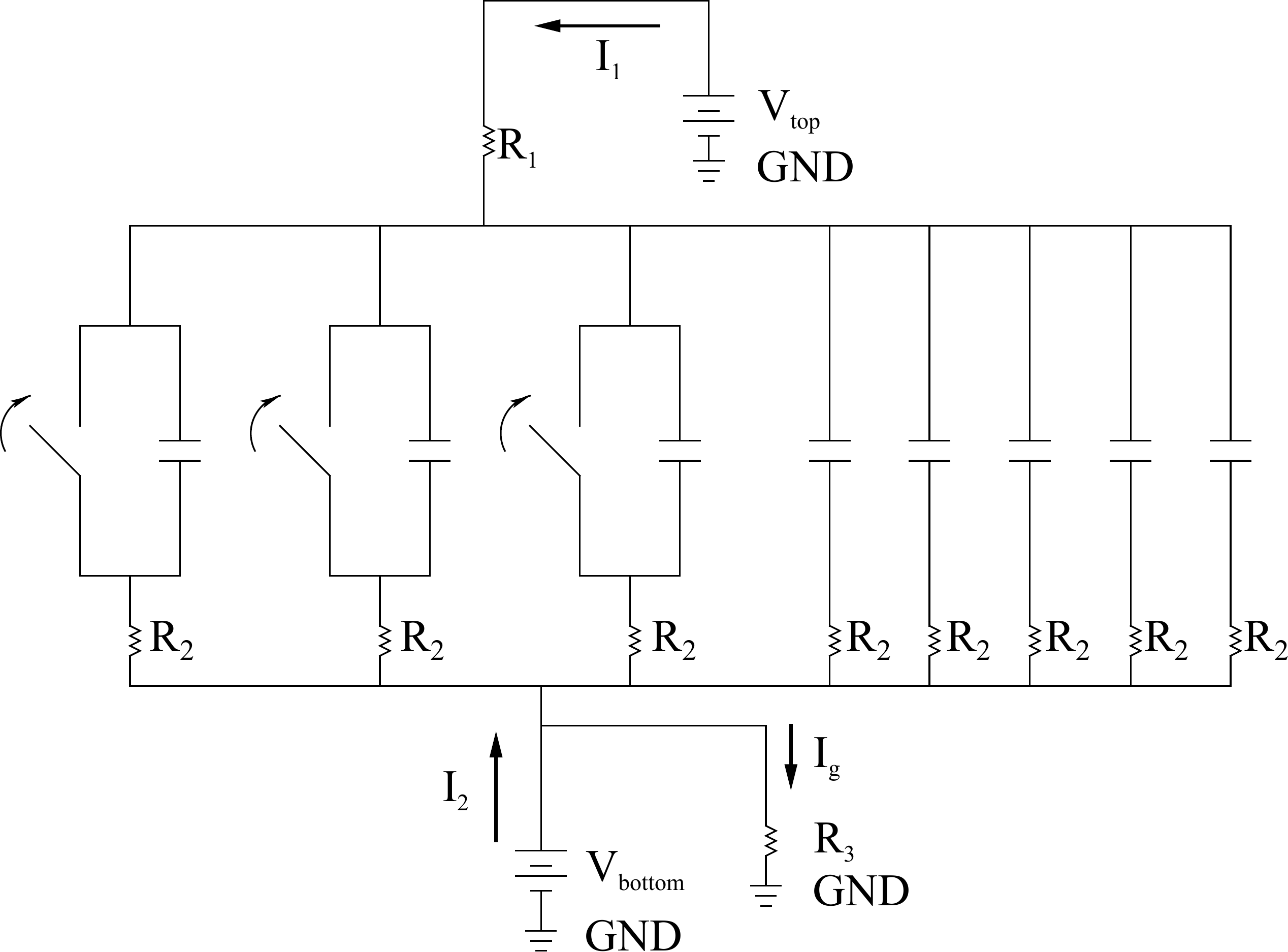}
  \caption{An equivalent circuit for a GEM foil. The switches
    illustrate the possibility of a shortcut sector due to a
    discharge.}
  \label{pic:gem-HV-scheme}
\end{figure}

In the nominal configuration, the three GEM foils in a stack are
mounted with the sectorized side facing the drift region
\cite{Altunbas:02a}.  Each sector is powered through a separate bias
resistor ($R_2$=\SI{10}{\mega\ohm},
\figref{pic:gem-HV-scheme}) in order to limit the current in case of a
discharge.  The upper limit of $R_2$ is determined by the
fact that the voltage drop across the GEM due to charge produced in
the detector should remain small, such that the gain is not affected
by the particle rate.
The unsectorized side of a foil does not have a bias resistor, but is
connected directly to the power supply.  This configuration has the
virtue of fixing the potential at the bottom side of each GEM to its
nominal value in case of a discharge, thus avoiding excessive potential
differences to the next stage or to the readout anode and effectively
preventing discharges from propagating to the next stage.  In
addition, such a scheme 
allows for an operation of the detector even in case of permanent
short circuits in several sectors.  For a TPC, however, such a scheme
is not directly applicable, because, at least for the top GEM directly
facing the drift volume, the homogeneity of the drift field would be
severely distorted by one sector at a potential which is different
from the others.  Therefore,
at least the top GEM has to be mounted with the sectorized side facing
the anode.  The other foils should be mounted in the standard
orientation, as shown in \figref{pic:gemstack}.

Due to an error in the design, however, a flipping of the GEM foils is
not possible for the detector described in this paper.  Therefore, all
foils have to be mounted with the sectorized side facing the anode.
To limit the increase of the potential on the lower side of the GEM in
case of a discharge, a \SI{10}{\mega\ohm} bias resistor is mounted
also for the unsectorized side ($R_1$).  The resulting
configuration, which is shown schematically in
\figref{pic:gem-HV-scheme}, puts 
additional constraints onto the operational scheme, which has to be
coped with during the operation: (i) $V_\mathrm{top}$ needs to be
increased when a short circuit appears (switches in
\figref{pic:gem-HV-scheme}) because of the additional voltage drop
over $R_1$, which limits the numbers of bad sectors due to
the voltage limit of the power supply.  (ii) In case of a high
particle rate, 
the resulting current produced in the detector leads to a larger
voltage drop over $R_1$ and $R_2$ than for the
nominal configuration with only $R_2$.  Each GEM foil is
powered by two independent channels of an ISEG EHS 8060n high
voltage module.  In case of a shortcut sector the voltage drop over
the grounding resistor $R_3$ prevents currents from flowing
into the high voltage supply.

In conclusion, the adopted power supply scheme, although flexible,
turned out to be not the optimal choice, and for stable operation at
fixed settings should be replaced by a simple voltage divider
(possibly exchangeable if flexibility is needed), or a stack of
floating power supplies.
 
\section{The Pad-Plane}
\label{sec:prototype.design.readout}
\noindent The charge cloud emerging from the last GEM induces a
negative, fast electron signal on the readout pads.  For single
incoming electrons, the FWHM of the induced signal is of the order of
\SI{400}{\micro\meter}, depending on the transverse diffusion between
the GEM stages and the spread inside the GEM holes. Consequently, a
pitch of the 
readout elements of the same order of magnitude would be required to
profit from a center-of-gravity reconstruction.  For a TPC, however,
transverse diffusion of the ionization electrons during their drift to
the amplification stage will dominate the precision with which its
original position can be reconstructed, except for very small drift
distances.  The pad size should then be chosen such that the
transverse diffusion for a given gas mixture and magnetic field
dominates the resolution, and not the pad size.  Furthermore the
geometrical configuration of the pads should prevent a deterioration
of the spatial resolution of the tracks due to angular
effects.
The pad-plane used for this GEM-TPC is designed following these
guiding principles.

\begin{figure}[tbp]
  \sbox\twosubbox{%
    \resizebox{\dimexpr.95\textwidth}{!}{%
      \includegraphics[height=3cm]{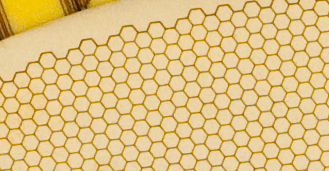}%
      \includegraphics[height=3cm]{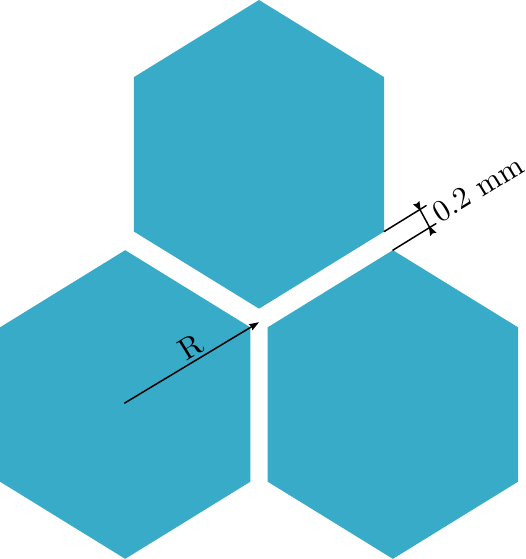}
    }%
  }%
  \setlength{\twosubht}{\ht\twosubbox}
  \begin{center}
    \subcaptionbox{}{\includegraphics[height=\twosubht]{./figures/readout/pad_geometry.pdf}}
    \subcaptionbox{}{\includegraphics[height=\twosubht]{./figures/readout/padplane_pads_closeup.png}}
  \end{center}
  \caption{(a) Geometry of the pads indicating the radius and the
    distance between two pads. (b) Closeup view of the pad-plane.}
  \label{pic:prototype.pads}
\end{figure}

\begin{figure}[tbp]
  \centering
  \includegraphics[width=.7\textwidth]{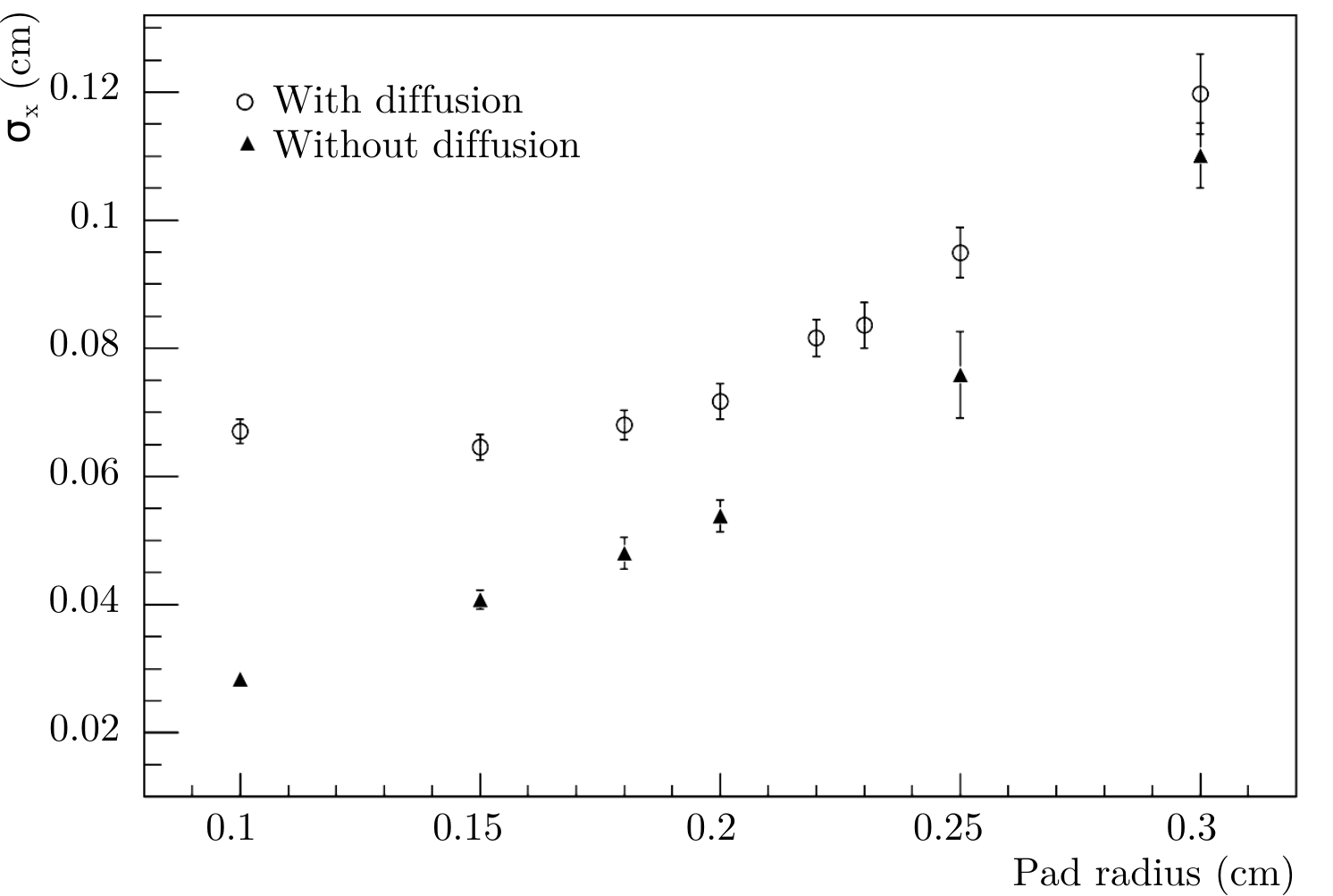}
  \caption{Standard deviation of residuals plotted against the pad
    radius with diffusion (circles) and without (triangles). A
    magnetic field of \SI{2}{\tesla} was applied to the simulation.}
  \label{fig:prototype.pad.simu}
\end{figure}

For a uniform charge distribution independent of the crossing angle of
tracks, pads with a hexagonal shape are chosen.  The optimum outer pad
radius ($R$) was determined to be \SI{1.5}{\mm} with the help of Monte
Carlo simulations assuming a Ne/CO$_2$ (90/10) gas mixture and a
magnetic field of \SI{2}{\tesla}, as originally foreseen for the TPC
of the PANDA experiment.  The definition of the pad radius is depicted
in \Figref{pic:prototype.pads} in the left panel.  The gaps
between the pads are \SI{200}{\micro\meter} wide.  As one can see in
\Figref{fig:prototype.pad.simu} these simulations show that the
spatial resolution and hence the momentum resolution does not improve
for outer radii of the hexagonal pads below \SI{1.5}{\mm}, as
diffusion in the drift volume is the dominant contribution to the
localization accuracy.

The pad-plane is realized as a printed circuit board (PCB) made of FR4
with \SI{35}{\micro\meter} thick gold-plated Cu
pads. \Figref{pic:prototype.pads} (b) shows a photographic image of
the pad-plane, which in total has $10,254$ readout pads. 
Since the pad-plane
is also used to close the gas volume it had to be designed in a
gas-tight way.  This is done by using a four-layer PCB with signal
tracks on all 
layers and staggered connections between neighboring layers.  On the
backside of the pad-plane the signal lines are routed to 42
high-density connectors (Samtec BTH-150-01-F-D-A) with 300 pins and a
pitch of \SI{0.5}{\mm} each.
The connectors are arranged radially.

To avoid crosstalk between different pads the connections from the
pads to the connectors are designed to minimize crossing of tracks or
narrow parallel tracks of neighboring pads.  
Field inhomogeneities on the outer part of the pad-plane are avoided
by placing copper areas, matching the shape of the pads, as shown in
\figref{pic:prototype.pads} (b). Further
details about the construction of the pad-plane can be found in
\cite{MB15}.

%% file: fee_daq/fee_daq_new.tex
\section{Front-end Electronics and Data Acquisition}
\label{sec:FEE_DAQ}
\subsection{General Requirements}
\noindent
The current signal 
seen by the charge-sensitive amplifier (CSA) connected to a readout pad is
determined by 
the convolution of two distributions. The first is the 
ionization-charge
distribution along the drift direction at the moment when it reaches
the GEM foils. The spread of the charge is caused by the longitudinal
diffusion of electrons during their drift through the TPC gas
volume. The second is the shape of the current pulse induced by
electrons drifting from the last GEM electrode to the readout pad,
which has a negative polarity with a fast risetime 
and a duration of about $50\,\ns$. The length of the convoluted signal
depends on the drift length. For the GEM-TPC considered here it is
about $100\,\ns$ on average. 
The CSA is followed 
by a shaper with a programmable peaking time between $100$ and
$200\,\ns$ for optimization of the signal-to-noise ratio and the
effect of pile-up. 
The sampling rate has to be adapted to the shaper output and should
thus be chosen between $10$ and $20\,\MHz$. 
A crucial parameter for a continuously operating TPC is the noise
level of the readout system, which has to be 
as low as possible ($< 1000\,e^-$) in order to operate the chamber at
the minimum possible gain, defined by the required 
signal-to-noise ratio. A low gain in turn minimizes the ion backflow
and the resulting track distortions.  
To allow for a continuous readout of the TPC, the front-end chip has
to be able to simultaneously sample the incoming data and transfer
the processed data to the next stage.  
Power consumption should be minimized to a level of 
$\sim 20\,\mW/\mathrm{channel}$ to avoid the necessity of 
excessive cooling of the front-end cards. 
Last but not
least, the chip has to be highly integrated to match the channel
density on the pad plane.\\
%
%
The SAMPA chip
\cite{sampa:2016} which is currently being developed for the upgrade
of the ALICE TPC will fulfill all of the above requirements. At the
time of the design and construction of the GEM-TPC no chip
that matched all these requirements was available. The
AFTER T2K chip \cite{T2K-LaserCal} was chosen as a good compromise
because it fulfils the low noise, low power consumption and high
integration requirements. It  
does not allow, however, on-chip zero suppression and to concurrently
sample and transfer the data. 
\subsection{The AFTER ASIC}
\label{sec:after}
\begin{figure}[tbp]
  \centering
  \includegraphics[width=0.5\textwidth]{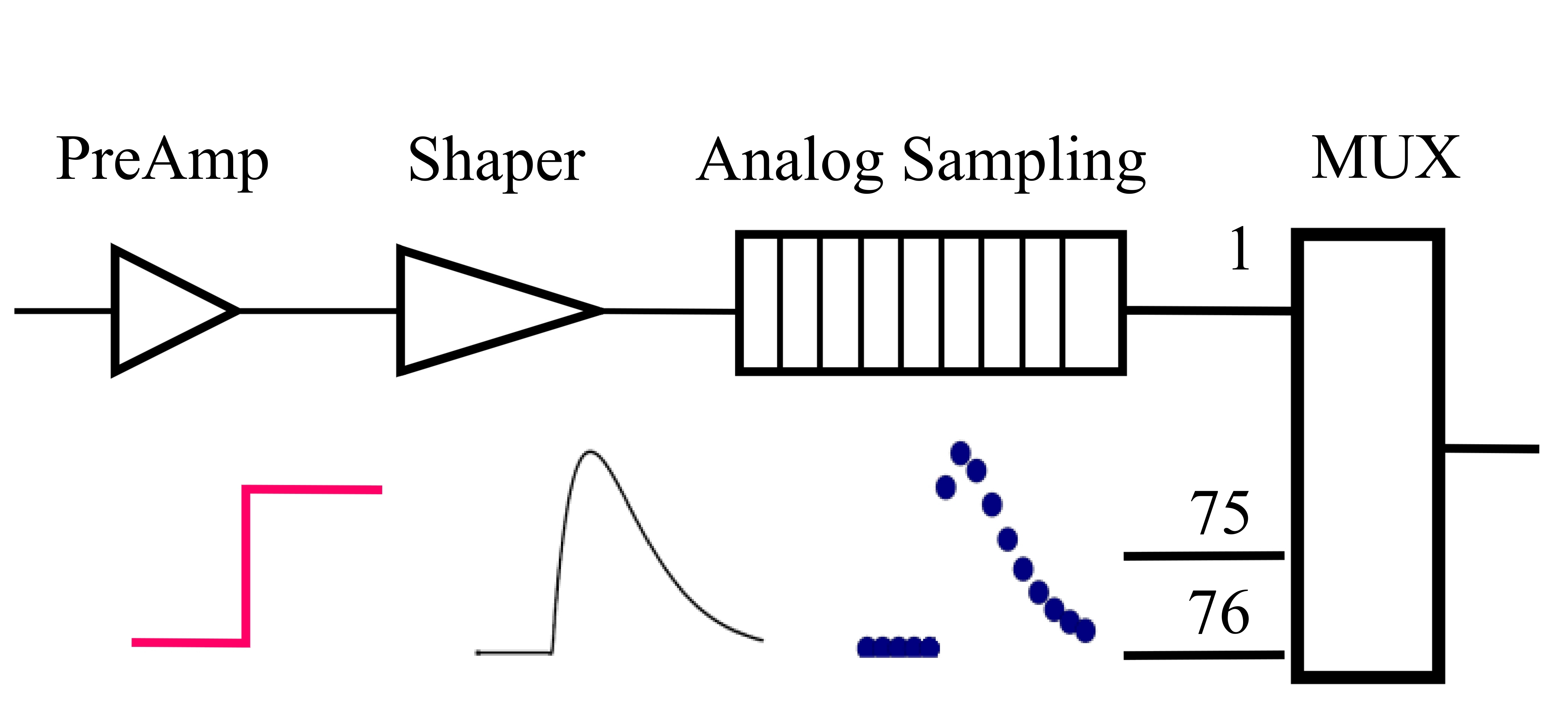}
  \caption{AFTER data flow.}
  \label{pic:after_scheme}
\end{figure}

\noindent The AFTER chip is a low-noise, low-consumption analog
pipeline ASIC developed for the T2K 
near detector TPCs and fabricated in $0.35\,\upmu m$
CMOS technology \cite{Baron:2008zz}. 
It has 76 channels, of which 72 are connected to input pins, each with
a tunable preamplifier/shaper followed by a 511-cell switched
capacitor array (SCA) analog memory.  
In \figref{pic:after_scheme} a simplified functional diagram of a
single channel of the AFTER 
ASIC is shown. 
The main parameters of the AFTER
ASIC can be found in Table \ref{tab:AFTER-Values}.
\begin{table}[tbp]
  \centering
  \begin{tabular}{lc}
    \hline\hline
    Parameter 			& Value 															\\ \hline
    Number of input channels 	& 72 																\\
    Samples per channel & 511 																\\
    Noise (nominal)		& 500 $\mathrm{e}^-$ at \SI{10}{\pico\farad}						\\
    Dynamic range 		& \SIrange[range-units=single]{120}{600}{\femto\coulomb} (4 values) \\
    Peaking time 		& \SI{116}{\nano\second} to \SI{1.9}{\micro\second} (16 values) 		\\
    Sampling frequency 	& \SIrange[range-units=single]{1}{50}{\mega\hertz}					\\
    Readout frequency 	& \SI{20}{\mega\hertz} 				\\
    Power Consumption	&
    \SIrange[range-units=single]{6.2}{7.5}{\milli\watt\per Channel}
    \\ 
    Size of the package & $7.8\times 7.4\,\mm^2$ \\\hline\hline
  \end{tabular}
  \caption{Main parameters of the AFTER ASIC \cite{Baron:2008zz} } 
  \label{tab:AFTER-Values}
\end{table}
%
The preamplifier charge range can be chosen to be
from \SIrange{120}{600}{\femto\coulomb}.  The shaper has an adjustable
peaking time in the range from \SI{116}{\nano\second} to
\SI{1.9}{\micro\second}.  After the shaper the signals are sampled
into an analog buffer which is implemented as two SCAs, each with 36
channels, and a depth of 511 samples.  Each 
of the SCAs has two internal channels not connected to the input which
can be used to correct for common mode noise and so-called fixed
pattern noise due to  
charge loss in the capacitors (SCA-leakage). The sampling clock is
supplied externally and can range from \SIrange{1}{50}{\mega\hertz}.
All the tunable parameters 
of the chip can be accessed via a custom serial slow
control protocol.\\
%
Upon a trigger the content of the SCA is multiplexed to a differential
line at
a frequency of \SI{20}{\mega\hertz}, time slice by time slice starting
from the oldest sample. The full reading cycle takes $2\,\ms$ during
which writing is disabled, which poses the main rate limitation of
this chip.  

\subsection{Front-end card}
\begin{figure}[htp] \centering
  \includegraphics[width=.8\textwidth]{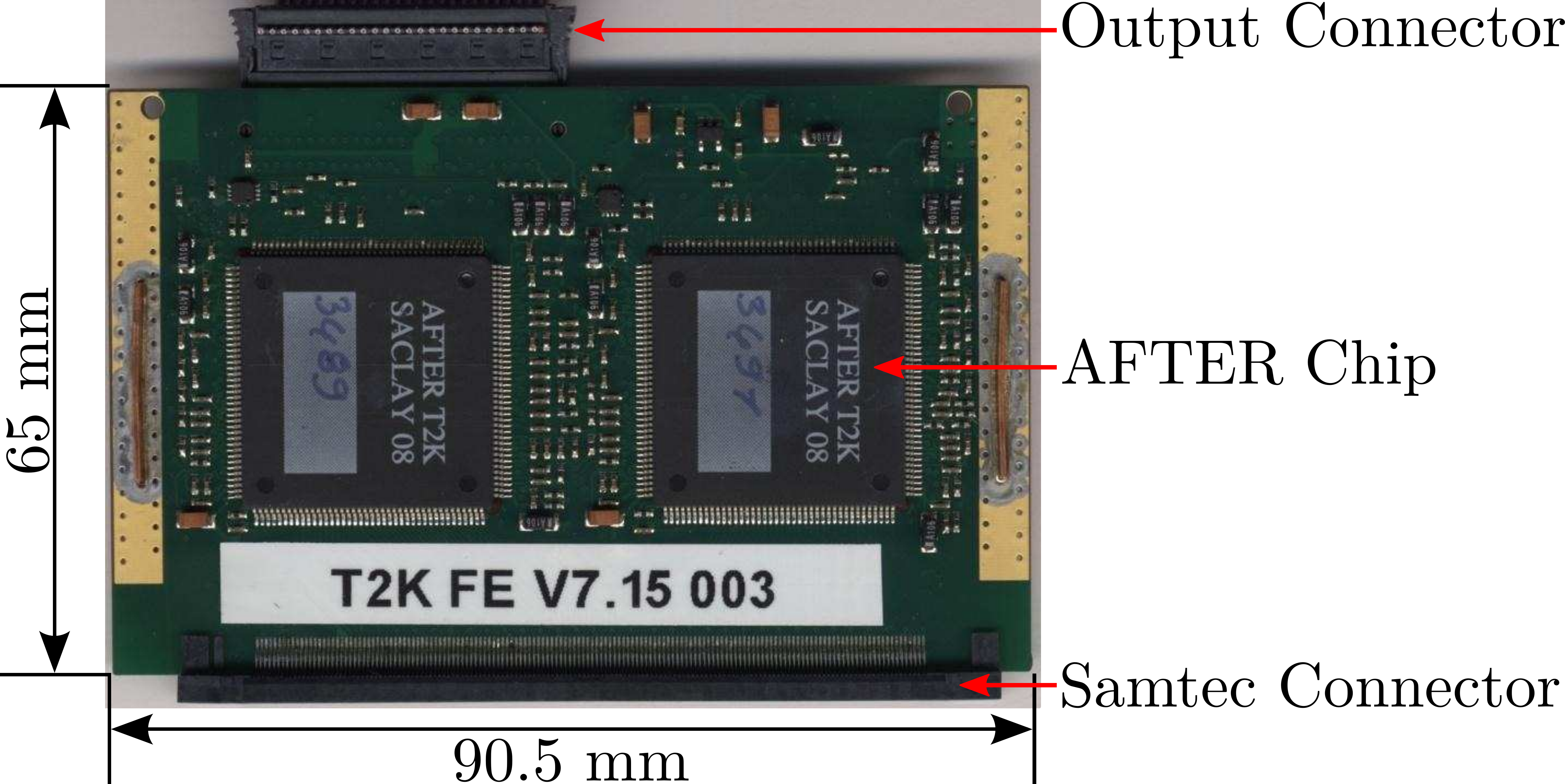}
  \caption{Photograph of the front-end card with two AFTER chips on
    each side. The Samtec connector is directly plugged to the
    backside of the pad plane. Each chip reads out 64 detector
    channels, 8 channels are not connected to the pad plane. The first
    36 channels are connected to the pins of the ASIC on the far side from the
    Samtec connector, the second 36 to the ones on the near side. The
    backside of the card looks identical.}
  \label{pic:fecard}
\end{figure}

\noindent One front-end card (\figref{pic:fecard}) houses four
AFTER-chips and is directly connected to the pad plane via a 300-pin
connector (SAMTEC BSH-EM-150) mounted sideways. Only 64 channels of
the available 72 per chip are connected to the pad plane,
such that one front-end card reads 
out a total of 256 detector
channels.  All four chips are controlled and read out in parallel.
The total power consumption per card is \SI{3.2}{\watt}.  As the TPC
is very sensitive to temperature variations, heat transfer
from the front-end cards to the gas volume is avoided by dissipating
the excessive heat with an active water-driven cooling system (see
\secref{sec:cooling}). Figure~\ref{pic:fecard} shows a photograph of a
front-end card. Detector channels 1 -- 32 are connected to the pins on the far
side of the Samtec connector and thus have longer signal paths than
detector channel 33 -- 64, which are connected to the pins facing the
Samtec connector (see also \figref{pic:noise} in \secref{sec:CalSys.pedestal}). 

\subsection{Readout Chain}
\label{sec:FEE_DAQ.readout}
\begin{figure}[htp] \centering
  \includegraphics[width=0.9\textwidth]{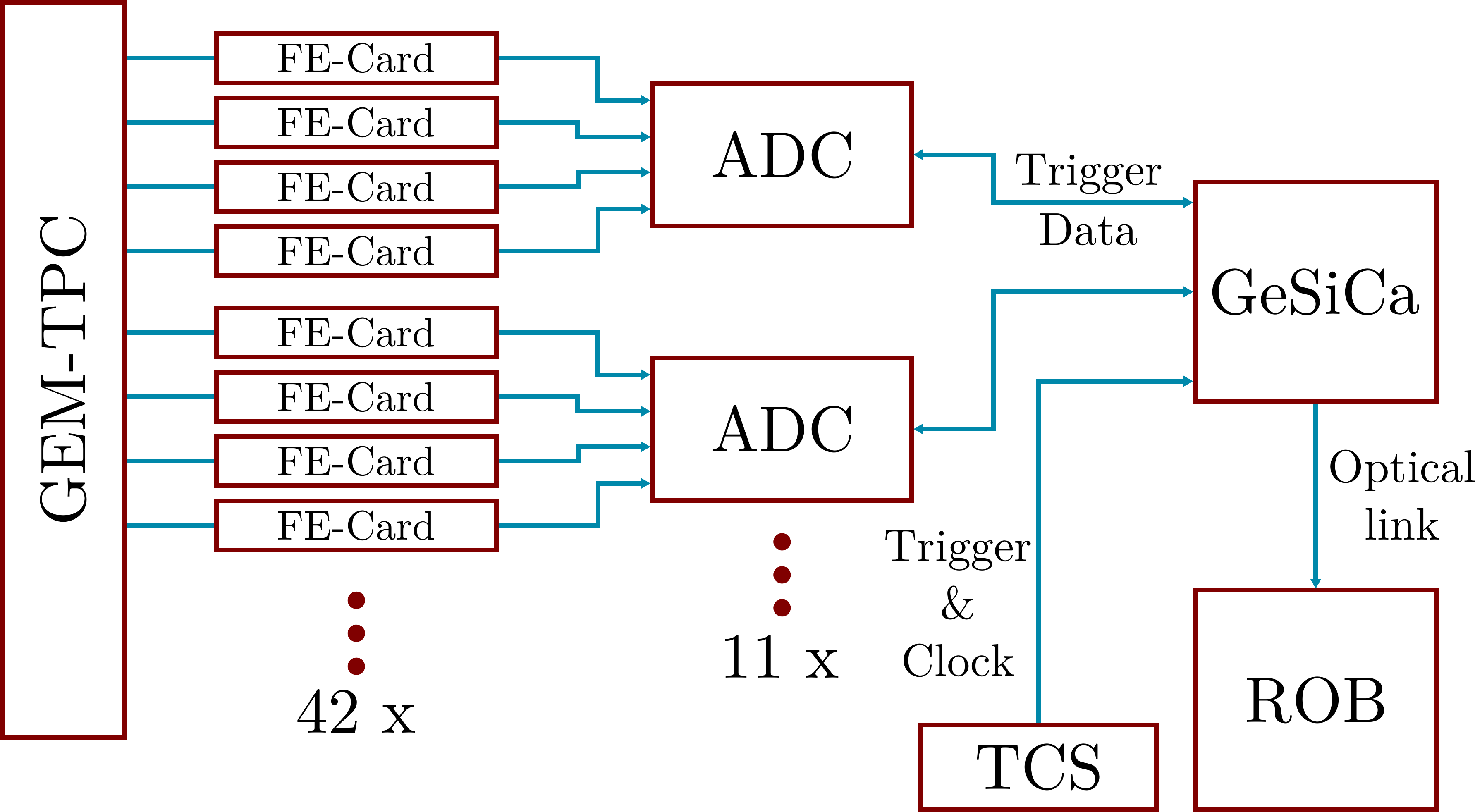}
  \caption{The GEM-TPC read out scheme. The signals from the GEM-TPC are
    sampled by the front-end (FE) cards and digitized by the
    analog-digital-converters (ADC). The ADCs send their data to the
    GEM-Silicon Control and Acquisition (GeSiCa) boards. Finally all data are
    collected in the read-out-buffer (ROB) computer. The trigger and
    clock is controlled by the Trigger Control System (TCS) and
    distributed via the GeSiCas to the ADCs and from there to the
    FE-cards.}
  \label{pic:read_out}
\end{figure}
\noindent
%
For the read-out, the scheme developed for the COMPASS experiment
\cite{COMPASS:96} at CERN has been adapted for the GEM-TPC. 
The read-out scheme is
displayed in Figure \ref{pic:read_out}.  
Clock ($38.88\,\MHz$) and
trigger signals are sent via optical fibers from the Trigger
Control System (TCS) to readout driver modules (so-called GeSiCA) and from there
they are distributed to the ADC modules. 
From there, 
each AFTER ASIC is supplied with two clock signals derived from the
TCS clock:
\texttt{WRITE\_CLK}, which defines the sampling frequency and which is
configurable as an integer division of a $77.67\,\MHz$, and
\texttt{READ\_CLK}, which defines the data readout to the ADC and which is
fixed to half the TCS clock ($19.44\,\MHz$).
During the commissioning phase (cf.\ \secref{sec:Com}, the detector was operated at an
average trigger rate of $\sim 140\,\s^{-1}$, determined mainly by the dead time needed to read out
the buffer of the AFTER chip (cf.\ \secref{sec:after}).   
Upon arrival of a 
trigger signal the AFTER chips read the values stored in the SCAs
and multiplex them via a differential line to
the ADC modules.  
One ADC module can handle
the data from up to four front-end cards (16 AFTER chips). 
The analogue
signals from the AFTER ASICs are sampled by pipelined \SI{12}{bit}
ADCs with a sampling rate matching the readout clock of the AFTER ASIC. 
A Xilinx Virtex-4 FPGA
on the ADC board performs online baseline subtraction and zero
suppression of the 
digitized data.  The zero suppression algorithm takes into account
individual thresholds on each channel, calculated from the measured noise
performance of the channels (see Sec.~\ref{sec:CalSys.pedestal}). To
correct for the  
fixed pattern noise in the AFTER chip another correction algorithm has
been implemented.  For each channel of one SCA the mean of the two
unconnected channels is subtracted to correct for the 
charge loss in the capacitors.  
After processing the data of 16 chips in
the ADC module they are sent via an optical fiber to the GeSiCA
module. 
One GeSiCA module can combine data from up to seven ADC modules and
transmit these sub-events via an optical link to a read-out buffer at
a maximum speed of \SI{160}{MB\per\second}.  The read-out buffer
performs the final event building for the TPC and stores the data to a
hard drive or sends it via TCP/IP to other computers for further
processing.  In total 42 front-end cards (168 AFTER chips) are used
to read out the 10,254 channels of the GEM-TPC. This leads to a
required number of 11 ADC and 2 GeSiCa modules.  
The total raw data size at the input of the ADC modules amounts
to $9.8\,\mathrm{MByte\ per\ event}$ and corresponds to a data rate of
$1.4\,\mathrm{GByte}/\s$ at a trigger rate of $140\,\s^{-1}$. After
zero suppression in the ADC modules, the total data rate is reduced to
$\mathcal{O}(10\,\mathrm{MByte}/\s)$, depending on the occupancy and
the duty cycle of the beam. For the experiment with a pion beam (cf.\
\secref{sec:Com}), the average occupancy was $0.05\%$ with a duty
cycle of $53\%$, while for the tests with heavy-ion beams the
occupancy was about 4 to 6 times higher with a similar duty cycle. 


%
%
%
\subsection{Gain Calibration of the AFTER-ADC System}
\label{sec:FEE_DAQ.gain}
\begin{figure}[htp] \centering
  \includegraphics[width=0.9\textwidth]{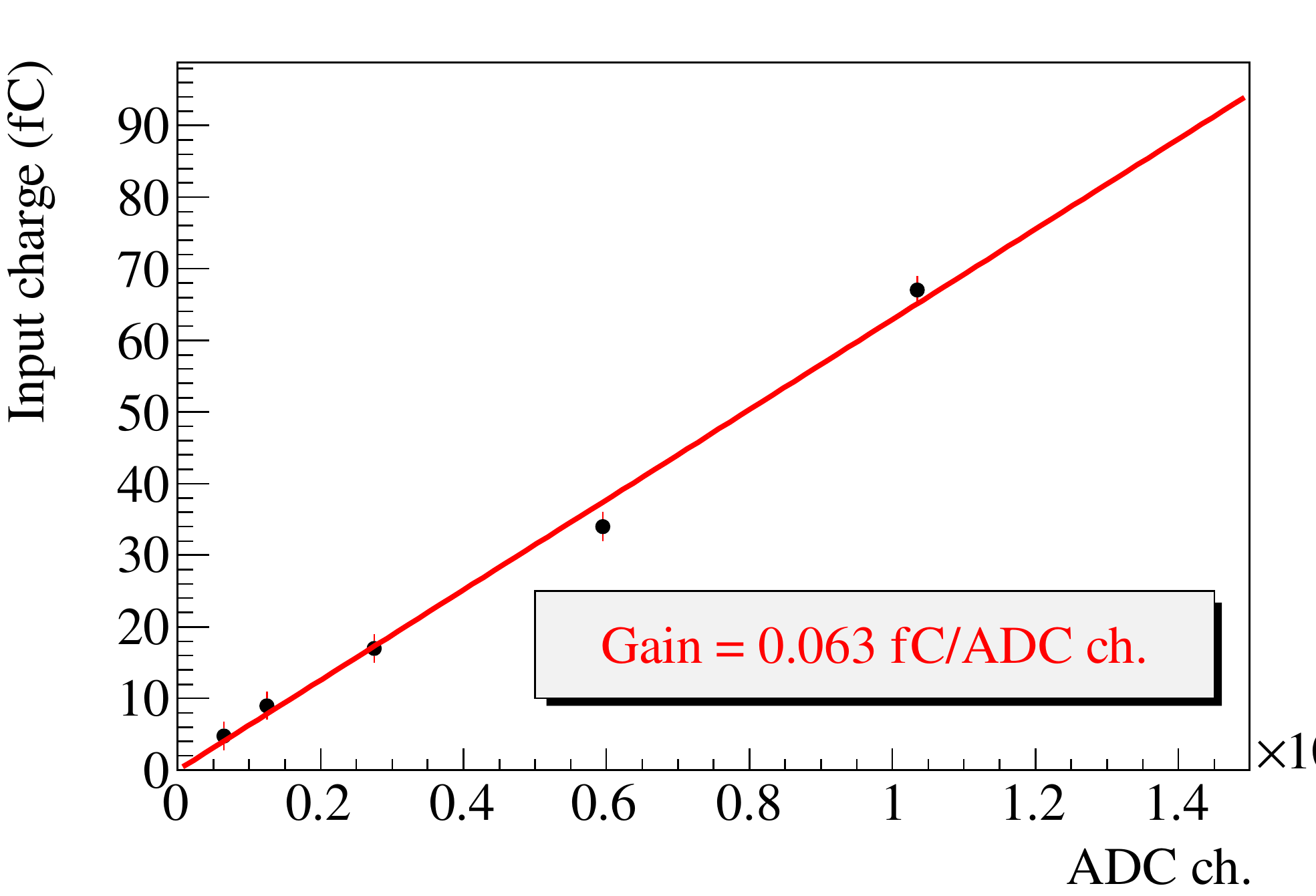}
  \caption{Gain calibration of the AFTER + ADC system for a peaking
    time of $116\,\ns$ and a charge range of $120\,\fC$.}
  \label{fig:after.conversion_gain}
\end{figure}

The conversion gain of the system AFTER + ADC has been measured by
injecting a known charge into a pin of the AFTER chip via a $1\,\pF$
SMD capacitor soldered to a readout pad of the pad plane. Several
different amplitudes of a step-like input pulse have been used as
shown in Fig.~\ref{fig:after.conversion_gain}. The conversion gain
derived from a linear 
fit is $(0.063\pm 0.003)\,\fC/\mathrm{ADC\ channel}$, or $(393\pm
19)\,e^-/\mathrm{ADC\ channel}$. 

%% file: cooling/cooling.tex
\section{Cooling}
\label{sec:cooling}
\begin{figure}[ht]
\centering
\includegraphics[width=\textwidth]{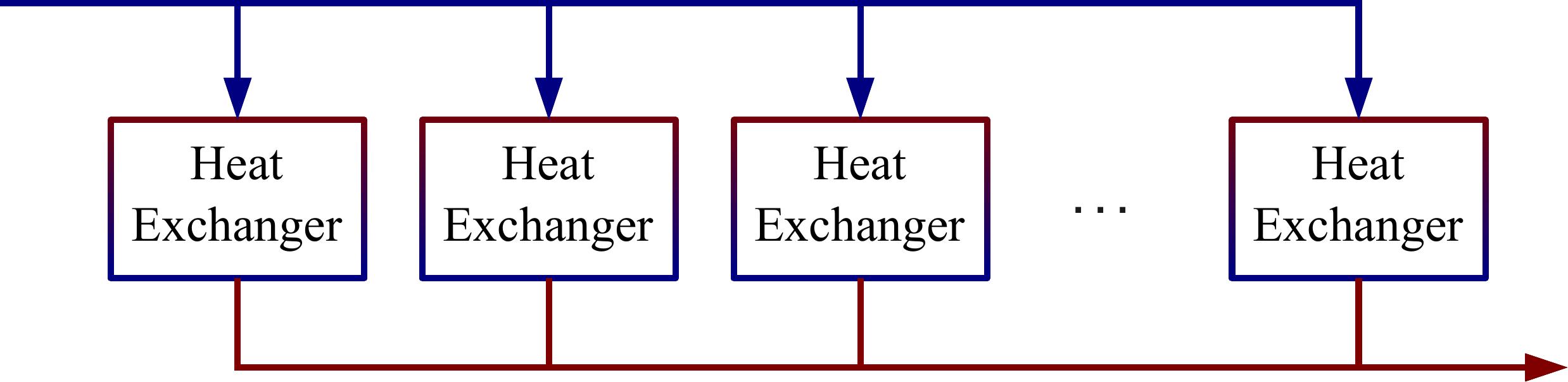}
\caption{Scheme of the cooling distribution.}
\label{subsec_cooling_schematiccooling}
\end{figure}

\begin{figure}[ht]
	\begin{center}
		\begin{subfigure}[b]{.49\textwidth}
		\centering
    	\includegraphics[width=\textwidth]{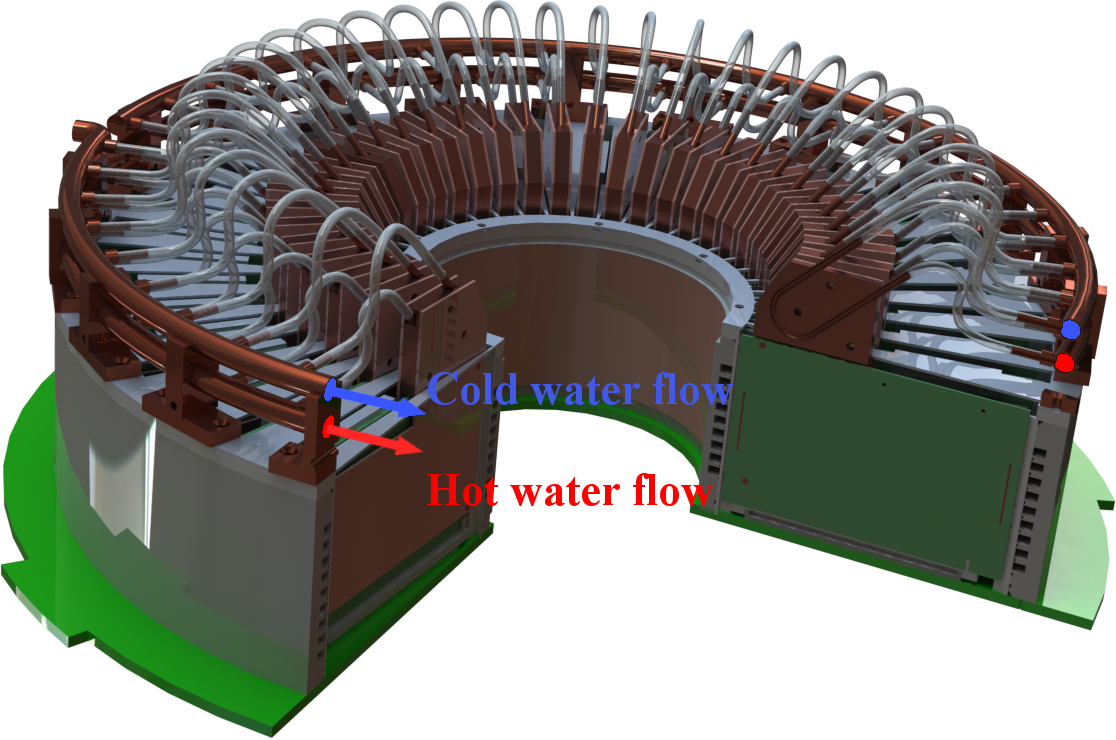}
    	\caption{}
  	\end{subfigure}
  	\begin{subfigure}[b]{.49\textwidth}
    	\centering 
		\includegraphics[width=\textwidth]{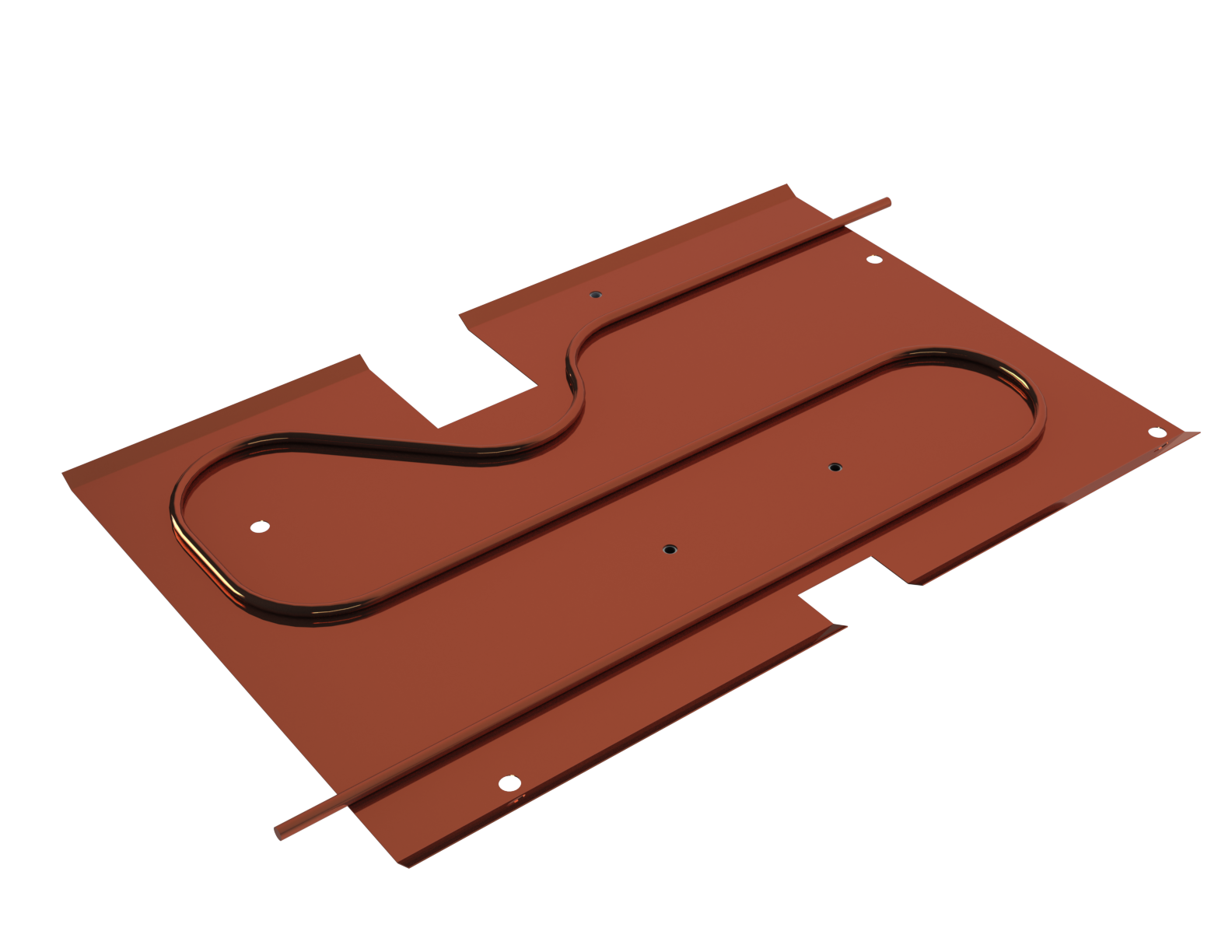}
    	\caption{}
    \end{subfigure}
    \caption{Panel a): CAD rendering of the complete cooling setup. For a better visibility a quarter of the setup has been cut away.
     Panel (b) shows a CAD rendering of the ADC cooling sheet metal.}
  	\label{subsec_cooling_fullsetup}
	\end{center}
\end{figure}

\begin{figure}
	\centering
 	\includegraphics[width=.8\textwidth]{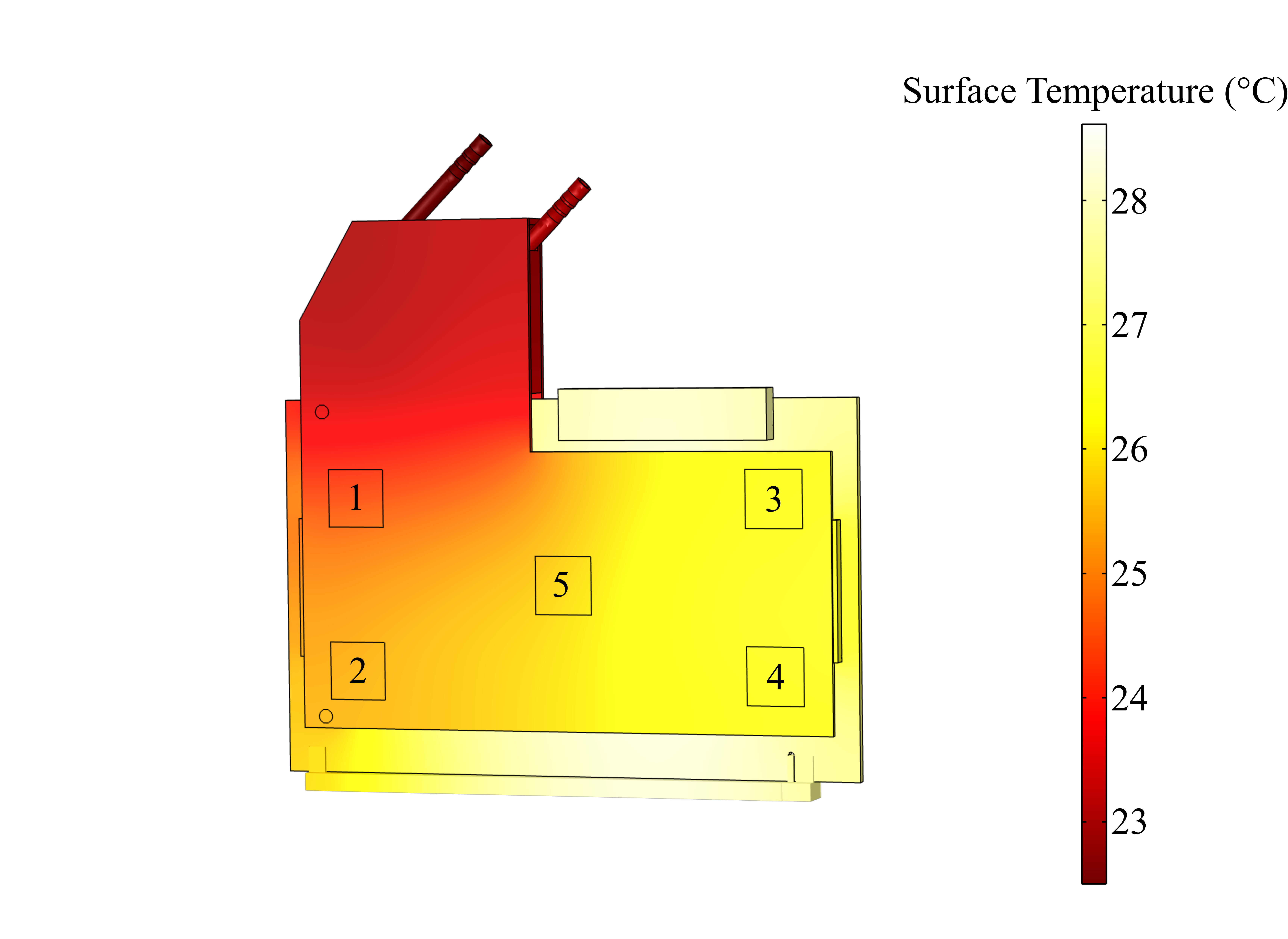}
	\caption{Simulated surface temperature distribution of one FE card. The 5 rectangles indicate the measurement positions.}
	\label{subsec_cooling_surftemp}
\end{figure}

\noindent 
With \SI{7.5}{\milli\watt} power consumption per channel a total of \SI{110}{\watt} created by all front-end cards has to be dissipated. 
Since the cards are directly plugged to the backside of the readout plane without a thermal barrier, it is important to actively cool the cards in order to avoid heat being introduced into the gas volume via the pad-plane, which would cause local variations of the drift velocity.
In addition, cooling is needed for the ADC modules in order to avoid overheating of electronic components installed in regions without air convection.
Both systems are cooled using a closed circuit water-driven cooling
system operated at overpressure and connected to a chiller (Huber
K\"al\-te\-ma\-schi\-nen UC080T-H Umw\"alzk\"uhler \cite{unichill}). 
The latter provides a coolant temperature of \SI{20}{\celsius}.\\
Each FE card is sandwiched between two copper plates put in direct contact with the four FE chips through heat conducting pads. 
\begin{figure}[ht]
	\begin{center}
	\begin{subfigure}[t]{0.45\textwidth}
	\includegraphics[width=\textwidth]{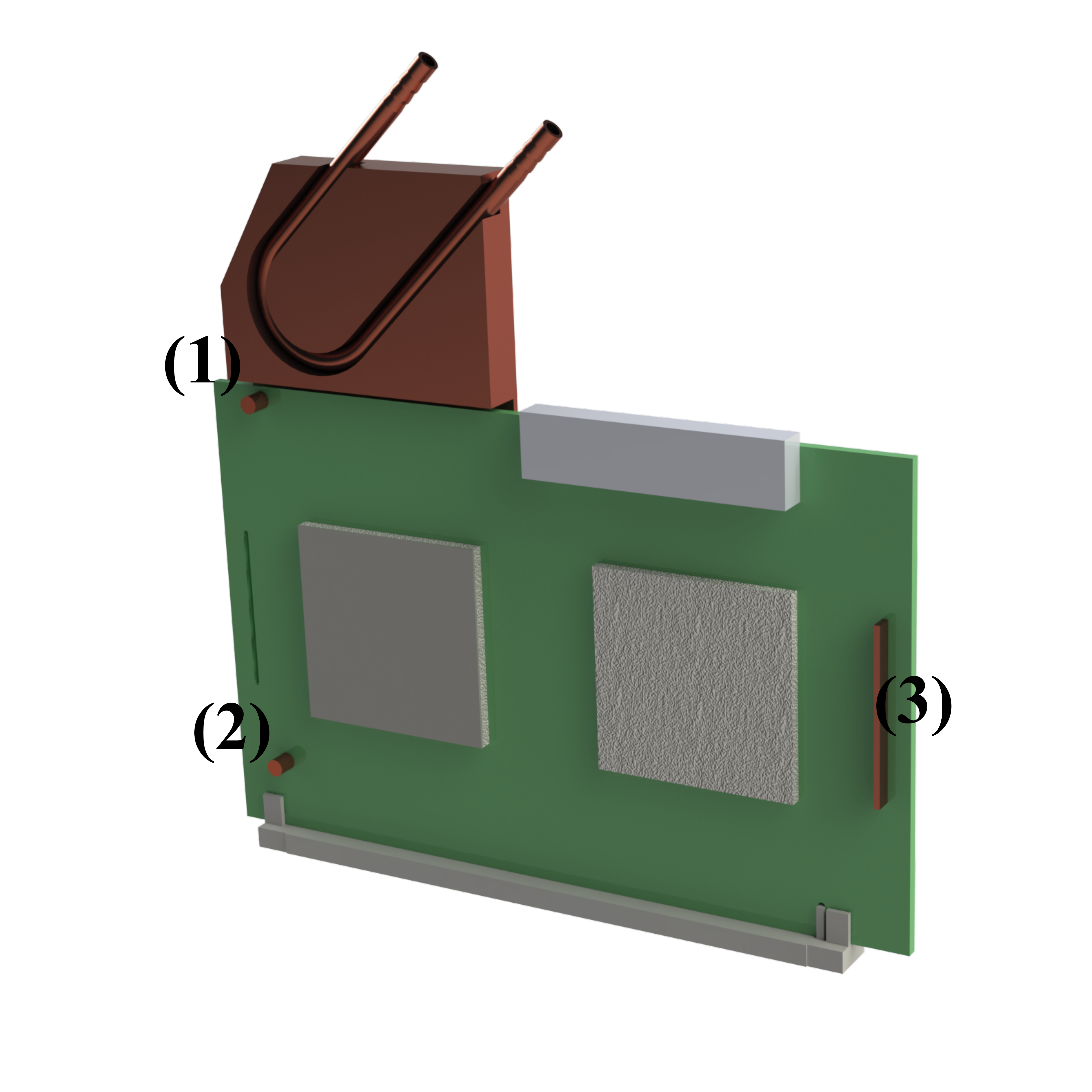}
\caption{}
\end{subfigure}
\hfill
\begin{subfigure}[t]{0.45\textwidth}
	\includegraphics[width=\textwidth]{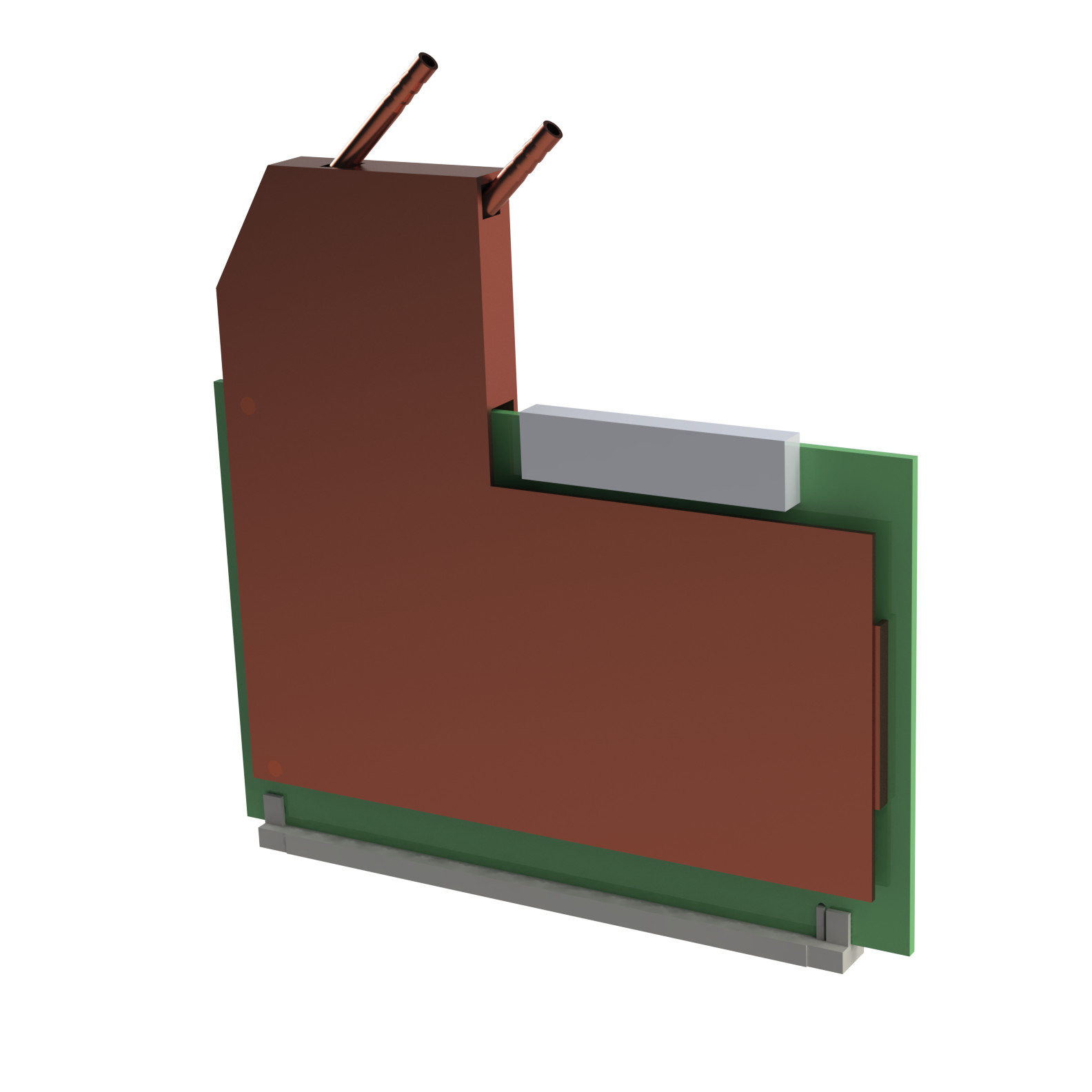}
\caption{}
\end{subfigure}
\caption[CAD rendering of one front-end card with its cooling components.]{CAD rendering of one front-end card with its cooling components. Panel a) shows the front-end card with partially removed cooling system. One copper plate for heat transfer as well as one heat conducting pad between the copper plate and T2K chips (grey squares) has been removed. (1), (2) and (3) indicate the soldering positions of the copper plate. The front-end card with the complete cooling system attached is depicted in panel b).}
\label{cooling_fig}
\end{center}
\end{figure}
The copper plates in turn are in contact with a heat exchanger connected to the cooling water circuit.\\
The FE cards are shown in \figref{cooling_fig} with the heat-conducting pads attached to the housing of the FE chips shown in panel a) and with the copper plates attached  to the pads as in panel b). 
The copper plates are mounted by soldering them at three positions 1-3 indicated in \figref{cooling_fig}, panel a). 
The two pins (1-2 in \figref{cooling_fig}, panel a)) are only soldered to the copper plates while the metal bar (3 in \figref{cooling_fig}, panel a)) is also soldered to the PCB. 
The heat exchanger is mounted by three screws to the copper plates. 
This way, the heat exchanger can be removed and this allows for an easy mounting of the FE cards. 
The pipe for the cooling water is soldered into a groove in the copper block and has an outer diameter of \SI{3}{\mm} and a wall thickness of \SI{0.5}{\mm}.\\
Since 42 FE cards have to be cooled at the same time, a homogeneous flow of cooling water through all heat exchangers has to be guaranteed.
This was achieved by attaching inlets and outlets, one pair for each FE card, to the supply line such that the flow resistance is the same through all heat exchangers. 
The scheme in \figref{subsec_cooling_schematiccooling} shows how this is realized. 
One can see that the path length and therefore the flow resistance of the cooling liquid is the same for each heat exchanger.
The cooling water is distributed by two rings made out of copper pipes to allow for soldering nozzles for the connection to the FE card heat exchangers by flexible polyurethane tubes.    
\Figref{subsec_cooling_fullsetup}(a) shows a CAD rendering of the complete cooling setup as it was used for the GEM-TPC.
To validate the uniformity of the cooling, simulations were taken out in two steps. In the first step the flow of the cooling liquid including the heat exchangers but not the FE cards were simulated.
The second step was the simulation of the heat distribution in the FE cards taking the results of the first step into account.\\
The simulation for the first step was performed with the FEM flow simulation framework of SolidWorks.
It was found that taking into account an inlet water flow of \SI{50}{\milli\liter\per\second} the flow of the cooling liquid through the heat exchangers is in average \SI{1.1}{\milli\liter\per\second} with a still rather large variance of \SI{0.56}{\milli\liter\per\second}.
As mentioned above, in a second step the temperature distribution of the FE card was calculated for all 42 liquid flow values found in the previous simulation. 
This calculation was carried out using the COMSOL FEM software and the model shown in \figref{subsec_cooling_fullsetup}.
Each chip introduces a power of \SI{0.625}{\watt} uniformly on its surface and only heat transport by conduction through the FE card and cooling system materials is assumed. 
The model was validated beforehand by a measurement showing that the maximal difference between measurement and calculation is below \SI{0.6}{\celsius}.
\Figref{subsec_cooling_surftemp} shows the outcome of this calculation where a flow of \SI{1e-5}{\milli\liter\per\second} and a water temperature of \SI{22.5}{\celsius} were used for the calculation.
Finally, an average temperature of \SI{21.3}{\celsius} with a variance of \SI{0.25}{\celsius} measured at position 5 in \figref{subsec_cooling_surftemp} was found for the FE cards taking into account the various flow values from the calculation mentioned above and a water temperature of \SI{18}{\celsius}.\\
The same system provides the cooling of the voltage regulators of the ADCs and the ADCs themselves through a serial connection of the pipes following the cooling ring for the FE cards.
In this case, a copper sheet with an soldered pipe was attached to each ADC module (\figref{subsec_cooling_fullsetup}(b)). 
In the same way as for the FE cards each chip was in thermal contact via an heat conducting pad with an aluminium block which again was screwed to the copper sheet. 
The bending of the pipe in contact with the copper sheet was chosen to maximize the contact area and to cross each aluminium block.

%% file: gas/gas.tex
\section{Gas System}
\label{sec:gas}
\subsection{Operational Requirements}
\label{sec:gas.requirements}
The gas system supplies the detector vessel with the drift gas
mixtures Ar/CO$_2$ (90/10) and Ne/CO$_2$ (90/10) and with a radioactive
Kr gas during calibration.  In order to test different gas mixtures
the system must be capable of switching between gases and modifying
the mole 
ratio to test different quencher contributions.
The system also must allow the user to change the detector gas to
nitrogen during 
stand-by mode in order to keep the detector volume dry and oxygen-free.

One of the main requirements to the GEM-TPC gas system is the
minimization of the oxygen and water vapour concentration inside the
detector vessel.  A contamination by an electro-negative gas
as O$_2$ leads to attachment of the primary electrons and thus to a
deterioration of the spatial and d$E$/d$x$ resolution.
The water vapour reduces the effective drift field and the drift
velocity since H$_2$O molecules can be easily polarized shielding the
electric field.  A contamination by oxygen or water vapour can 
originate from badly sealed joints and permeation through the TPC
wall material.  For the large ALICE TPC, where the volume-to-surface
ratio is about \SI{0.7}{\m}, an oxygen level lower than \SI{5}{ppm}
was reached.  This value can be considered as a reference for this
GEM-TPC where the volume-to-surface ratio is only \SI{0.1}{\m}.  In
addition to the gas purity, the accuracy and stability of the CO$_2$
content is crucial, as it determines important parameters like the
drift velocity and diffusion within the gas.
\subsection{Implementation}
\begin{figure}[htp]
  \centering
  \includegraphics[width=0.9\textheight,angle=90]{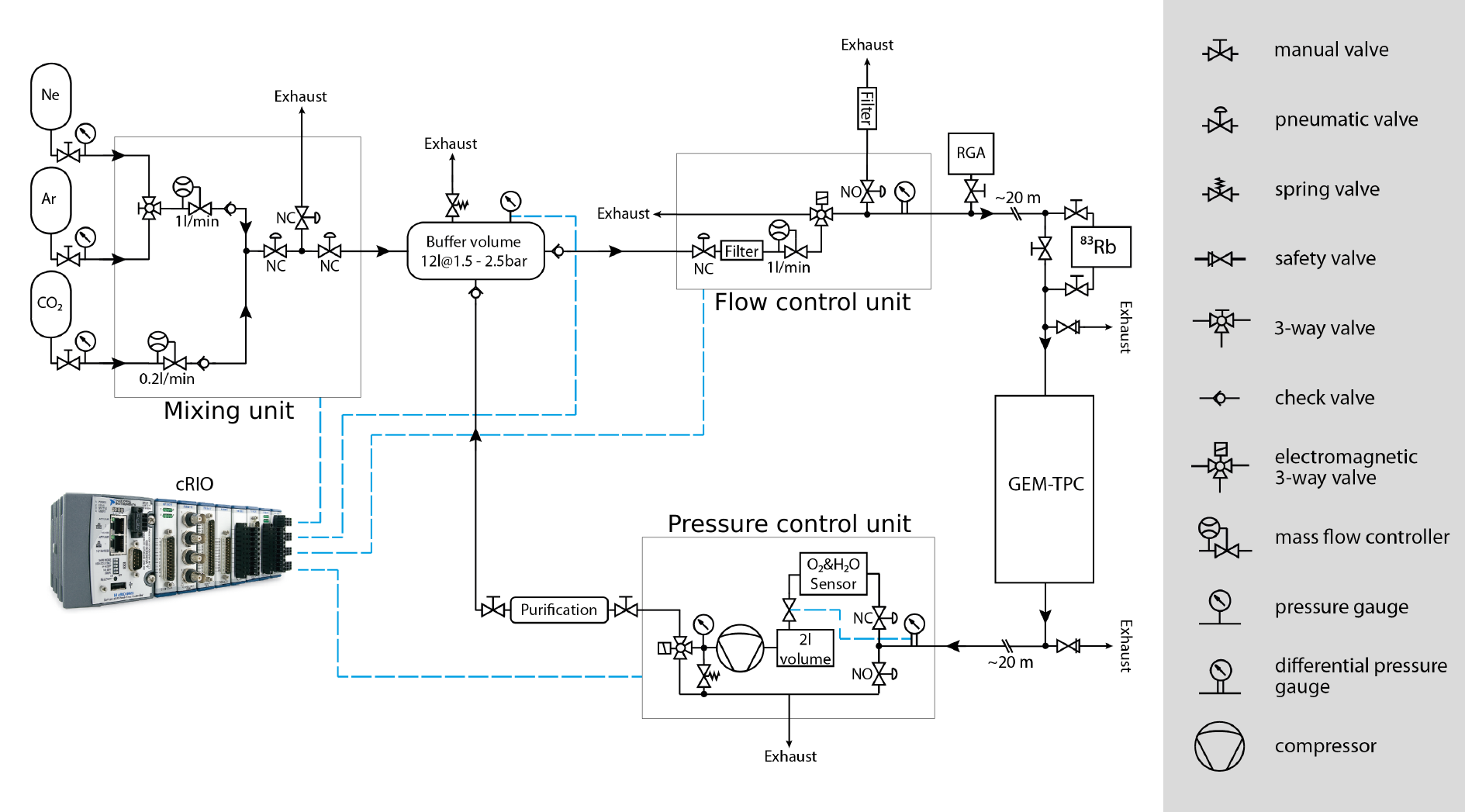}
  \caption{Flow chart of the gas system for the GEM-TPC. The symbols
    are  
  explained in the legend on top; NC/NO is shorthand for normally
  closed/open; residual gas analysis (RGA) is performed using a
  quadrupole mass spectrometer (QMS). Black lines indicate stainless
  steel or copper pipes, 
  blue lines 
  electrical connections.}
  \label{pic:prototype.opengas}
\end{figure}

\begin{figure}[htp]
\begin{center}
  \includegraphics[width=.9\textwidth]{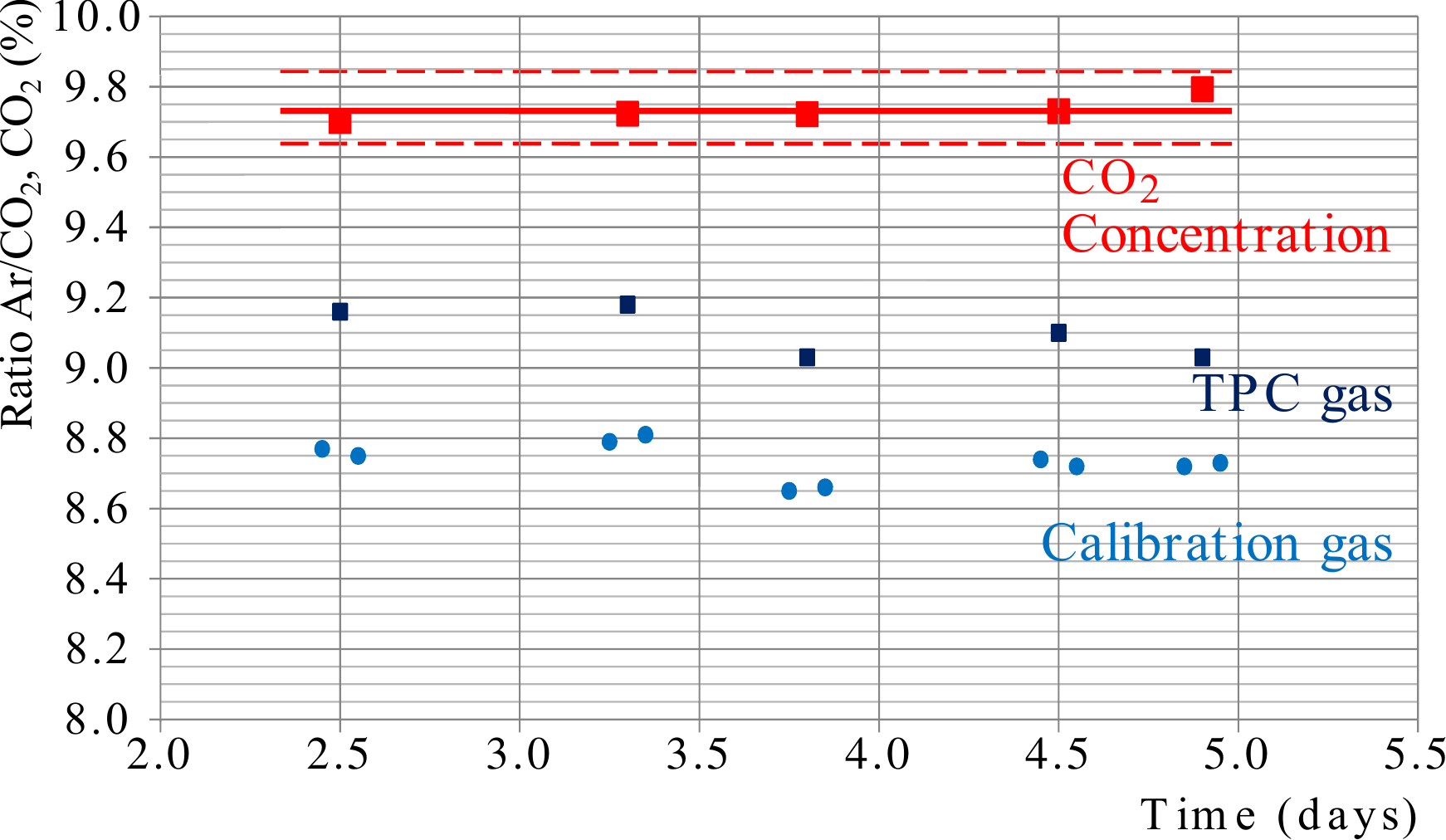}
  \caption[labelInTOC]{QMS (quadrupole mass spectrometer) analysis of the CO$_2$ content in an 
    Ar/C0$_2$ (90/10) gas mixture over 2.5 days of data taking.}
  \label{pic:gas_purity_a}
\end{center}
\end{figure}

\noindent A closed gas circulation system is employed for the \gt
(see Figure \ref{pic:prototype.opengas}).
The system is composed of a mixing unit, a flow control unit, and a
pressure control unit.  For the mixing of the gas two SLA5850S mass
flow controllers (MFC) are used.  By regulating the gas flow of the
individual mass flow controllers the desired mixing ratio of Ar/Ne and
CO$_2$ is achieved.  The gas mixture is filled into a 
\SI{12}{\liter}
buffer volume which serves as a reservoir to refill the gas system if
needed.  During operation the pressure inside the buffer volume is
kept between \num{1.5} and \SI{2}{\bar}.  One of the outputs of the
buffer volume is connected via a \SI{10}{\micro\meter} particle filter to
the flow control unit which controls the total gas flow to the TPC.
The total volume of the GEM-TPC is about \SI{45}{\liter} and a
flow rate of \SI{45}{\liter\per\hour} is used for both Ar/CO$_2$ and
Ne/CO$_2$ gas mixtures.  A differential pressure gauge is installed in
the flow control unit to measure overpressure up to
\SI{100}{\milli\bar}.  A residual gas analyzer (RGA) is installed in
order to monitor the gas composition after the flow control unit using
a quadrupole mass spectrometer (QMS).
\Figref{pic:gas_purity_a} shows an example of this measurement for a
time interval of 2.5 days.  The ratio Ar to CO$_2$ is shown for the
analyzed gas sample and a calibration sample composed of a premixed
Ar/CO$_2$ gas.  The measured CO$_2$ concentration with an error band
corresponding to \SI{1}{\percent} fluctuations is shown as well.

\begin{figure}[htp]
\begin{center}
  \includegraphics[width=.9\textwidth]{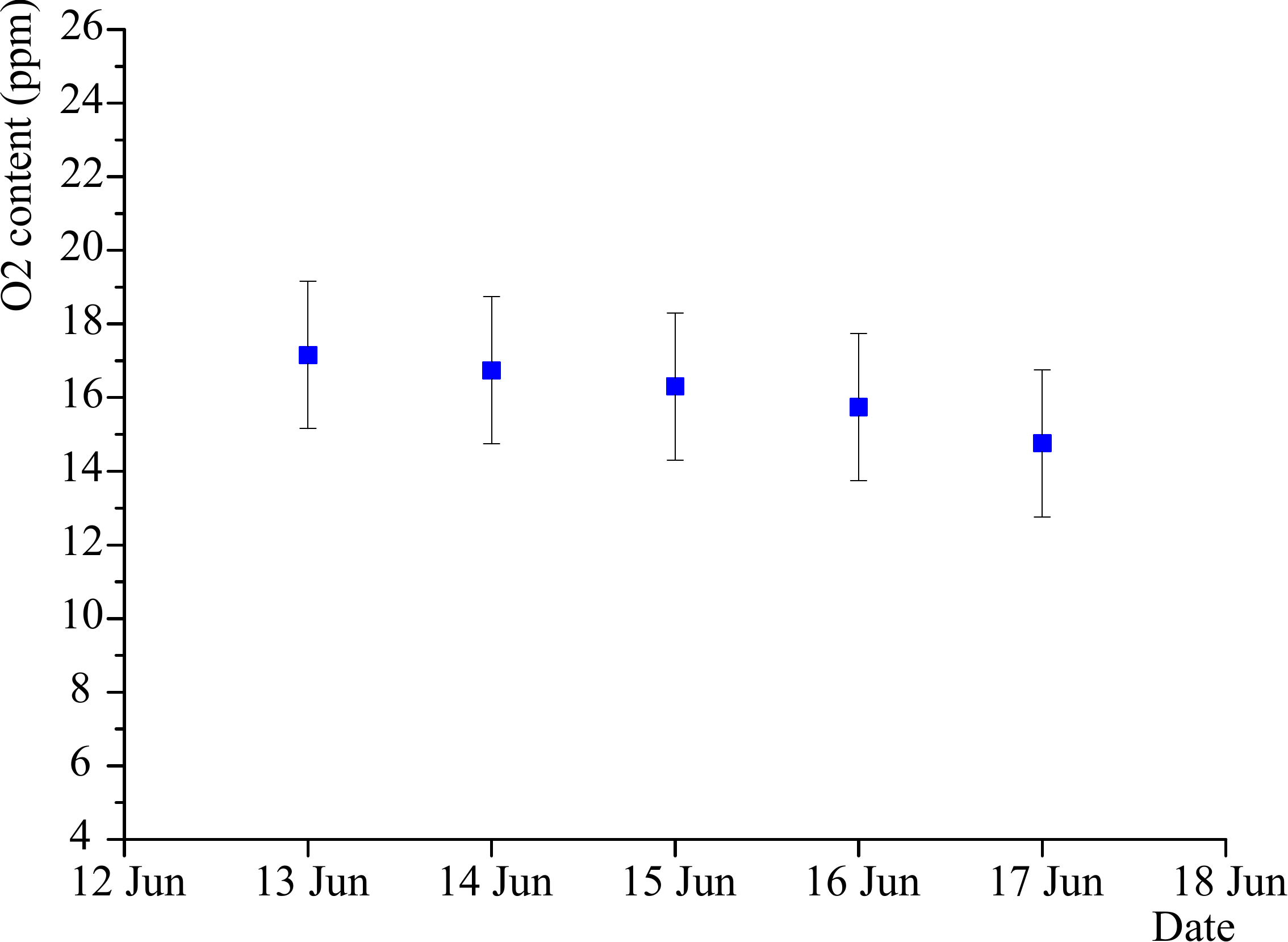}
  \caption[labelInTOC]{Development of the mean oxygen content over a period of five days.}
  \label{pic:gas_purity}
\end{center}
\end{figure}
%
%
%
For calibration of the gain uniformity 
(see \secref{sec:calibration.krypton}), radioactive $^{83m}$Kr can be
introduced into the TPC by flushing the 
gas through a vessel containing $^{83}$Rb. A pressure control unit is
located at the gas outlet of the TPC.  This unit controls the total
overpressure and measures the oxygen and water vapour content of the
gas.  For the oxygen measurement a Teledyne Model 3190 system with a
measurement range from \SIrange{0}{100}{ppm} and a relative accuracy
of \SI{2}{\percent} was used.  In \figref{pic:gas_purity_b} the
oxygen content for a period of five days is shown.  The mean oxygen
content within this time drops slightly from \num{17} to \SI{15}{ppm}.
This still rather large number is probably due to leakages in the field
cage sealing but it did not lead to
any significant signal losses. 

To measure the water vapour content a EE375 dew point transducer was
used.  After leaving the pressure control unit the gas enters the
purification to remove oxygen and water vapour from the gas.  The
oxygen is removed with activated copper by oxidizing to copper-oxide
while the water vapour is removed by molecular sieves with a pore size
of \SI{3}{\angstrom}.  The whole gas system is controlled by a
National Instruments CompactRIO
module which consists of:
\begin{itemize}
\item a NI-9265 current output module to control the set points of the
  MFCs,
\item a NI-9203 to read the MFC output signals,
\item a NI-9205 to read the outputs of the pressure gauges, the oxygen
  sensor and the water vapour sensor,
\item a NI-9237 to read out the pressure gauge of the flow control
  module,
\item Two NI-9485 to control the electromagnetic three-way valves.
\end{itemize}
All valves and flow MFCs were controlled by a LabView-based slow
control system which also monitored and recorded all values.  Further
details about the gas system can be found in \cite{PM}.

%% file: Control/Control.tex
\section{Slow Control System}
\label{sec:slowcontrol}
\noindent
For a safe detector operation it is crucial to monitor and control all
parameters in a convenient and reliable way.  In the case of the TPC the
main parameters to be controlled are the high-voltage potentials of
the cathode, the last strip, and the GEM foils and the low voltage of
the front-end 
powering.  Furthermore various temperature and pressure sensors
located inside the TPC have to be monitored.
\subsection{High-Voltage System}
\noindent
A safe operation of the TPC requires a supervision of every individual
high voltage that is applied to the GEM foils and the drift cathode.  
In addition, the possibility to operate all high voltage channels
simultaneously, especially during ramping, is required.  To avoid any
damages to 
the chamber, a fast emergency shutdown of the high voltage system is
indispensable.
Furthermore the system should allow the configuration of this
emergency shutdown.\\ 
The high voltage system for the GEM stack requires a voltage stability
below \SI{50}{\milli\volt}, current measurement with a resolution of
\SI{\approx 1}{\nano\ampere}, adjustable ramp speeds, full remote
controllability and output voltages of up to \SI{6}{\kilo\volt}.  For
the drift cathode a high voltage system with a voltage of up to
\SI{30}{\kilo\volt} and currents up to \SI{1}{\milli\ampere} is
needed.  An ISEG EHS 8060n HV module and an ISEG HPn300, controlled by
a W-Ie-Ne-R MPOD crate were employed as this modules satisfy all
demands.
Both systems have a fast hardware based over-current trip switch with a channel-wise adjustable current limit and they can be controlled via Simple Network Management Protocol (SNMP) commands over Ethernet or directly over CAN bus.\\
The current architecture of an emergency high voltage shutdown system
offers two solutions.  A purely hardware-based approach and a hardware
plus software approach.  The hardware approach couples the HV system
of the cathode and the field-cage with the HV system of the GEM stack
by an interlock cable.  In case of a trip in one of the two systems
both are shutdown immediately.
In this scenario a trip in the GEM stack also causes the drift voltage to trip and vice versa.\\
The long ramping time of the cathode after such a trip introduces
hereby a significant dead time of several hours.  For the hardware
plus software approach the HV systems are not coupled by an interlock
cable and therefore trip independently.  To minimize the the potential
danger of sustained discharges between the first GEM foil and the last
field-cage strip in case of a trip of the GEM stack, a software
adjustment of the drift cathode voltage to \SI{60}{\percent} of the
nominal value is implemented.  In case of a trip of the cathode HV no
additional actions have to be taken.  Besides the trip behavior, two
additional security functions are implemented: if the measured
voltages rises above a threshold voltage or if the current steps up
unexpectedly the TPC is ramped down with \SI{100}{\volt\per\second} to
prevent severe damages.  The moving average over the last 10 values is
calculated continuously and compared with the set value.  If the
deviation is larger than the limit an alarm rings.
The limit can be adjusted via the slow control interface (\secref{sec:slowcontrol.gui}) with a default value of \SI{5}{\percent} of the set voltage of the respective channel.\\
To detect over-currents, the average of the last ten measured values
is compared with the average of ten values from ten seconds earlier.
Especially the currents are good indicators for a possible failure of
the detector, because any short-cut between the GEM foils or at the
field-cage will result in a sudden increase of the current.  The
default for the over-current check is an increase of the current by a
factor of 1000.  This value can also be configured via the slow
control interface.  It has been found that this check is only
necessary during the ramping.  As soon as the desired potentials are
reached the hardware over-current protection of the high voltage
modules is preferable as it is much faster.
\subsection{Temperature, Pressure and Gas Flow}
\noindent
Various sensors are directly attached to the TPC, namely temperature
sensors on the pad-plane and the outside field-cage walls and gas flow
and pressure sensors.  For the temperature measurement at the outer
surface of the TPC 210 Dallas 18B20U 1-wire temperature sensors are
used.  The 12 PT100 temperature sensors located on the pad-plane are
digitized by AD7998 12-Bit ADCs and read out via an Ethernet-to-I$^2$C
adapter.  The ASF1400 gas flow sensor is read out via RS232 while the
pressure sensor is read out by a custom-made ATMega32 based read out.
\subsection{User Interface}
\label{sec:slowcontrol.gui}
\begin{figure}[htp]
  \begin{center}
    \includegraphics[width=\textwidth]{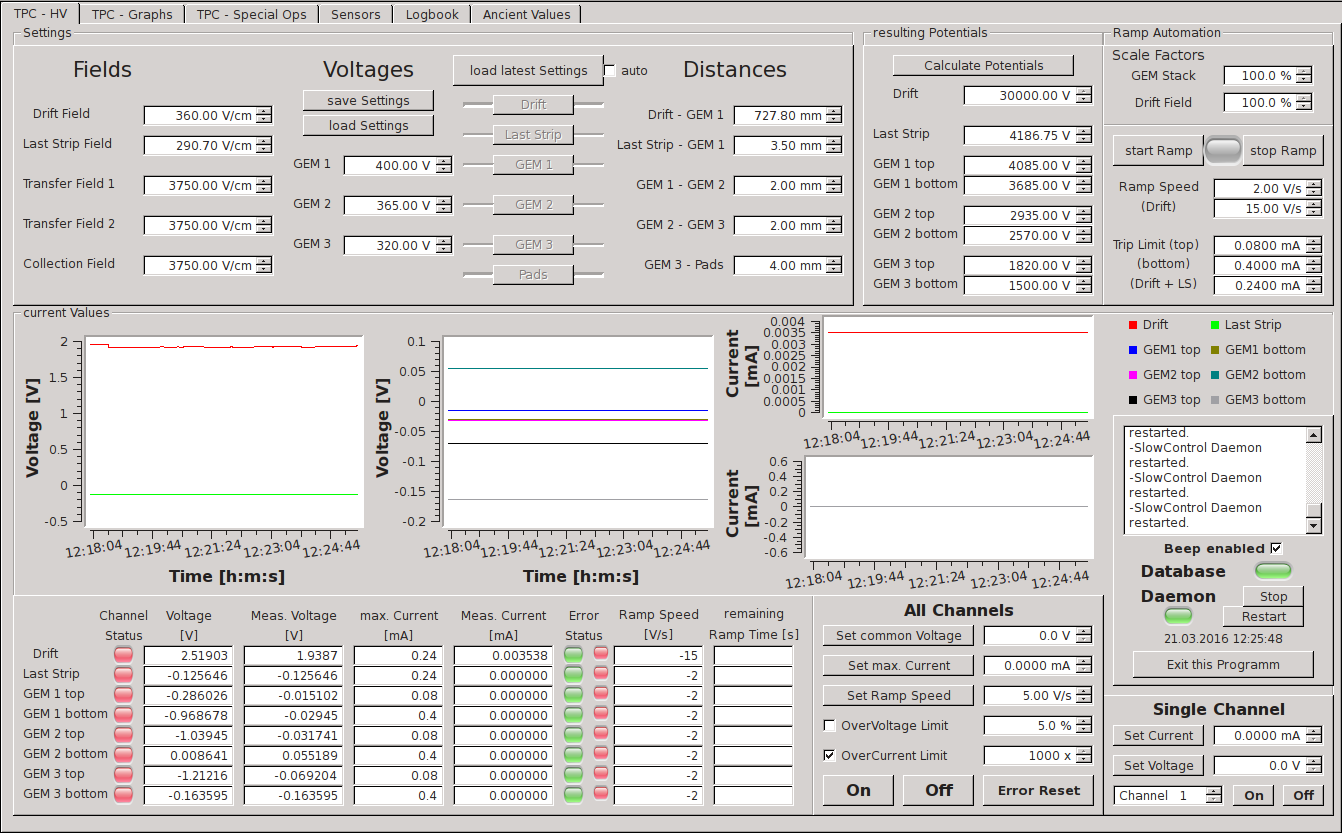}
    \caption[labelInTOC]{Screenshot of the slow control user
      interface. For details see text.}
    \label{pic:slowGUI}
  \end{center}
\end{figure}

\noindent
The control and read out of the high voltage modules, temperature,
flow and pressure sensors is realized by a C++ based graphical user
interface (GUI).  In order to communicate between the GUI and the high
voltage modules a background daemon is used.  Additionally all data
such as the actual voltages are collected by this daemon and written
to a SQL based database where they are saved with a time-stamp in
order to associate them to the detector measurement data files and for
a later use during the data analysis.  The GUI directly accesses this
database to display all recent values.  Furthermore it is used to
calculate all needed electric potentials from user defined electric
fields and potential differences across the GEMs (\figref{pic:slowGUI}
upper part) taking into account a linear scaling factor
(\figref{pic:slowGUI} upper right area).  If all values are set the
ramping of all channels to the desired values with given ramp speeds
can be started from the GUI.  The lower part of the interface display
is mainly to monitor the crucial voltages and currents.  There is also
the possibility to change specific settings such as trip limits, or
ramping speeds for all or single HV channels.  Further details
concerning the slow control system can be found in
\cite{DKaiser:2014}.
\subsection{Low-Voltage System}
\label{sec:slowcontrol.lv}
\noindent
Besides the high voltage also the low voltage of the 42 front-end
cards as well as the 11 ADCs has to be controlled.  The low voltage
for both systems is supplied by two pairs of Agilent 6031a and two
pairs of Agilent E3634 power supplies which are controlled by a
LabView based slow control software.  To communicate with the GPIB
based power supplies a GPIB-to-Ethernet converter is used.  The power
supplies are floating with their ground being defined by the detector
to avoid ground loops.  In order to distribute the low voltage to the
individual front-end cards and ADCs break-out-boxes are used.  The
sense wires for measuring the actual voltage are placed inside these
boxes.

%% file: commissioning/commissioning.tex
\section{Commissioning at the FOPI spectrometer}
\label{sec:Com}
\begin{figure}[!ht]
  \centering
  \includegraphics[width=0.7\textwidth]{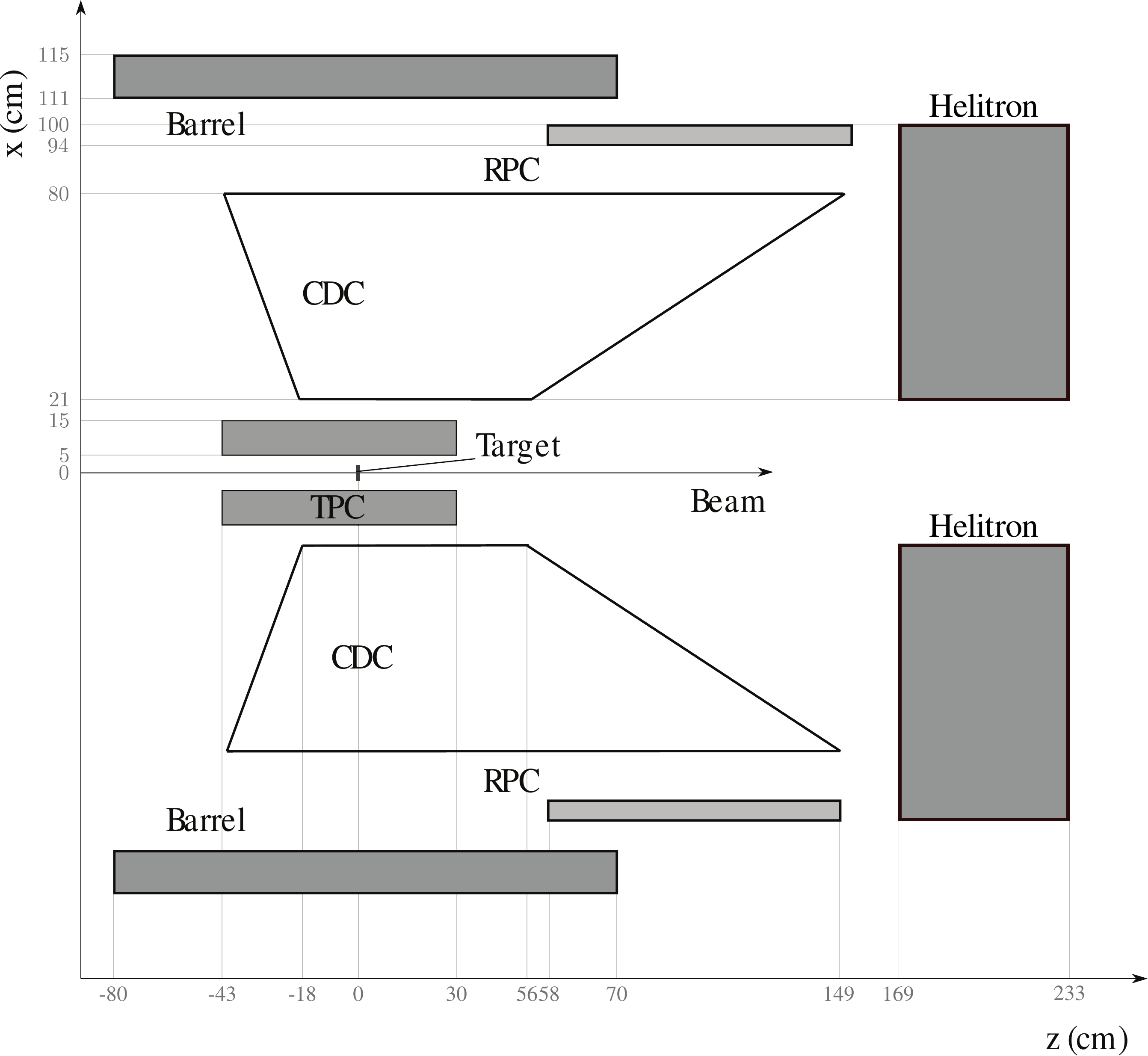}
  \caption{Schematic view of the sub-detector arrangement within the
    FOPI spectrometer.} 
  \label{fig:prototype.scheme}
\end{figure}

\begin{figure}[ht]
  \centering
  \includegraphics[width=0.8\textwidth]{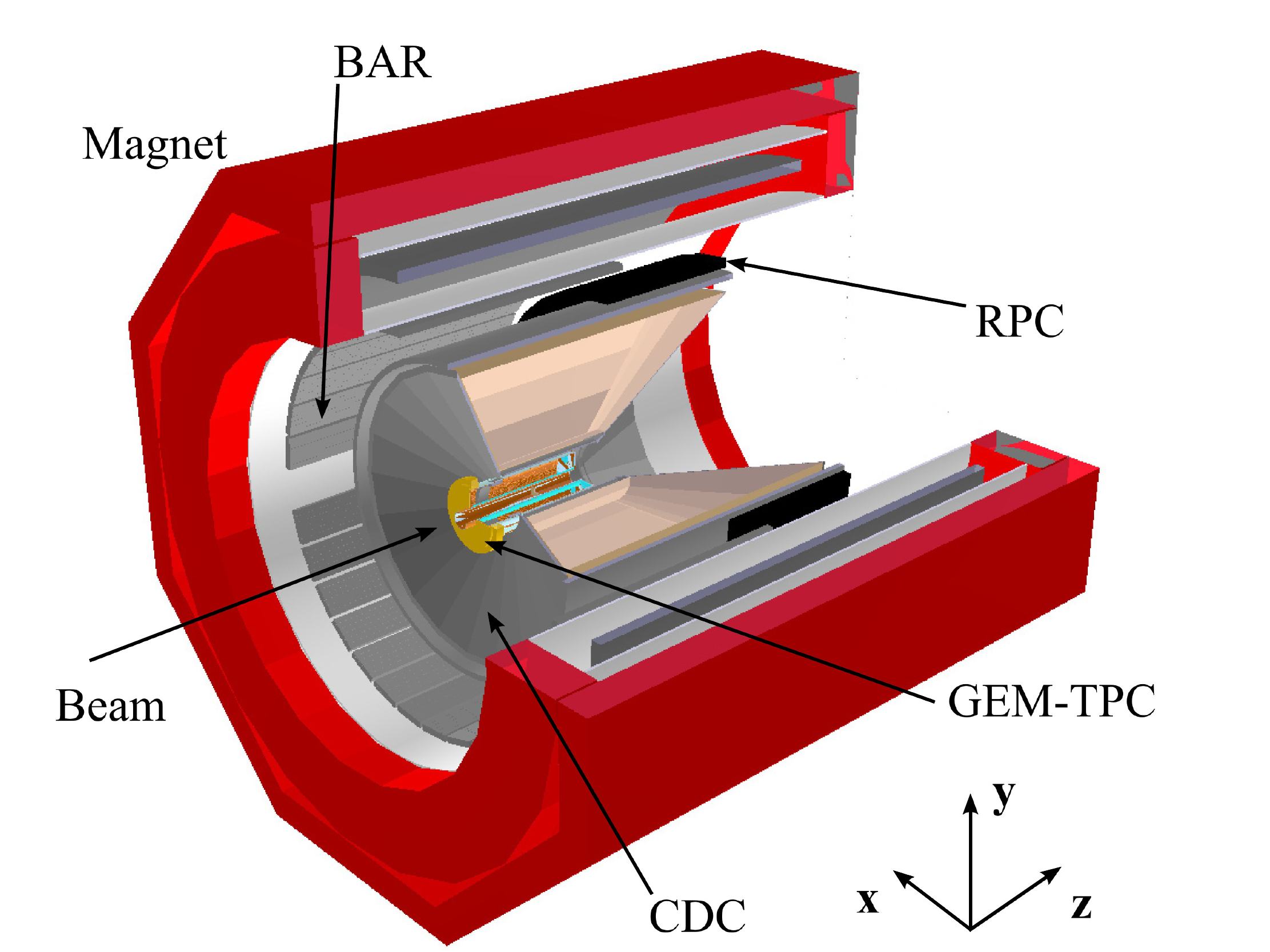}
  \caption{Model of the barrel part of the FOPI spectrometer showing,
    with increasing radial distance from the beam axis, the GEM-TPC,
    the central drift chamber (CDC), the resistive plate chambers
    (RPC), the scintillator barrel (BAR), and the $0.6\,\T$ solenoid
    magnet.} 
  \label{fig:fopi3D}
\end{figure}

\noindent
To test the performance of the large \gt it was installed inside the
FOPI \cite{FOPI1} experiment at GSI (Darmstadt, Germany),   
a large-acceptance spectrometer 
designed to study the properties of compressed nuclear matter produced
in heavy-ion collisions at energies from $0.1\,A\GeV$ to
$2.0\,A\GeV$. 
The detector system has an almost complete azimuthal 
symmetry and nearly 4$\pi$ coverage of the solid angle. 
It consists of a central drift chamber (CDC), a
scintillator barrel,  
and resistive plate chambers (RPC) arranged cylindrically around the
target. The setup is   
operated inside a $0.6\,\T$ superconducting solenoid
magnet. The barrel part is augmented by a system of forward detectors
including a planar drift chamber (Helitron) and a plastic scintillator
wall (PLAWA). 
The \gt was mounted inside the inner bore of the CDC, supported only
on the upstream side at the
media flange by a light-weight ring-shaped structure made of carbon
fiber. 
A schematic cross-section of all FOPI barrel sub-detectors, including the
TPC, is shown in \figref{fig:prototype.scheme}. 
\Figref{fig:fopi3D} shows a 
three-dimensional  
model of the FOPI barrel system. 
FOPI is able
to identify light charged particles like pions, kaons and protons and
intermediate mass fragments.  Hadron resonances and neutral hadrons
can also be reconstructed from their decay products.\\
Before the installation of the GEM-TPC, the FOPI spectrometer
delivered a vertex resolution of about \SI{5}{\mm}  
in the $x-y$ plane and \SI{5}{\cm} along the beam axis.  The 
momentum resolution for charged particles was
\SIrange[range-phrase={~-~}]{4}{10}{\percent}.  The motivation to
install the \gt in FOPI was threefold: (i) the barrel detectors
provided a full-fledged magnetic spectrometer and thus and ideal
reference environment to test the \gt in different beams including heavy-ion
interactions, (ii) the \gt was expected to improve substantially the
vertex and 
secondary vertex resolutions of the spectrometer, and (iii) 
with its large
geometrical acceptance covering both barrel and forward detectors, the
\gt was supposed to  
improve the inter-connection between these different sub-detector
elements.

The commissioning and operation of the \gt installed in FOPI was done
in several phases. 
In the
first phase, which started in October 2010, the TPC was
commissioned with cosmic muons without and 
with the solenoid magnetic field. Data with cosmics were also taken in
2011 and 2012 
for more detailed studies of the detector performance
\cite{MB15}, which will be 
reported on in a forthcoming publication. 
In the 
second phase the stability of the \gt in collisions of different
heavy-ion beams with nuclear 
targets was tested in November 2010 and April 2011. The third phase
consisted of a three-week period of 
stable operation of the \gt during the FOPI S339 physics campaign in
June 2011 with  
a $1.7\,\GeV/c$ pion beam impinging on solid targets. 
For this secondary beam only a 
rate of 25.000 $\pi^-$/spill with a total spill length of
\SI{3.5}{\second} and a duty cycle factor of \SI{42}{\percent} was
available. The major aim of this experiment was to measure charged
and neutral kaons produced almost at rest in a nuclear medium.  
The data taken
during this period resulted in the first measurement of the specific
energy loss over a wide momentum range with a large \gt
\cite{Boehmer:2014hna}. The physics results from this beam time
\cite{Boehmer:2015}  
will be the subject of a forthcoming publication. 

During the first two phases, two different gas mixtures, 
Ar/CO$_2$ (90/10) and a Ne/CO$_2$ (90/10), and 
several different field configurations
were tested, both for the drift field and for the GEM stack. For the
GEM stack, two categories of settings were used: the \emph{Standard
  Setting}, optimized for stable operation in hadron beams and derived
from the COMPASS experiment \cite{Altunbas:02a}, and the so-called
\emph{Ion Backflow Settings}, which were found to minimize the ion
backflow in a triple-GEM setup \cite{Ball:2012xh}. The values for the
fields and GEM potential differences corresponding to these two
categories for a gas gain of $\sim 1\EE{4}$, respectively, are
displayed in \tabref{tab:GEM_Settings}. Since the \gt is normally
operated at a lower gain of $1-2\EE{3}$, all potentials except the
drift field are scaled
down from the values given in the table by an overall scaling factor.  
\begin{table}[!ht]
  \centering
  \begin{tabular}{ c| S s |S s }
    \hline\hline
    & \multicolumn{2}{c}{Standard Setting} 	& \multicolumn{2}{c}{Ion Backflow Setting}	\\ \hline
    Drift Field 					& 400	&\volt\per\centi\meter		& 250	&\volt\per\centi\meter					\\
    $\Delta$U$_\mathrm{GEM1}$ 	& 400	&\volt 						& 330	&\volt									\\ 
    E$_{\mathrm{trans1}}$		& 3750	&\volt\per\centi\meter		& 4500	&\volt\per\centi\meter					\\
    $\Delta$U$_\mathrm{GEM2}$ 	& 365	&\volt			 			& 375	&\volt		 							\\ 
    E$_{\mathrm{trans2}}$		& 3750	&\volt\per\centi\meter		& 160	&\volt\per\centi\meter					\\ 
    $\Delta$U$_\mathrm{GEM3}$	& 325	&\volt			 			& 450	&\volt		 							\\ 
    E$_{\mathrm{Ind}}$			& 3750	&\volt\per\centi\meter		& 5000	&\volt\per\centi\meter					\\ \hline\hline
  \end{tabular}
  \caption{Standard and Ion Backflow settings for drift field and GEM
    amplification system. See \figref{pic:gemstack} for a sketch of
    the HV powering scheme of the GEM-TPC.}
  \label{tab:GEM_Settings}
\end{table}
\begin{itemize}
\item Several drift fields have been applied:
  \SIlist[list-units=single]{150;200;300;360}{\volt\per\cm} with the
  corresponding drift velocities:
  \SIlist[list-units=single]{0.9;1.4;2.2;2.9}{\centi\meter\per\micro\second}
  \cite{sauli:77a}.
\item Different voltage settings for the GEM-stack resulting in
  different gains have been tested ranging from
  \SIrange[range-phrase={~to~}]{80}{100}{\percent}.
\item Two gas mixtures, namely Ar/CO$_2$ and Ne/CO$_2$ have been used,
  both in a $90/10$ mixture by weight.
\end{itemize}
The third phase (physics campaign) was conducted with the Ar-based gas
mixture at a drift field of $234\,\V/\Cm$ and a GEM gas gain of $\sim
1\EE{3}$ achieved at a scaling factor of $81\%$ of the standard GEM
settings (second column in \tabref{tab:GEM_Settings}). 
\subsection{Tracks from Cosmic Muons}
\begin{figure}[ht]
  \centering
  \includegraphics[width=0.9\textwidth]{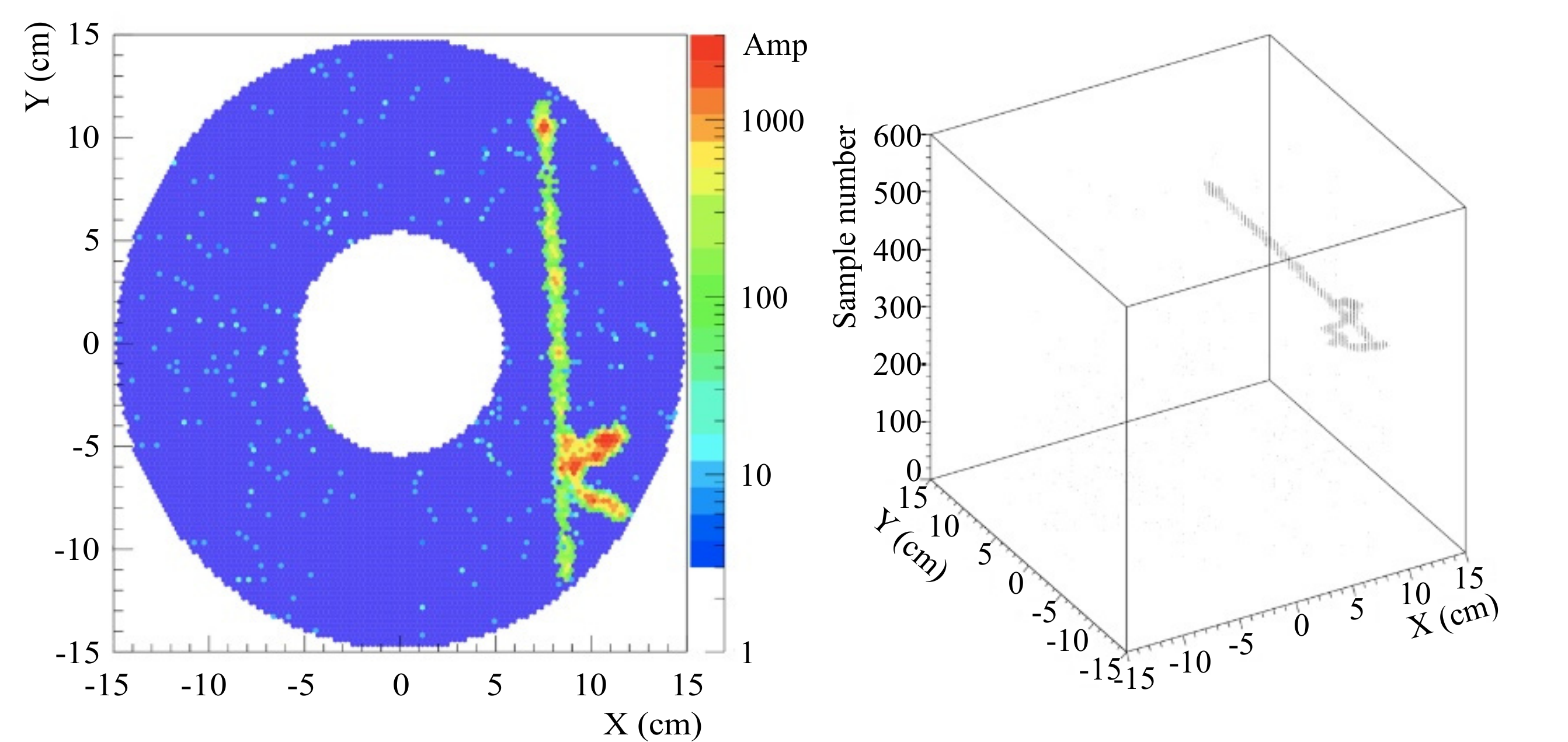}
  \caption{Online display of a cosmic muon track producing two 
    $\delta$-electrons in the \gt. The left panel displays
    the signal 
    amplitudes registered on the electrodes of the 
    pad plane in units of ADC channels, i.e.\ the 2-D projection of
    the track. The right panel 
    shows the full 3-D spatial information with the sample number on
    the $z$-axis.}
  \label{pic:prototype.realtpc-cosmic}
\end{figure}

\noindent
In the first phase, tracks from cosmic rays were measured using the
scintillator barrel of FOPI as a trigger.
\Figref{pic:prototype.realtpc-cosmic} shows the online event display for a
typical cosmic muon track traversing the chamber and producing
additional $\mathrm{\delta}$-electrons.  This event was registered
without a magnetic field and 
with the following settings for the TPC: a drift field of
\SI{360}{\volt\per\centi\meter} with an Ar/CO$_2$ (90/10) mixture,
which translates into a drift velocity of \SI{\approx
  2.9}{\centi\meter\per\micro\second}, and a GEM potential scaling factor of
\SI{85}{\percent} for standard settings. 
Various drift, gain, and gas settings were measured with cosmic muons
in order to get a comprehensive picture of the detector performance,
especially in terms of spatial resolution. 
%
\subsection{Tracks from Beam Interactions}
Several beam species were utilized to collect data for the joint 
GEM-TPC/FOPI system during two test experiments at GSI in
November 2010 and April 2011: $^{84}$Kr, $^{197}$Au and $^{22}$Ne
beams at 
$1.2\,A\GeV$, $1.0\,A\GeV$, and $1.7\,A\GeV$ 
kinetic energies, respectively, 
impinging on an Al target with \SI{2}{\percent} of a nuclear
interaction length placed in the inner bore of the GEM-TPC.  The beam
parameters 
were set to an average 
particle rate of 5$\cdot$10$^6$ particles/spill with a total spill
length of about \SI{10}{\second} and a duty cycle factor of
\SI{50}{\percent}. 
\begin{figure}[!ht]
  \begin{center}
    \begin{subfigure}[c]{.45\textwidth}
      \includegraphics[width=\textwidth]{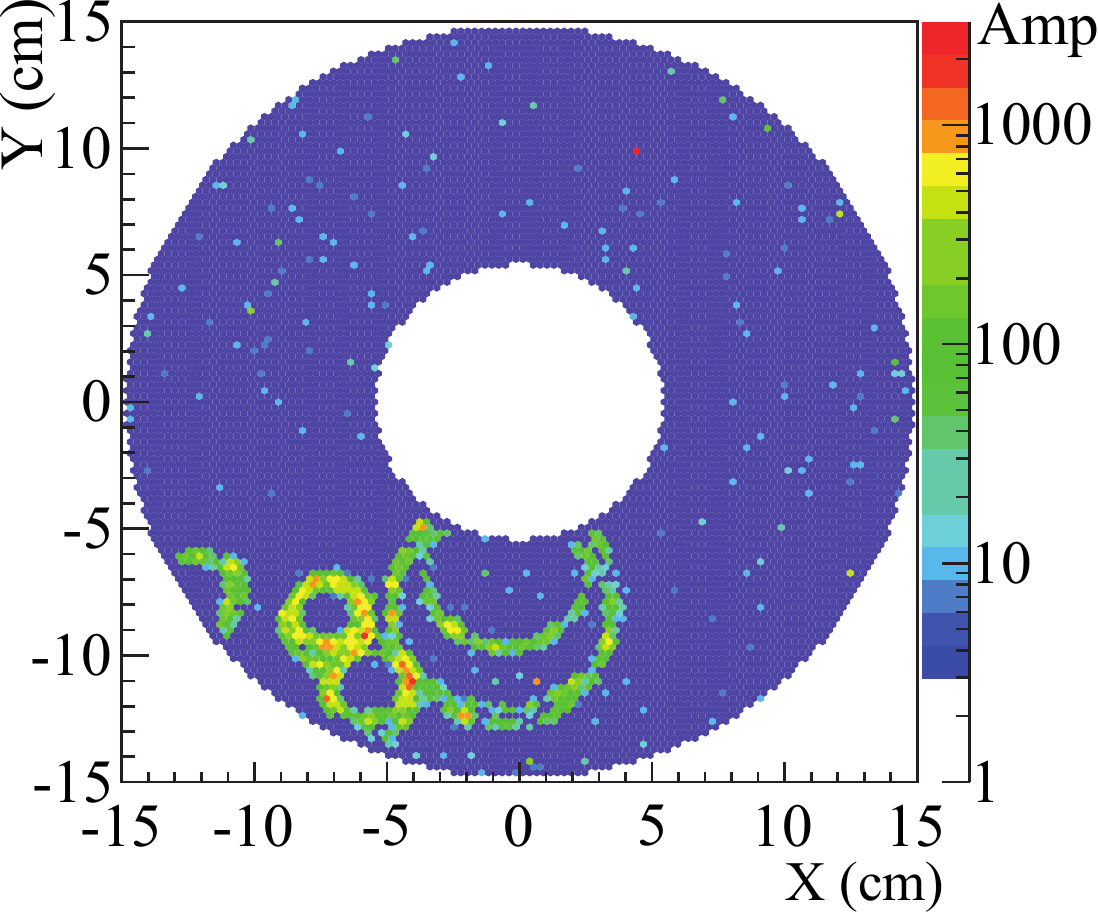}
      \caption{}
    \end{subfigure}
    \begin{subfigure}[c]{.45\textwidth}
      \includegraphics[width=\textwidth]{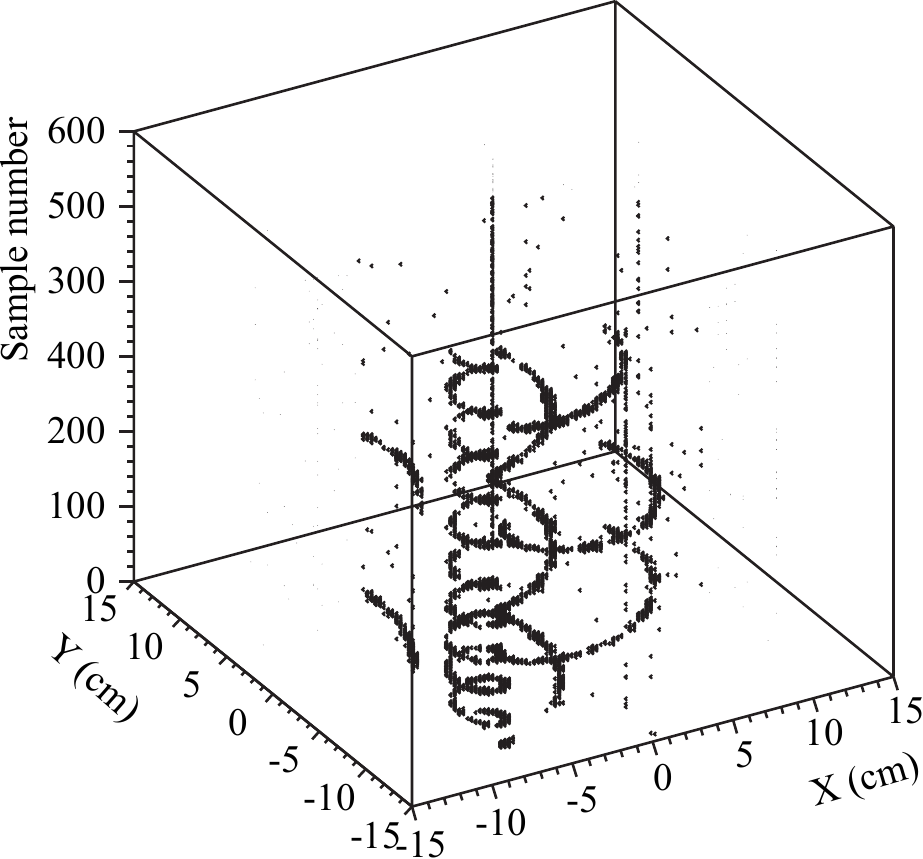}
      \caption{}
    \end{subfigure}
    \\
    \begin{subfigure}[c]{.45\textwidth}
      \includegraphics[width=\textwidth]{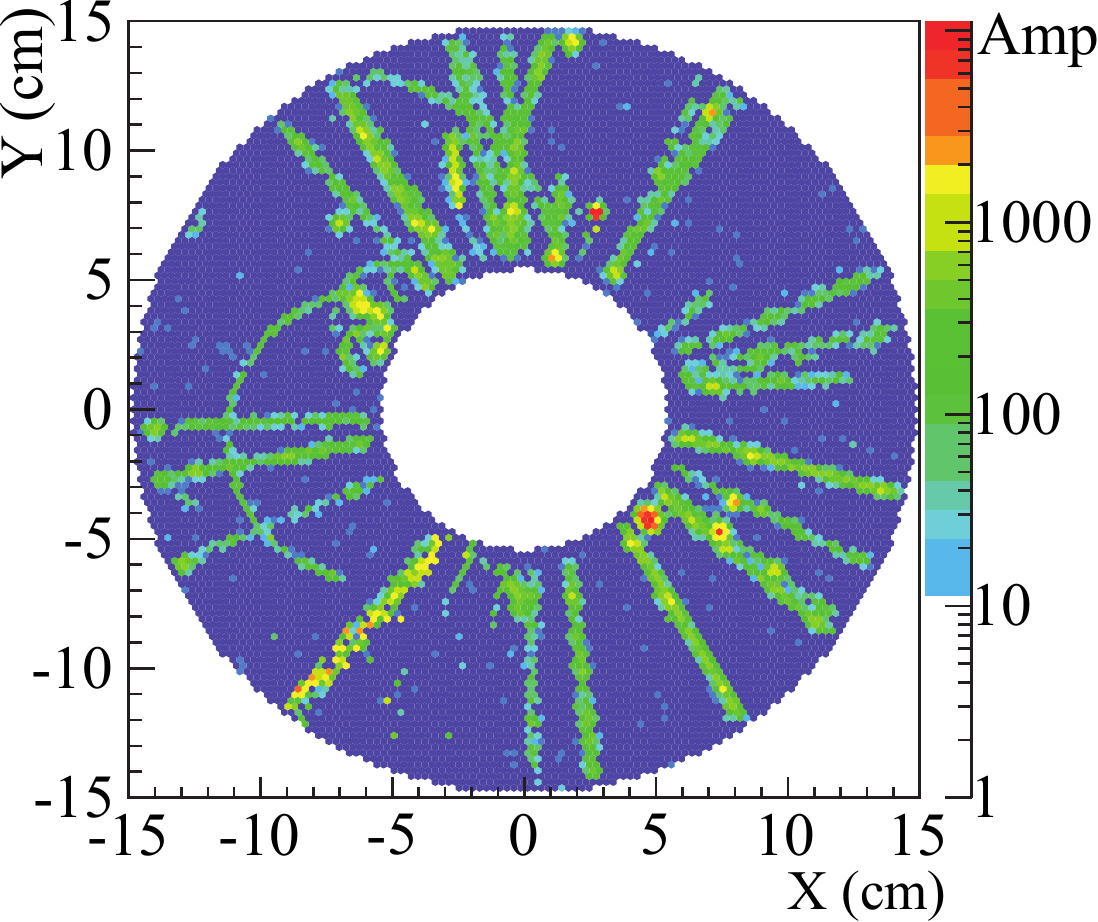}
      \caption{}
    \end{subfigure}
    \begin{subfigure}[c]{.45\textwidth}
      \includegraphics[width=1.1\textwidth]{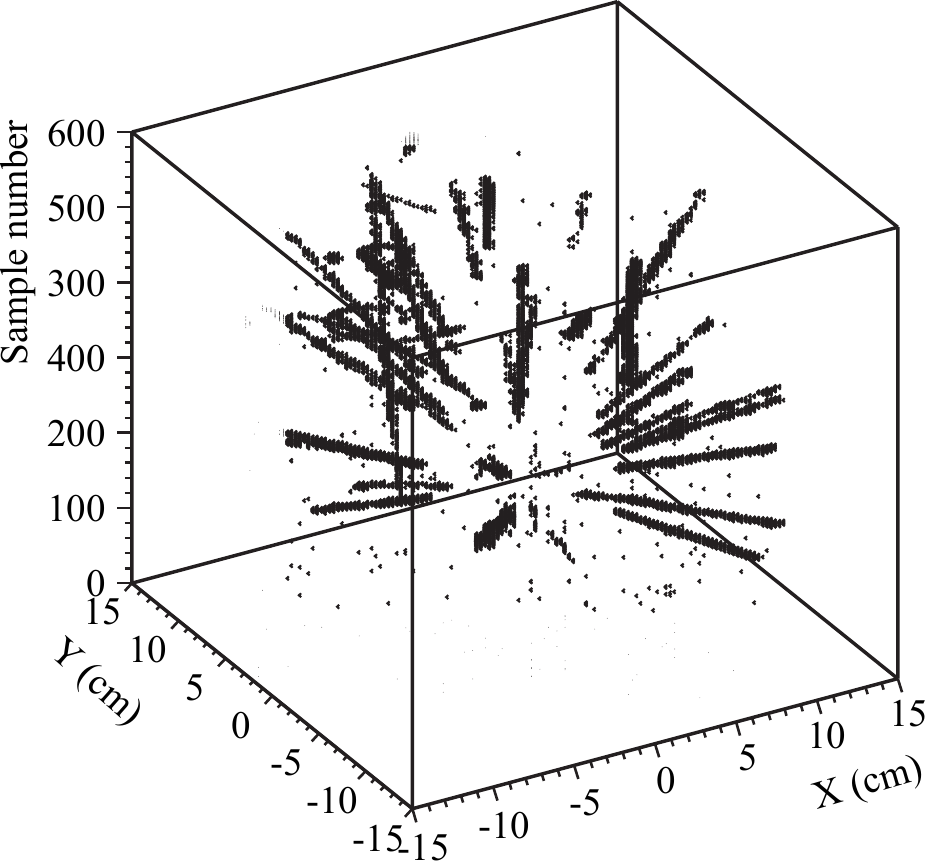}
      \caption{}
    \end{subfigure}
  \end{center}
  \caption{Online displays from the test with $^{22}$Ne ions with
    $1.7\,A\GeV$.   
    The left panels shows the signal amplitudes registered on the
    readout pads in units of ADC channels, the right panels the full
    3-D information with the sample number on the $z$-axis. 
    In the top row, an event is displayed where the beam ion
    presumably 
    interacted upstream of the target and produced several
    low-momentum particles curling in the magnetic field.  
    The bottom row shows a high-multiplicity event from an interaction
    of the beam in the Al target.}
  \label{pic:prototype.beam.event}
\end{figure}

\Figref{pic:prototype.beam.event} shows two typical events in the
GEM-TPC for $^{22}$Ni+Al reactions. 
The chamber settings during this test were
\SI{360}{\volt\per\centi\meter} drift field in the Ar/CO$_2$ (90/10)
mixture and a GEM scaling factor of  
\SI{86}{\percent} for standard settings.  
The top row shows an event where the beam particle presumably
interacted upstream of the target, producing tracks almost parallel to
the beam axis and several low-momentum particles curling in the
\SI{0.6}{\tesla} magnetic field. 
In the bottom row an interaction in the Al target is displayed which
produced a large number of charged particles.

\begin{figure}[tbp] \centering
  \includegraphics[width=0.8\textwidth]{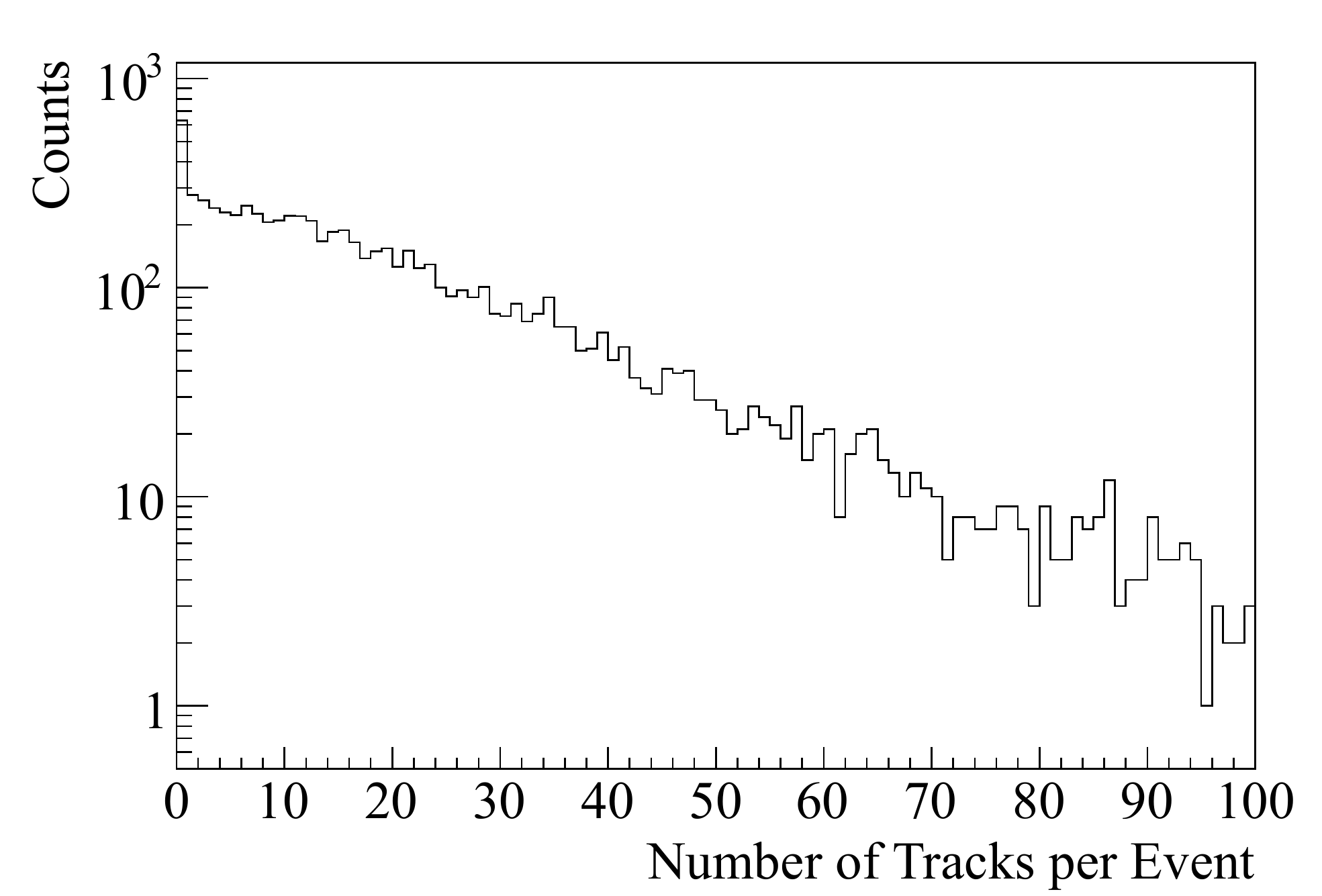}
  \caption{Track multiplicity distribution of the reconstructed events in
    the GEM-TPC detector for $^{22}$Ne+Al collisions at $1.7 
    \,\mathrm{AGeV}$.}
  \label{fig:prototype.beam.trackNum}
\end{figure}

These online event displays provide direct evidence for the excellent
3-D tracking capability of a TPC, where the large density of hits
along a particle track facilitates pattern recognition even in
high-multiplicity events. 
\Figref{fig:prototype.beam.trackNum} shows the track multiplicity,
i.e.\ the distribution of the number of reconstructed tracks per
event, 
in the \gt for $^{22}$Ne+Al collisions at $1.7\,A\mathrm{GeV}$. As one
can see, events with up to 100 tracks were successfully reconstructed.

For track fitting, the track segments in the TPC with an
average length of \SI{14}{\centi\meter} are combined with the longer
segments in the CDC. This allows the determination of
particle momenta exceeding $1\,\GeV/c$, for which
the curvature in the \gt alone is too small. 
%
%

%% file: calibration/calibration.tex
\section{Calibration procedures}
\label{sec:CalSys}
\noindent
A TPC requires a number of calibration procedures to be performed at
regular intervals in order to achieve the design performance.  For the
GEM-TPC, the following calibration steps were implemented and applied:
\begin{itemize}
\item pedestal and noise determination,
\item gain determination and equalization,
\item drift velocity calibration.
\end{itemize}
They will be described in the following subsections.

While the first procedure is done on a sample-by-sample basis for each
channel separately, the other two require a reconstruction of signals
from the digitized data.  The full set of algorithms to reconstruct
particle tracks in the TPC will be described in a forthcoming
publication 
 ; here, we limit ourselves to the
description of the algorithms to generate so-called
\emph{reconstruction clusters}, or simply \emph{clusters}, which then
also provide the input for the tracking algorithms.
\subsection{Cluster Reconstruction}
\label{sec:ClusterReco}
As a first step, consecutive samples of each pad with amplitudes above
a given threshold are combined to so-called \emph{pad hits} by using a
simple pulse shape analysis (PSA) technique.  Samples belonging to one
pulse are identified by a search for local minima.  The pad hit gets
assigned a total signal charge (the sum of all associated sample
amplitudes above threshold) and a signal time (time of the peak sample
minus a constant offset of \SI{149.5}{\nano\second} which takes into
account the rise time of the signal).  In the second step,
accumulations of pad hits in 3-D space are further compressed into
clusters in order to reduce the amount of data and further suppress
noise hits.  The clustering procedure starts from the pad hit with the
highest amplitude in a given drift frame and adds pad hits both from
neighboring pads on the pad plane and along the drift direction if
they are sufficiently close in time.  The hit association to a given
cluster stops at local minima of the hit amplitude and a new cluster
is started at the remaining pad hit with the highest amplitude until
all pad hits are taken into account.  Then a cut is applied on
clusters consisting of only one pad hit, which effectively removes
electronic noise hits.  The total amplitude of a cluster is given by
the sum of the amplitudes of the associated pad hits, its position is
calculated as the center of gravity of the associated pad hits.
%
\subsection{Pedestal and noise determination}
\label{sec:CalSys.pedestal}
\begin{figure}[ht] \centering
  \includegraphics[width=0.45\textwidth]{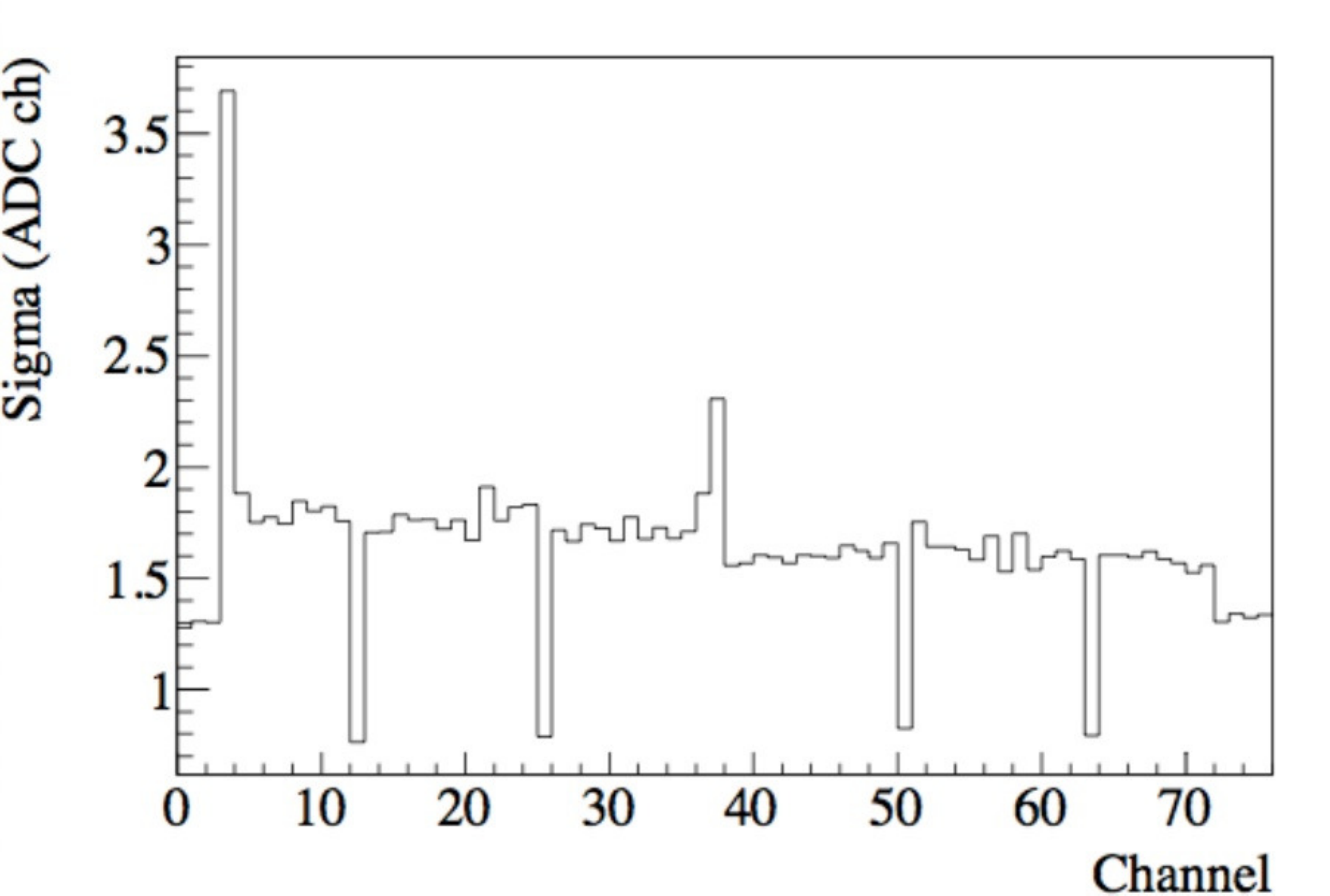}
  \hspace{0.5cm}
  \includegraphics[width=0.45\textwidth]{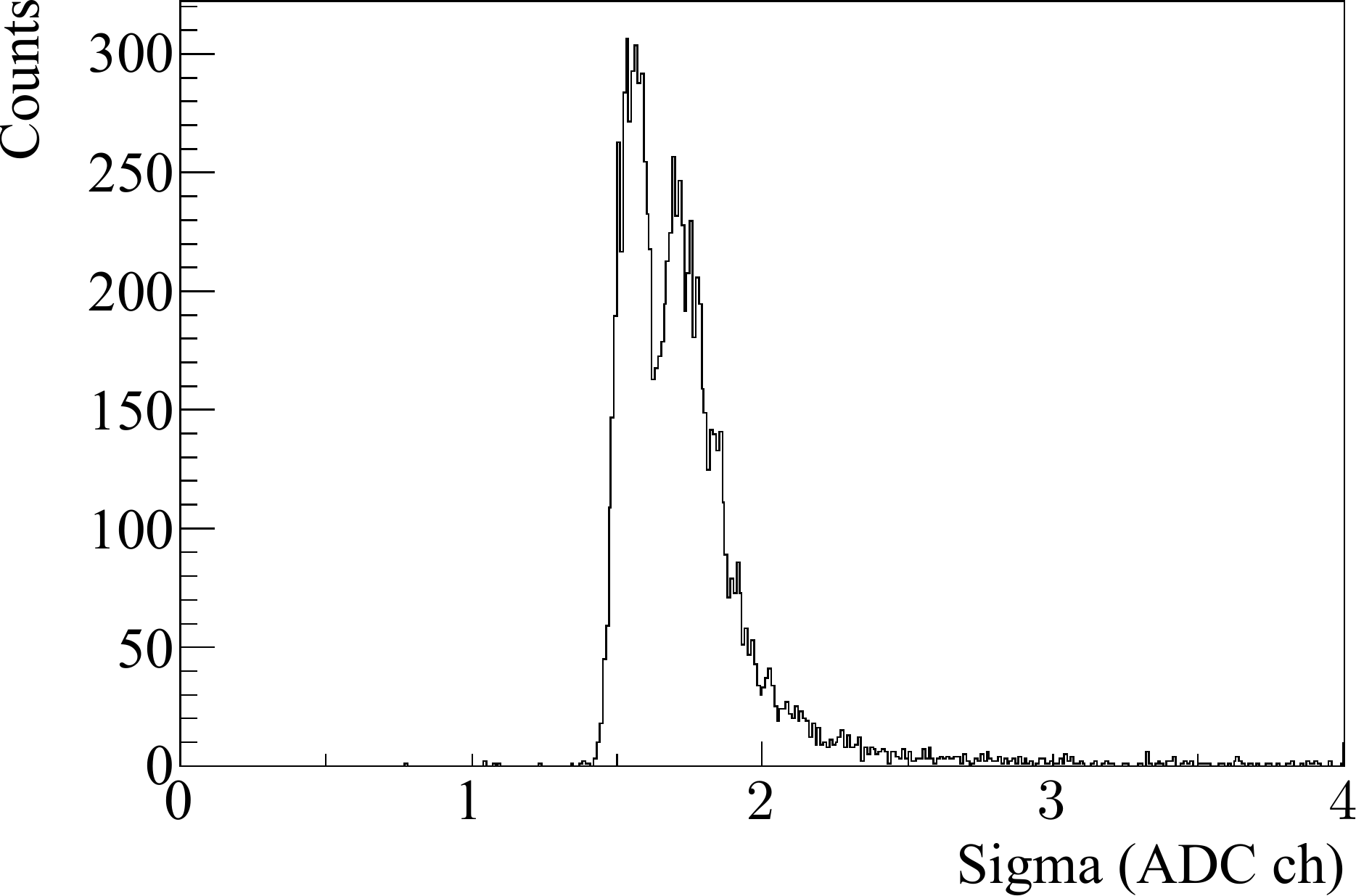}
  \caption{Left panel: noise values in ADC channels for all 76
    channels of one front-end chip connected to the GEM-TPC. The first
    and the last four channels are not connected to the pad plane. The
    four intermediate low-noise channels are used for the correction
    of fixed-pattern noise caused by leakage of the switched capacitor
    array. The first 36 channels have longer input lines on the
    front-end cards (see Fig.~\ref{pic:fecard}) and a correspondingly
    higher noise. Right panel: distribution of the noise values for
    all connected channels of the GEM-TPC.}
  \label{pic:noise}
\end{figure}

\begin{figure}[ht]
  \begin{center}
    \begin{subfigure}[b]{.495\textwidth}
      \centering
      \includegraphics[width=\columnwidth]{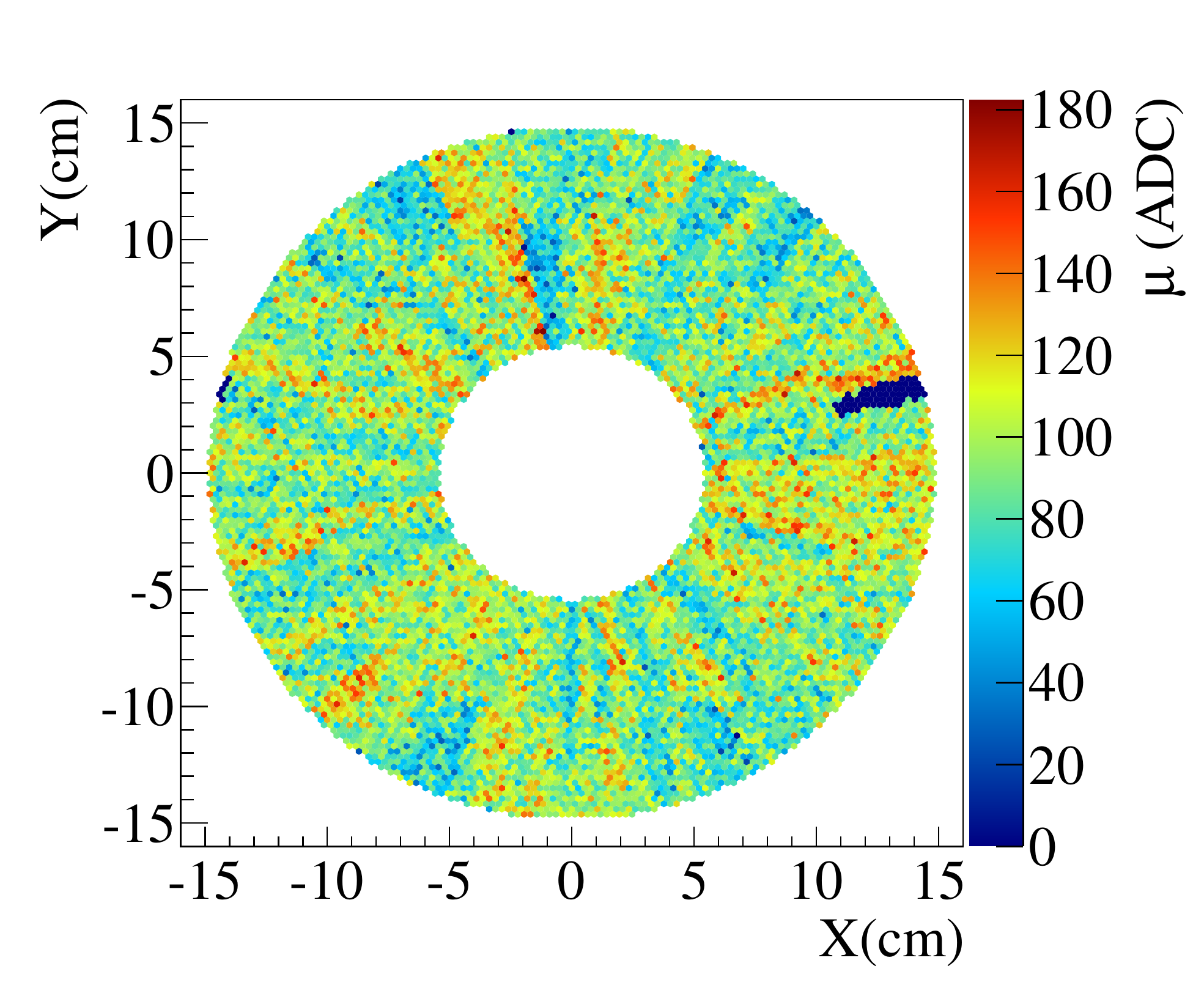}
      \caption{}
      \label{label2}
    \end{subfigure}
    \begin{subfigure}[b]{.495\textwidth}
      \centering
      \includegraphics[width=\columnwidth]{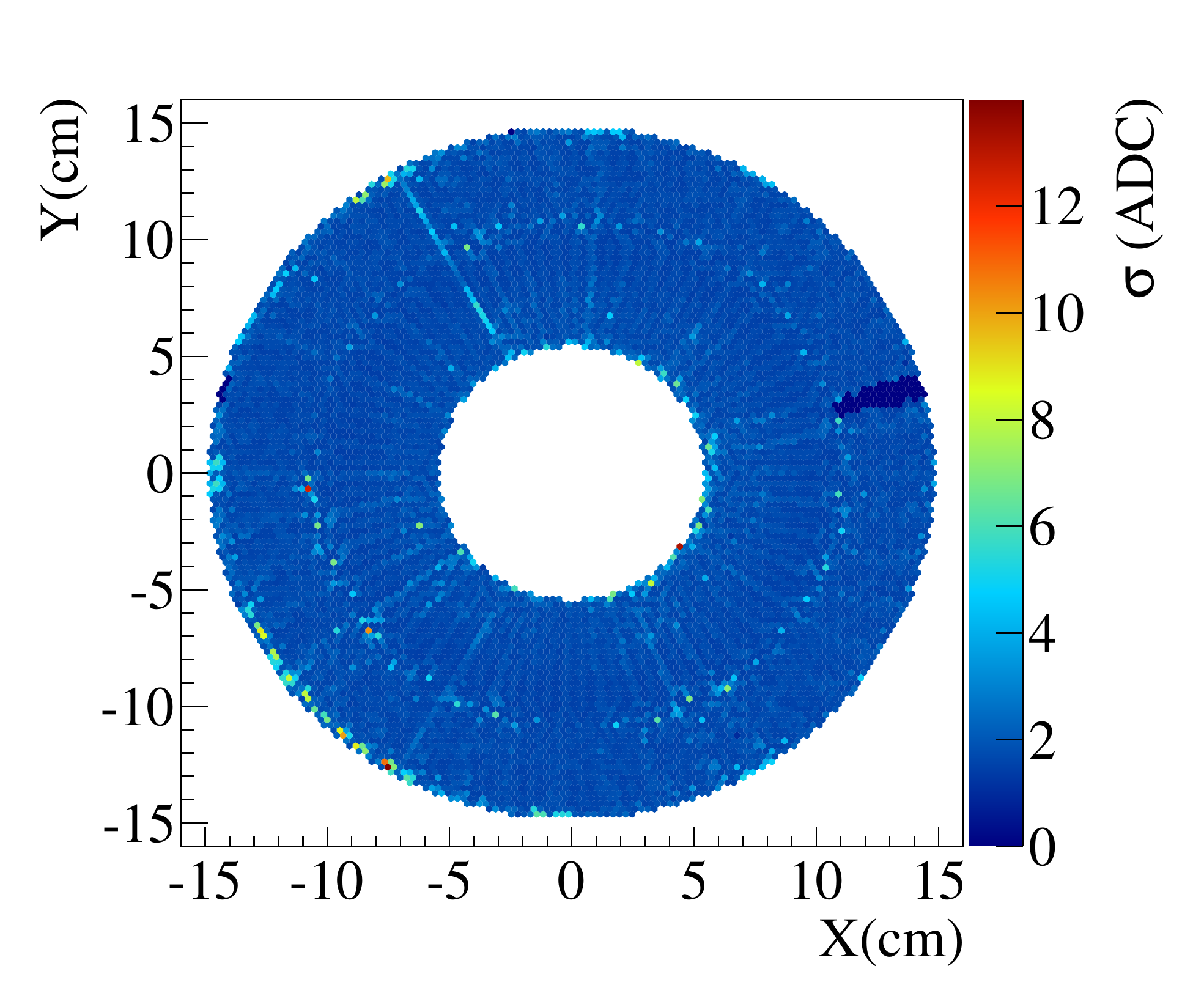}
      \caption{}
      \label{label1}
    \end{subfigure}
    \caption[shortCaption]{(a) Pedestals and (b) noise values in units
      of ADC channels as a function of the pad position. The dark blue
      areas correspond to defect electronic channels.}
    \label{pic:noise2d}
  \end{center}
\end{figure}

In order to cope with the large amount of data from the TPC, noise
hits are suppressed already during data taking after digitization in
the ADC, as described in Sec.~\ref{sec:FEE_DAQ.readout}. The zero
suppression logic is implemented in a Xilinx Virtex-4 FPGA on the
ADC card and consists of baseline (pedestal) subtraction with
individual values for each channel, common mode and fixed pattern
noise correction for all channels connected to one side of the chip
packaging, and suppression of all channels with amplitudes smaller
than a threshold defined by a programmable multiple (typically between
$3$ and $5$)
of the individual noise value of the given channel.\\
The pedestal and noise values for each channel are determined in
regular so-called pedestal runs. In these runs, which are performed at
nominal chamber conditions but without beam, a random trigger is used
to read out a small number of samples of each channel without zero
suppression. The number of samples per channel is limited to typically
6--10 by the bandwidth of the readout chain. From this data sample,
the average (pedestal) and the rms width (noise in ADC channels or
equivalent noise charge, ENC, if converted to charge) of the recorded
baseline values are determined offline for each channel and saved in a
file. The values in the file are then uploaded to the ADC card using
the I2C protocol and applied to the raw data when zero suppression is
activated. \\
%
The noise values of all channels of one front-end chip can be seen in
\figref{pic:noise} (left).  The distribution of the noise values for
all 10254 channels connected to the pad plane is shown in
\figref{pic:noise} (right). There are two groups of channels with a
slightly offset noise distribution, related to the signal paths on the
front-end card. The first 36 channels of each chip have longer input
lines than the second 36 channels. This reflects in a slightly
different noise level of these groups of channels, as can be seen from
\figref{pic:noise}. The outlier channels with higher noise are
typically the ones on the edges connected to the
far side of the chip pins.\\
The average noise (including the tails) of all channels is 1.83 ADC
channels, corresponding to an equivalent noise charge of
$720\,e^-$. The most probable values of the ENC for the two groups
of channels are $616\,e^-$ and $674\,e^-$, for input
capacitances of \SI{13}{\pico\farad} and \SI{16}{\pico\farad},
respectively.
\Figref{pic:noise2d}(a) shows the pedestals values for all pads of the
pad-plane in units of ADC channels, while \figref{pic:noise2d}(b)
displays their noise values.
The ring structure with a higher noise level, which can be seen in
\figref{pic:noise2d}(b) is caused by channels on the edges of the
block with longer track lengths from the connector to the chip on the
front-end PCB.
\subsection{Calibration of the gas gain}
\label{sec:calibration.krypton}
\begin{figure}[ht]
  \begin{center}
    \begin{subfigure} [t]{.43\textwidth}
      \includegraphics[width=0.9\textwidth]{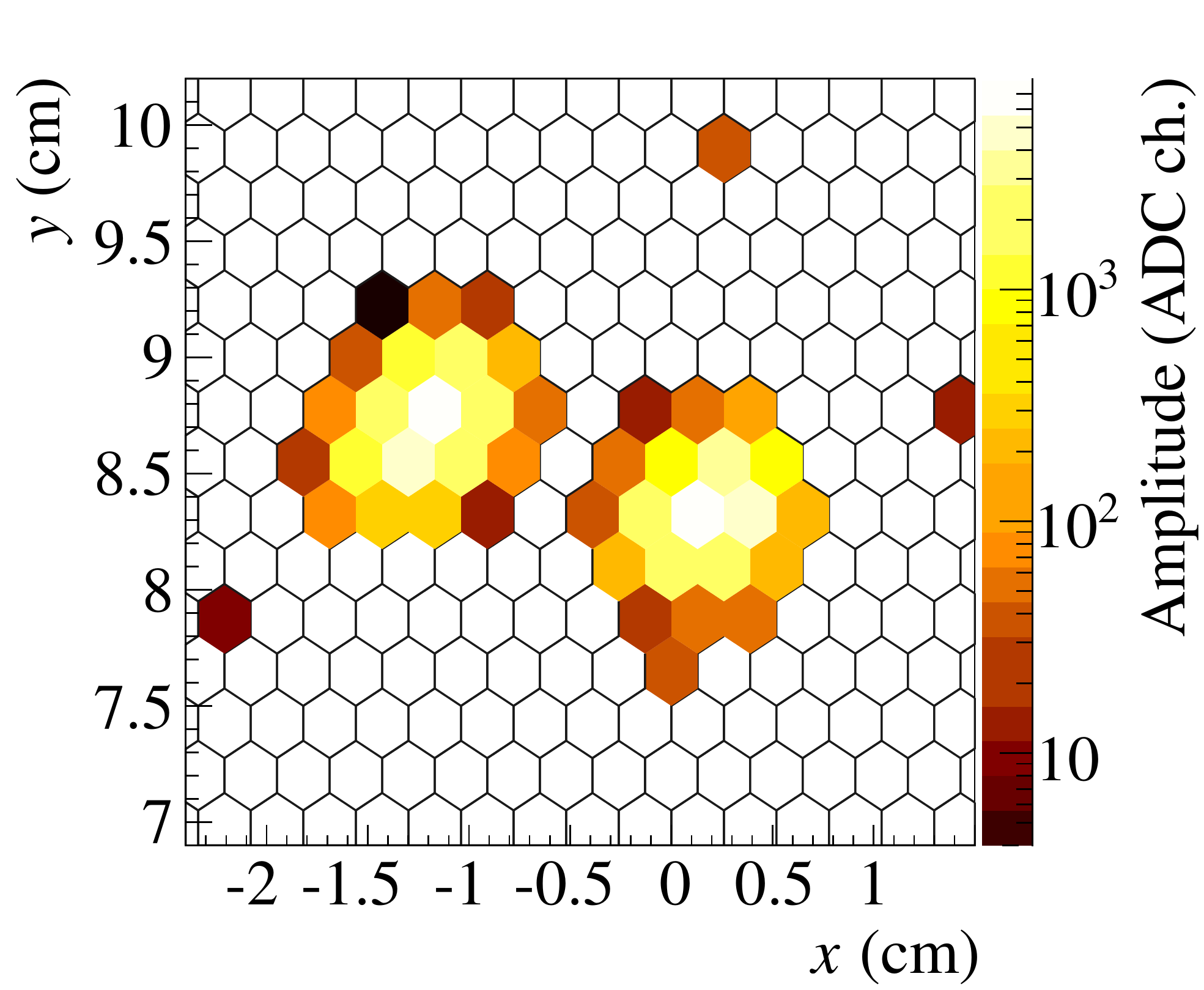}
      \caption {}
      \label{fig:kr.event}
    \end{subfigure}
    \hfill
    \begin{subfigure} [t]{.53\textwidth}
      \includegraphics[width=0.9\textwidth]{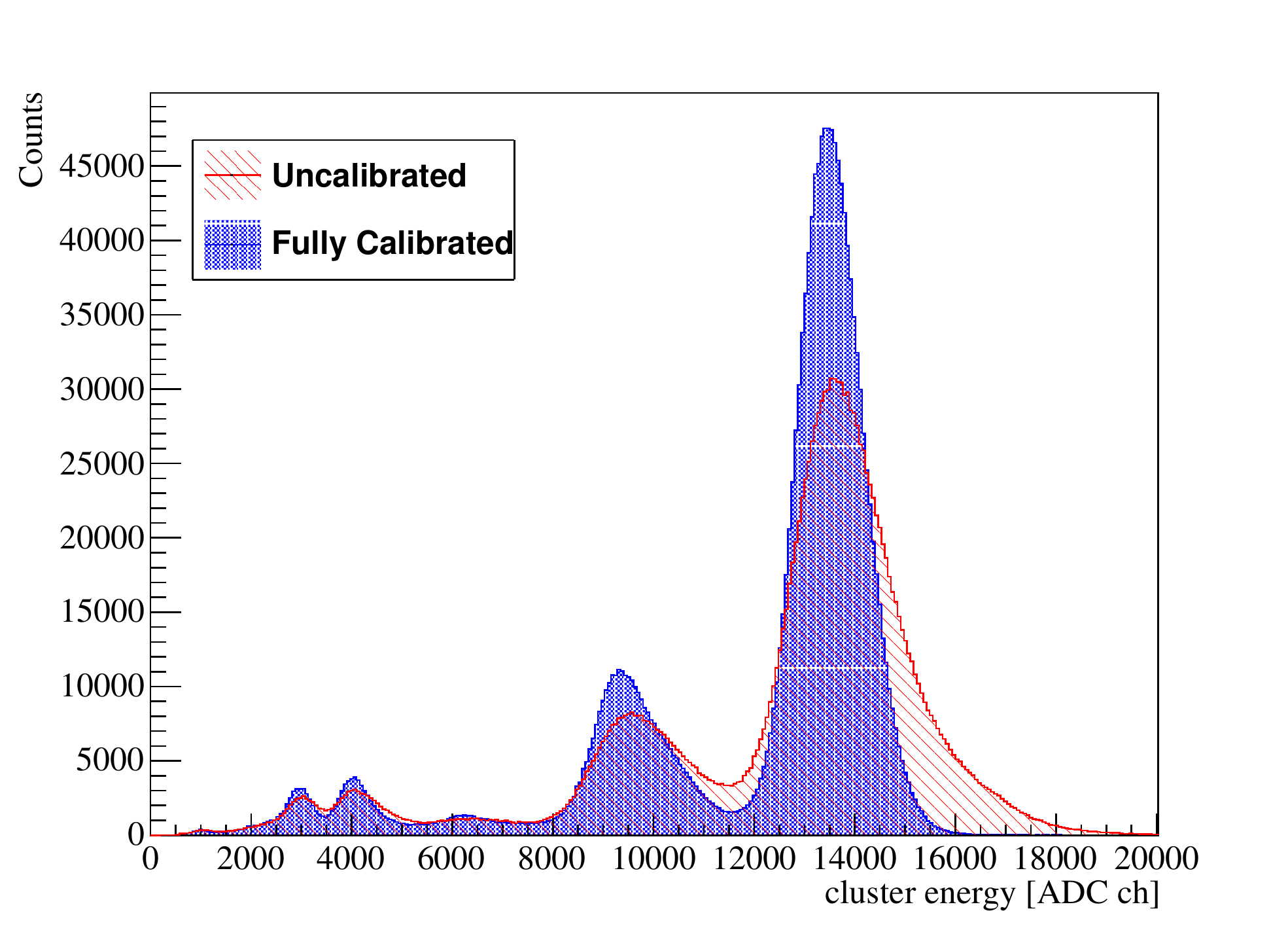}
      \caption{}
      \label{fig:spec_arco_83p}
    \end{subfigure}
  \end{center}
  \caption{(a) Clusters from $^{83m}$Kr decays recorded in the GEM-TPC
        with a random trigger. (b) Cluster amplitude spectrum for Ar/CO$_2$ (90/10)
        recorded with the GEM-TPC at a voltage scaling factor of
        \SI{81}{\percent}, (red) without and (blue) with pad-wise
        relative gain calibration.}
  \label{fig:spec-reso}
\end{figure}

\noindent
The gas gain is one of the key parameters which decisively influences
the performance of the TPC in terms of spatial and energy
resolution. For GEM detectors, the effective gain of the GEM stack is
the important quantity, because it takes into account charge losses in
the GEM structures. Using an X-ray source with known rate, it may be
determined e.g.\ from the current measured on the readout pads. In
addition to the absolute value of the effective gain, which determines
the signal-to-noise ratio, local relative variations of it strongly
impact the energy resolution of the detector and thus the
$\diff{E}/\diff{x}$ resolution. In multi-GEM detectors, such
variations may arise from non-uniformities of the hole diameters or
the foil spacings across the active area, and have been observed to be
as large as \SI{20}{\percent} \cite{Hallermann:10}. Contributions to
local variations of the gain can also originate from slightly varying
electronic conversion gains for individual channels of the front-end
chips. In order not to compromise the
$\diff{E}/\diff{x}$ resolution of the detector, a gain homogeneity of
\SI{1}{\percent} across the GEM surface has to be achieved by a proper
calibration. An elegant method to monitor and equalize the total gain
(effective gas gain and electronic conversion gain) for each pad
during data taking is by introducing metastable radioactive
$^{83m}\text{Kr}$
into the drift volume. This technique does not require any physical
invasion of the detector and allows full coverage of its active
volume, and has already been applied in various large TPCs (e.g. ALEPH
\cite{Decamp:90-1}, DELPHI \cite{DeMin:1995tk}, HARP \cite{Dydak:04},
ALICE
\cite{Alme:2010ke}, STAR \cite{SN0424}). \\
%
%
In order to introduce $^{83m}$Kr into the GEM-TPC, a dedicated
$^{83}$Rb source with an activity of \SI{2.5}{\mega\becquerel} was
produced at the HISKP\footnote{Helmholtz-Institut f\"ur Strahlen- und
  Kernphysik, Nussallee 14-16, D-53115 Bonn} cyclotron
\cite{Rasulbaev:08}.
$^{83}$Rb decays with a half-life of
\SI{86.2}{\day} via electron capture into $^{83}$Kr, mostly populating
an isomeric, metastable state at $41.55\,\keV$ with a half-life of
\SI{1.83}{\hour}. This state decays into the ground state via a
short-lived excited state at \SI{9.4}{\kilo\electronvolt}. Both
transitions occur predominantly via internal conversion, producing
short-ranged electrons. The resulting shell excitation is removed by
X-rays or Auger electrons.  The decay spectrum observed in the TPC
thus shows several peaks between \SI{9.4}{\kilo\electronvolt} and
\SI{41.6}{\kilo\electronvolt}.\\
%
%
\begin{figure}[ht]
  \sbox\twosubbox{%
    \resizebox{\dimexpr.9\textwidth}{!}{%
      \includegraphics[height=3cm]{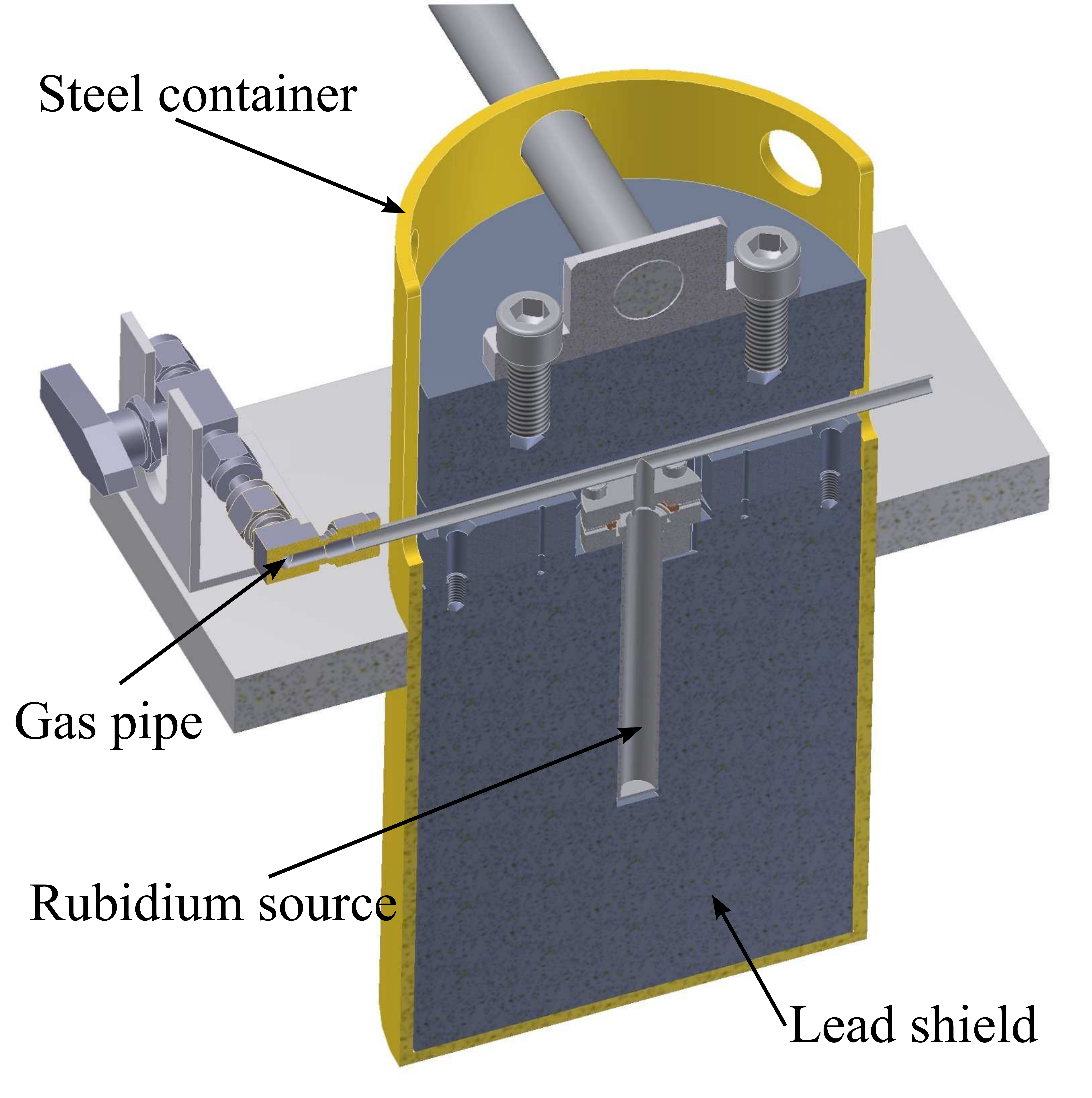}%
      \includegraphics[height=3cm]{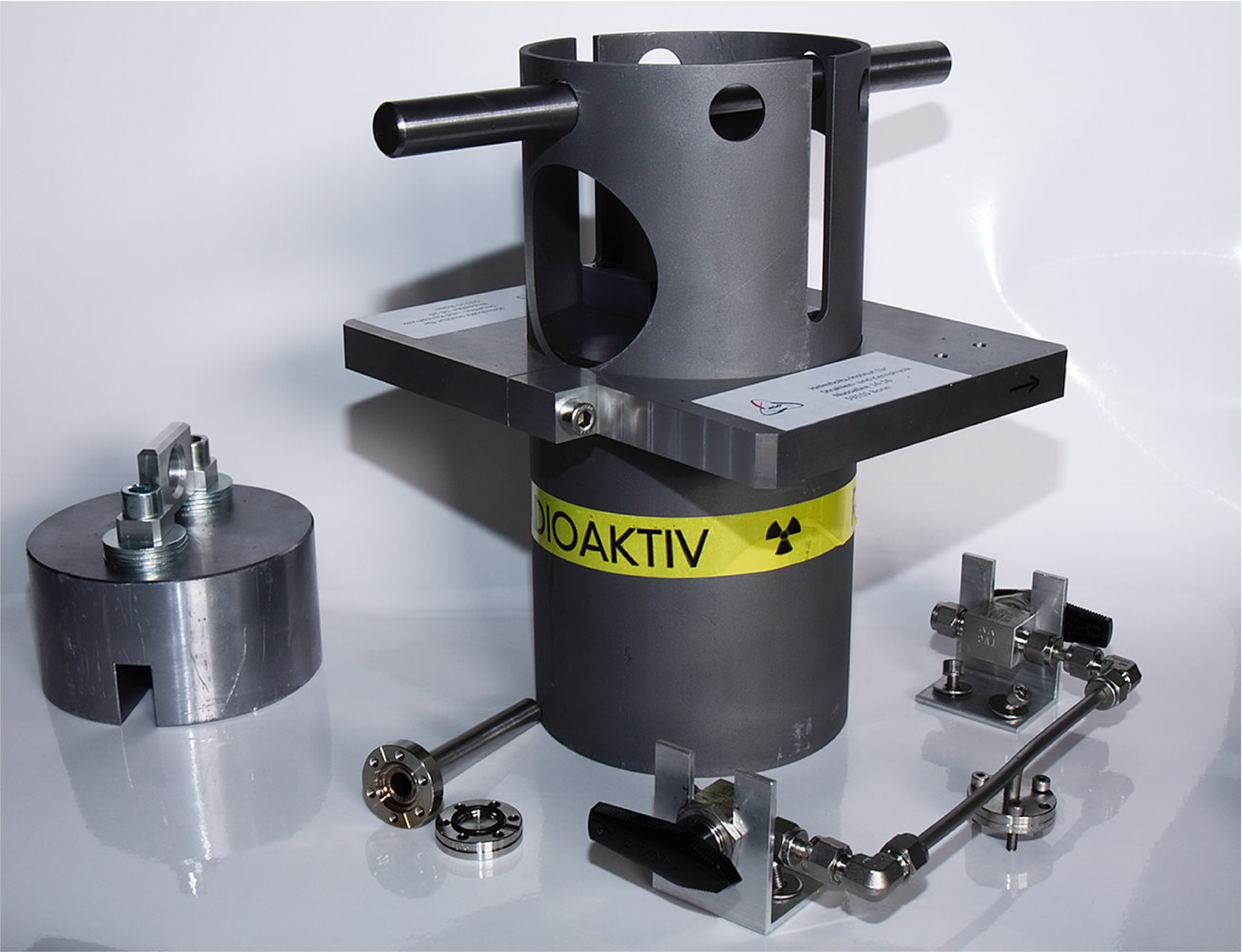}
    }%
  }%
  \setlength{\twosubht}{\ht\twosubbox}
  \begin{center}
    \subcaptionbox{}{\includegraphics[height=\twosubht]{./figures/calibration/rb_container}}
    \subcaptionbox{}{\includegraphics[height=\twosubht]{./figures/calibration/rb_container_foto.jpg}}
    \caption {\label{fig:Rb-container} Lead-shielded container housing
      the $^{83}$Rb source. The source is contained in a steel finger
      that can be attached to the gas inlet via a bypass system. The
      outer shielding absorbs higher energetic decay photons that are
      emitted during the decay.}
  \end{center}
\end{figure}
The container for the $^{83}$Rb source can be seen in
\figref{fig:Rb-container}.  It is connected to the gas input line of
the GEM-TPC such that, when the corresponding valves are opened, the
gas flows past the inner steel tube holding the radioactive material,
carrying the evaporated $^{83m}$Kr into the chamber volume.  The
container shielding consists of \SI{13}{\cm} lead that absorbs higher
energetic decay photons.
\begin{figure}[ht]
  \begin{center} \centering
    \includegraphics[width=0.6\textwidth]{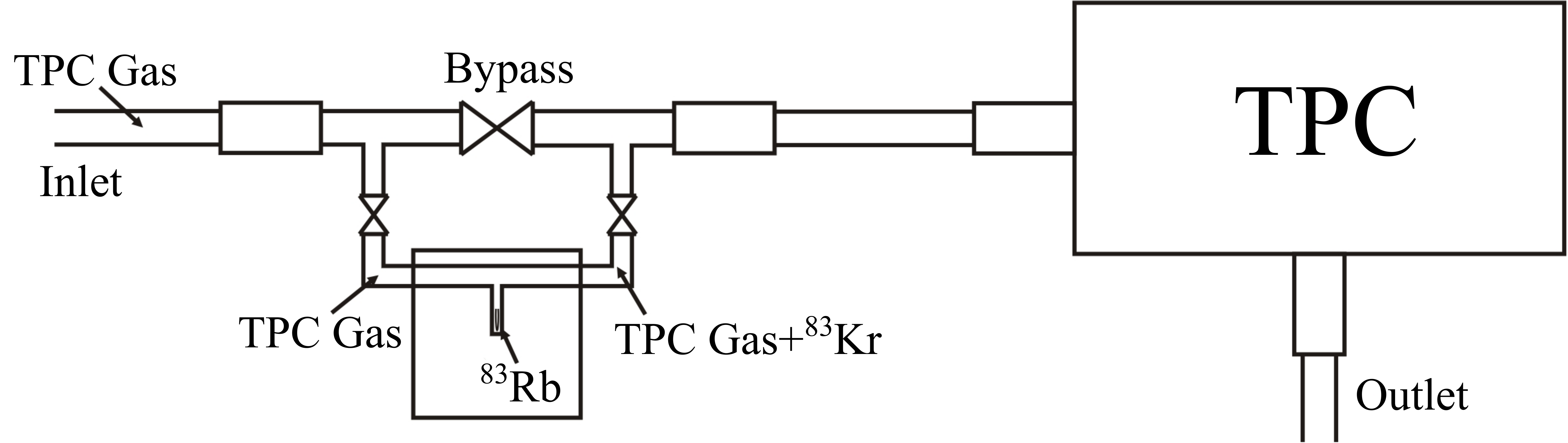}
    \caption {\label{fig:Kr-gas-connection} Sketch of the $^{83}$Kr
      container connected to the TPC gas system.}
  \end{center}
\end{figure}
During normal operation of the TPC
the Rb container is bypassed.\\
%
Dedicated Kr calibration runs have been performed regularly during
beam times using a random trigger. The electrons emitted in the decay
process are stopped quickly in the gas and lead to large but spatially
confined clusters which are reconstructed using the reconstruction
techniques described at the beginning of this
section. \Figref{fig:kr.event} shows an example of typical clusters
created by Kr decays.\\

\begin{figure}[ht]
  \begin{center} \centering
    \includegraphics[width=0.6\textwidth]{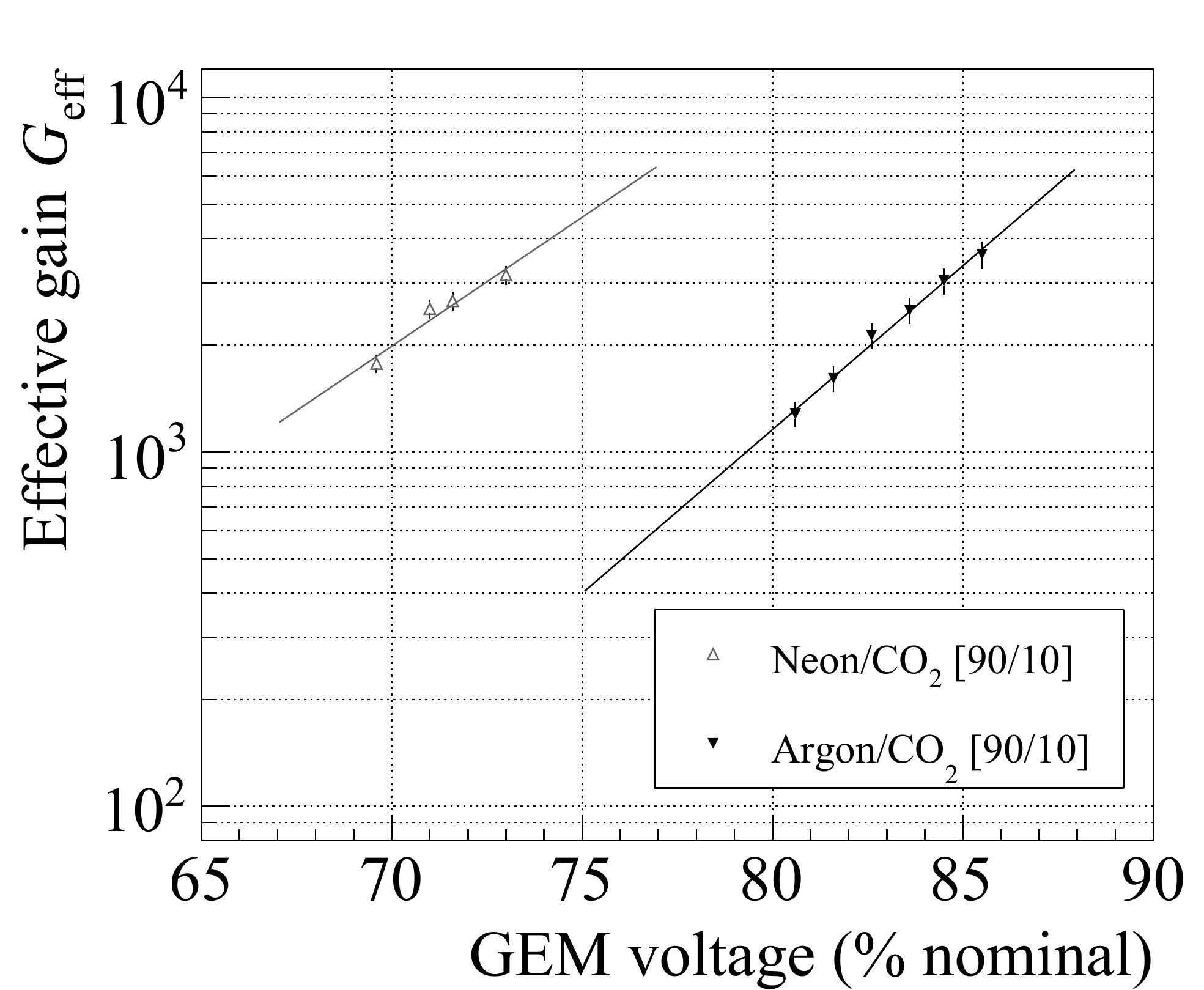}
    \caption {\label{fig:kr.Geff} Effective gain $G_\mathrm{eff}$ of
      the GEM-TPC from Kr calibration runs as a function of the GEM
      voltage settings (cf. \tabref{tab:GEM_Settings}).}
  \end{center}
\end{figure}

\begin{figure}[ht]
  \begin{center}
    \begin{subfigure}[t]{0.45\textwidth}
      \includegraphics[width=\textwidth]{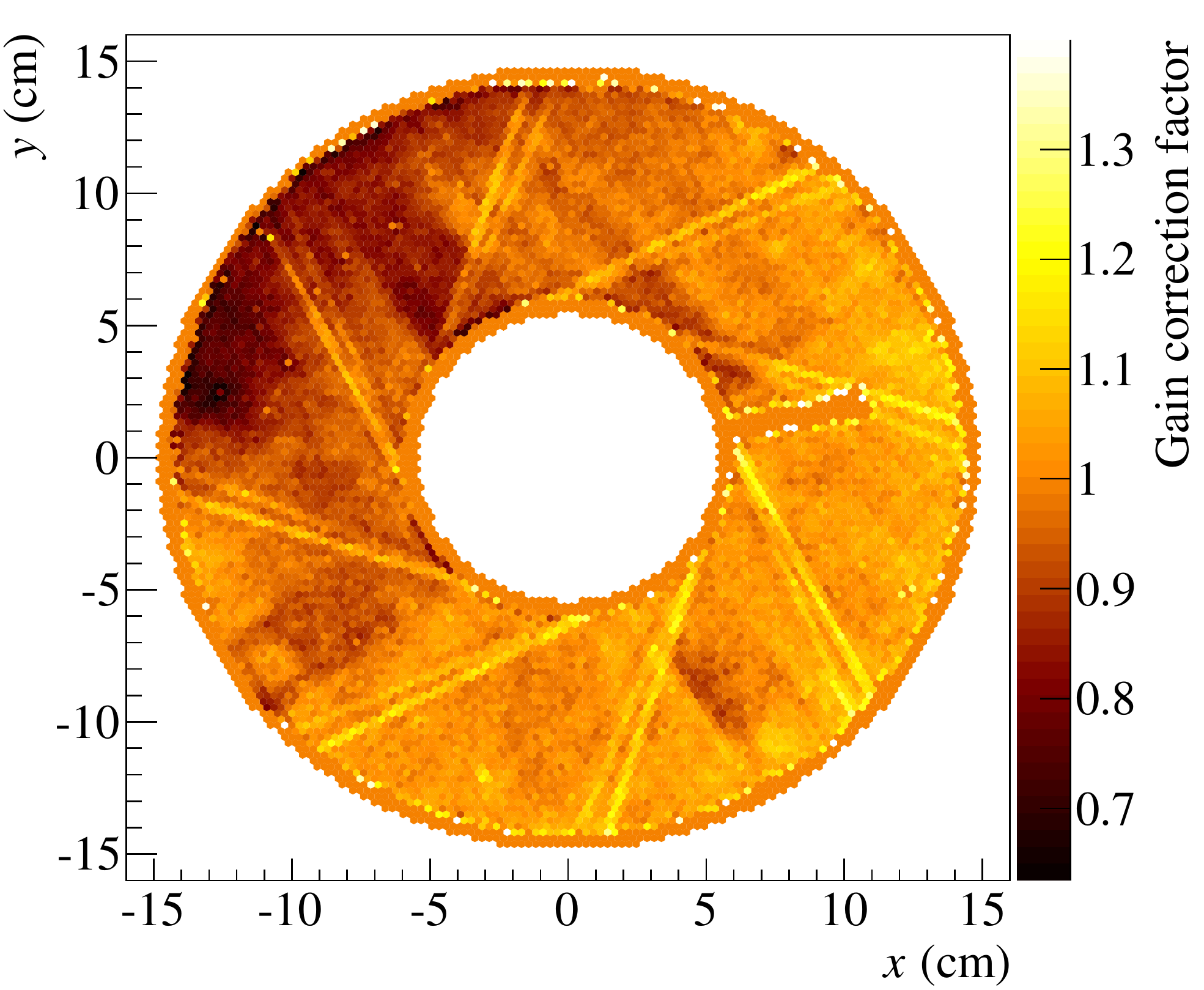}
      \caption{}
      \label{fig:Peaks_uni}
    \end{subfigure}
    \begin{subfigure}[t]{0.45\textwidth}
      \includegraphics[width=\textwidth]{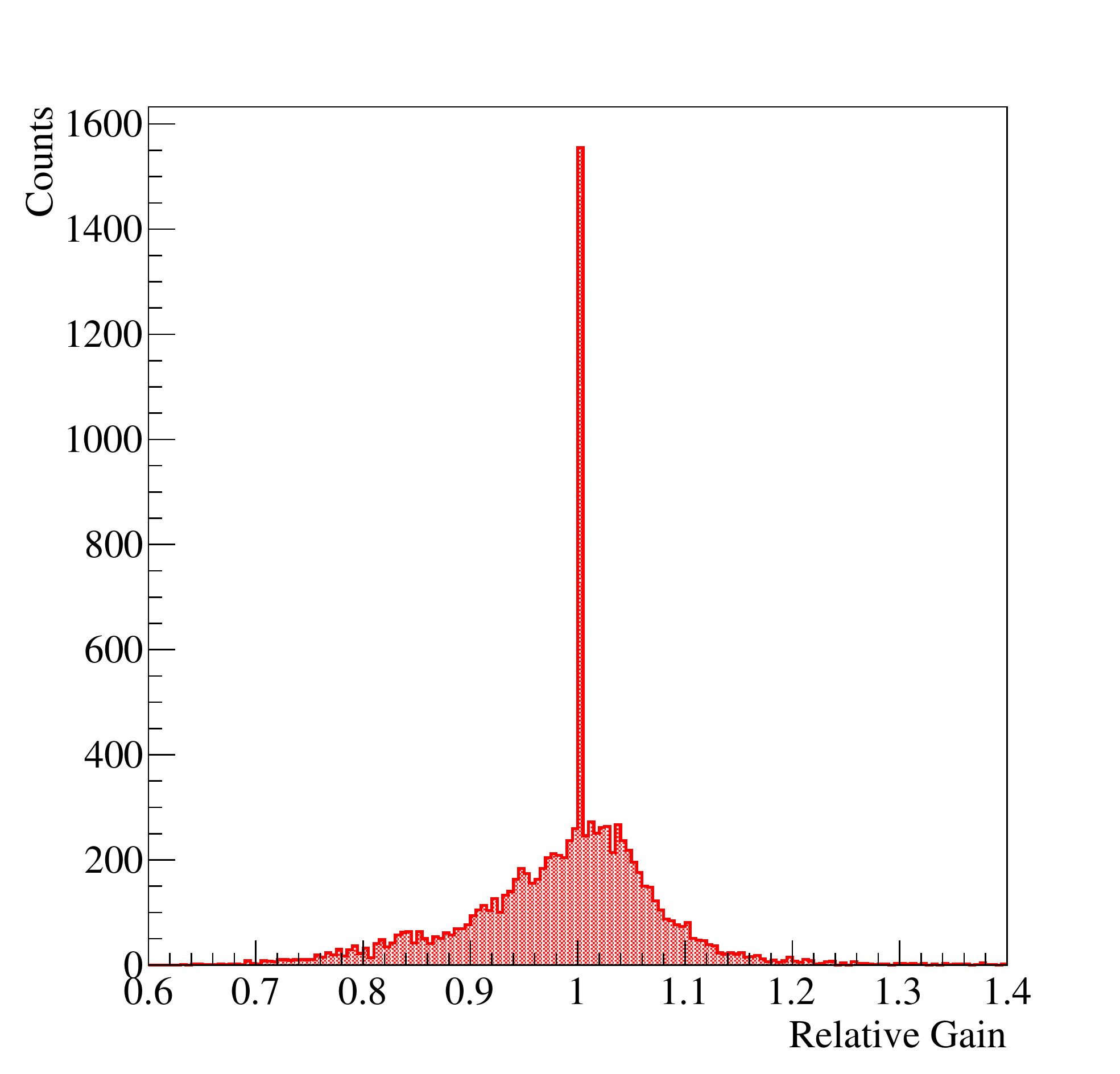}
      \caption{}
      \label{fig:gainmap}
    \end{subfigure}
    \caption{Figure (a) shows the relative gain calibration factors
      for all pads of the GEM-TPC as function of their
      position. Figure (b) shows the distribution of the calibration
      factors, which is centered at 1 has an RMS of 0.09. The peak at
      1 arises from channels for which no calibration was possible and
      which were consequently set to 1.}
    \label{fig:calib-fac}
  \end{center}
\end{figure}

\begin{figure}[!ht]
  \begin{center}
    \begin{subfigure}[b]{.49\textwidth}
      \centering
      \includegraphics[width=\textwidth]{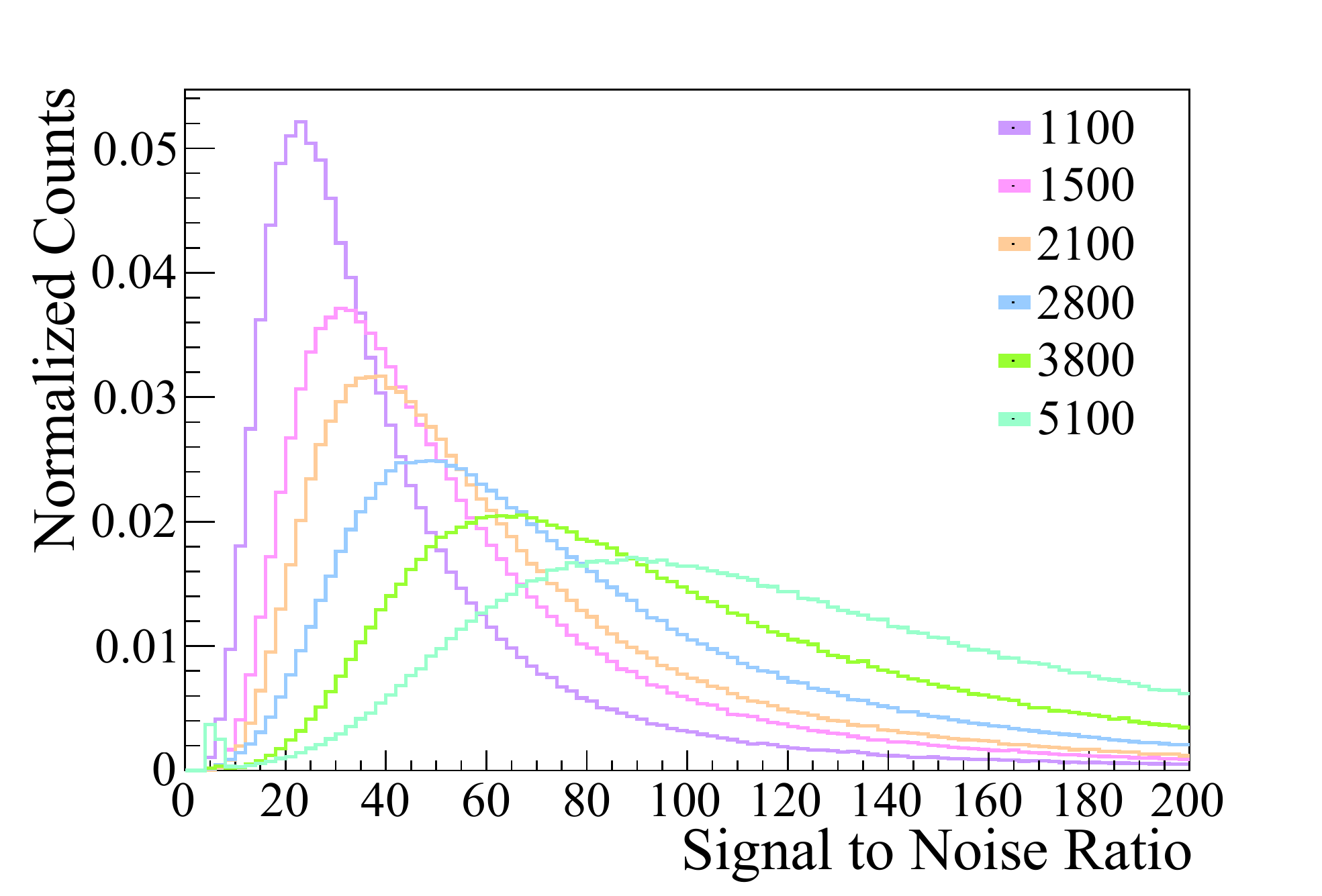}
      \caption{}
      \label{label1}
    \end{subfigure}
    \begin{subfigure}[b]{.49\textwidth}
      \centering
      \includegraphics[width=\textwidth]{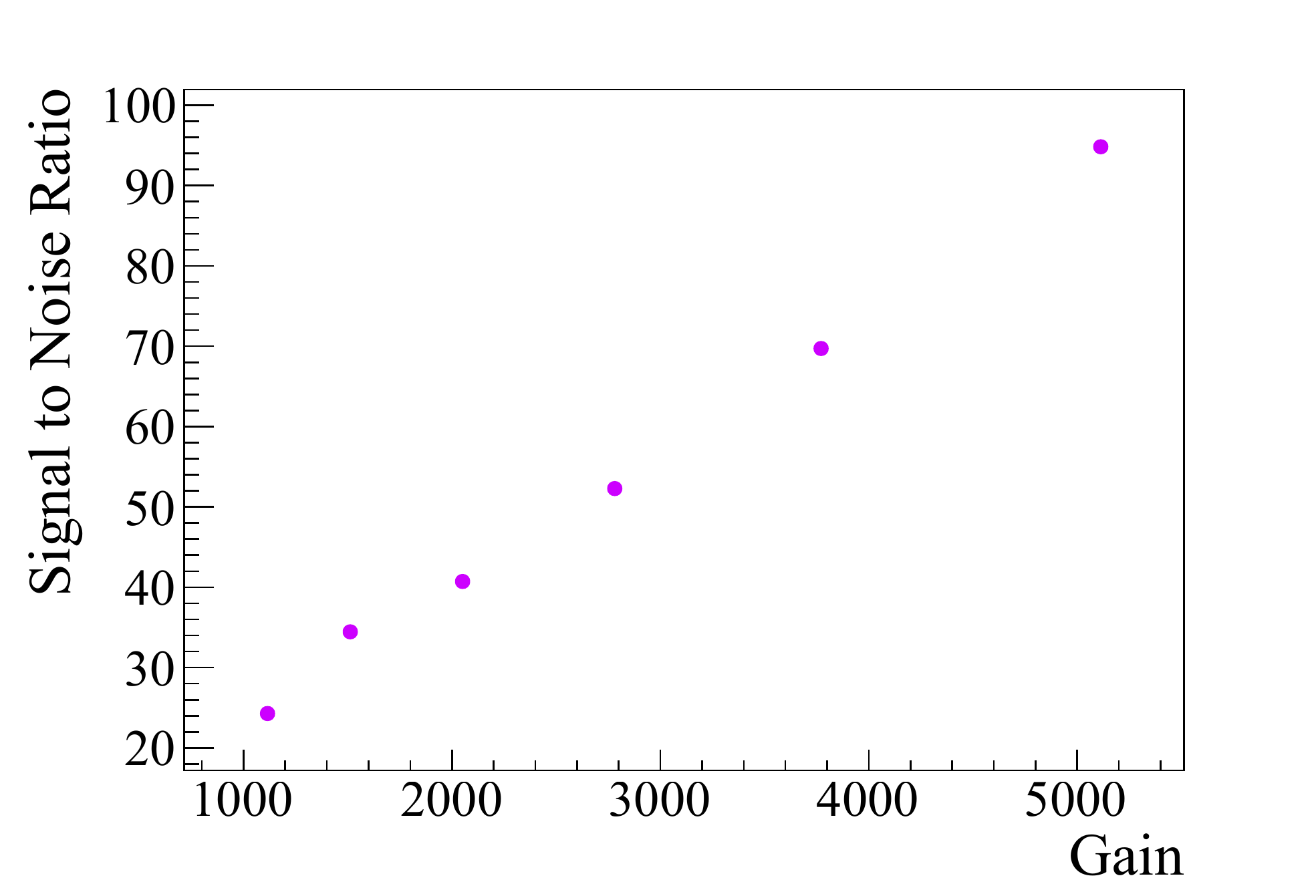}
      \caption{}
      \label{label2}
    \end{subfigure}
    \caption[shortCaption]{Signal to noise ratio calculated by
      dividing the highest pad amplitude in a cluster by the noise of
      the corresponding pad. The left plot shows the distributions for
      various gain settings in Ar/CO$_2$ (90/10) while the right panel
      shows the position of the most probable value of these
      distributions as a function of the gain. Note that the error
      bars are smaller than the points.}
    \label{pic:s2n}
  \end{center}
\end{figure}

The total cluster amplitude spectrum (in units of ADC channels)
measured in Ar/CO$_2$ (90/10) is shown in \figref{fig:spec_arco_83p}.
It depends on the deposited energy $E$, the mean energy required to
create an electron-ion pair in the gas $W$, the effective gain
$G_\mathrm{eff}$ and the conversion gain $A$ of the readout system,
measured in $e^-/\mathrm{ADC\ ch.}$ (see \secref{sec:FEE_DAQ.gain}):
\begin{equation}
  \label{eq:kr.amplitude}
  A_\mathrm{Kr} = \frac{E}{W}\cdot\frac{G_\mathrm{eff}}{A}\quad,
\end{equation}
With $W$ and $A$ known, the absolute effective gain can be calibrated
by reconstructing the full energy ($41.5\,\keV$) deposit of Kr
decays. The resulting gain curve is displayed in \figref{fig:kr.Geff}
as a function of the scale factor applied to all GEM potentials
(cf. \tabref{tab:GEM_Settings} for the definition of the nominal
setting corresponding to $100\%$).
For Ar/CO2 (90/10), a gain of approx.\ 2000 is achieved at a scale
factor of $83\%$.\\
For a sufficiently large data sample this procedure also allows to
extract pad-wise gain variations.
To this end, the total charge of a reconstructed Kr cluster is
assigned to the pad which contributed the pad hit with the largest
amplitude, and an amplitude histogram is generated for every single
pad.  Only pads/channels with more than 500 entries are used for
calibration.  Pads with a distance to the outer or inner field-cage
smaller than \SI{5}{\mm} are not taken into account as edge effects
introduced by the field-cage walls deteriorate the measurement of the
total charge.
Finally, the gain equalization is performed by moving the median
position of the main decay peak (\SI{41.5}{\kilo\electronvolt}) for
all individual channels to the overall median position.  From the
relative shift a relative gain map is generated which is normalized to
1.  The median was found to be less sensitive to outliers or values
from failed fits than the mean.
%
The relative calibration is done iteratively re-running the whole
reconstruction and recursively correcting the gain factors.\\
\Figref{fig:calib-fac} shows the relative gain correction factors as a
function of the pad positions obtained after 3 iterations.  Pads with
insufficient or no data (e.g.\ pads close to the field cage walls or
broken electronic channels) are set to have correction factors of
1. Clearly visible are the sector boundaries of the last two GEM
foils, slightly rotated with respect to each other, with a lower
effective gain (i.e.\ correction factor $>1$).  In addition, a darker,
cross-shaped pattern is visible, most likely caused by stress applied
during the glueing or mounting of the GEM stack. In these regions of
larger foil tension, an increase in effective gain is observed
(correction factors $<1$).
\Figref{fig:gainmap} shows a histogram of all gain calibration
factors.  The underlying distribution centered at 1 has an RMS of
$0.086$. The sharp peak at 1 comes from channels whose
calibration factor was set to 1.\\
The Kr cluster amplitude spectrum measured with the GEM-TPC in
Ar/CO$_2$ (90/10) gas at a voltage scaling factor of $81\%$ is shown
in \figref{fig:spec_arco_83p} before (red) and after (blue) applying
the pad-wise relative gain calibration.  A clear separation of the
main peaks can be observed while the double-peak-structure at
\SI{29/32.2}{\kilo\electronvolt} can not be clearly separated.  An
energy resolution of \SI{7.2}{\percent} before and \SI{4.1}{\percent}
after calibration in the main \SI{41.5}{\kilo\electronvolt} peak is
achieved, i.e.\ the pad-wise calibration enhances the energy
resolution of the detector at $41.5\,\keV$ by more than $40\%$.  This
translates into an improvement of $14.3\%$ on the specific-energy
resolution of the GEM-TPC \cite{Boehmer:2014hna,Boehmer:2015}.\\
%
%
%
An important quantity which influences both the position and the
energy resolution of the detector is the signal-to-noise ratio
(SNR). It is defined as the maximum pad signal amplitude in a cluster
divided by the corresponding pad noise value.  Typically, the SNR is
required be larger than 20 -- 25 for TPCs \cite{ALICE-TPC:2000}.  The
signal-to-noise ratio for the GEM-TPC for different values of the gas
gain in Ar/CO$_2$ (90/10) is shown in \figref{pic:s2n}.
\subsection{Determination of the drift velocity}
\label{sec:CalSys.vdrift}
\begin{figure}[ht]
  \begin{center} \centering
    \includegraphics[width=0.9\textwidth]{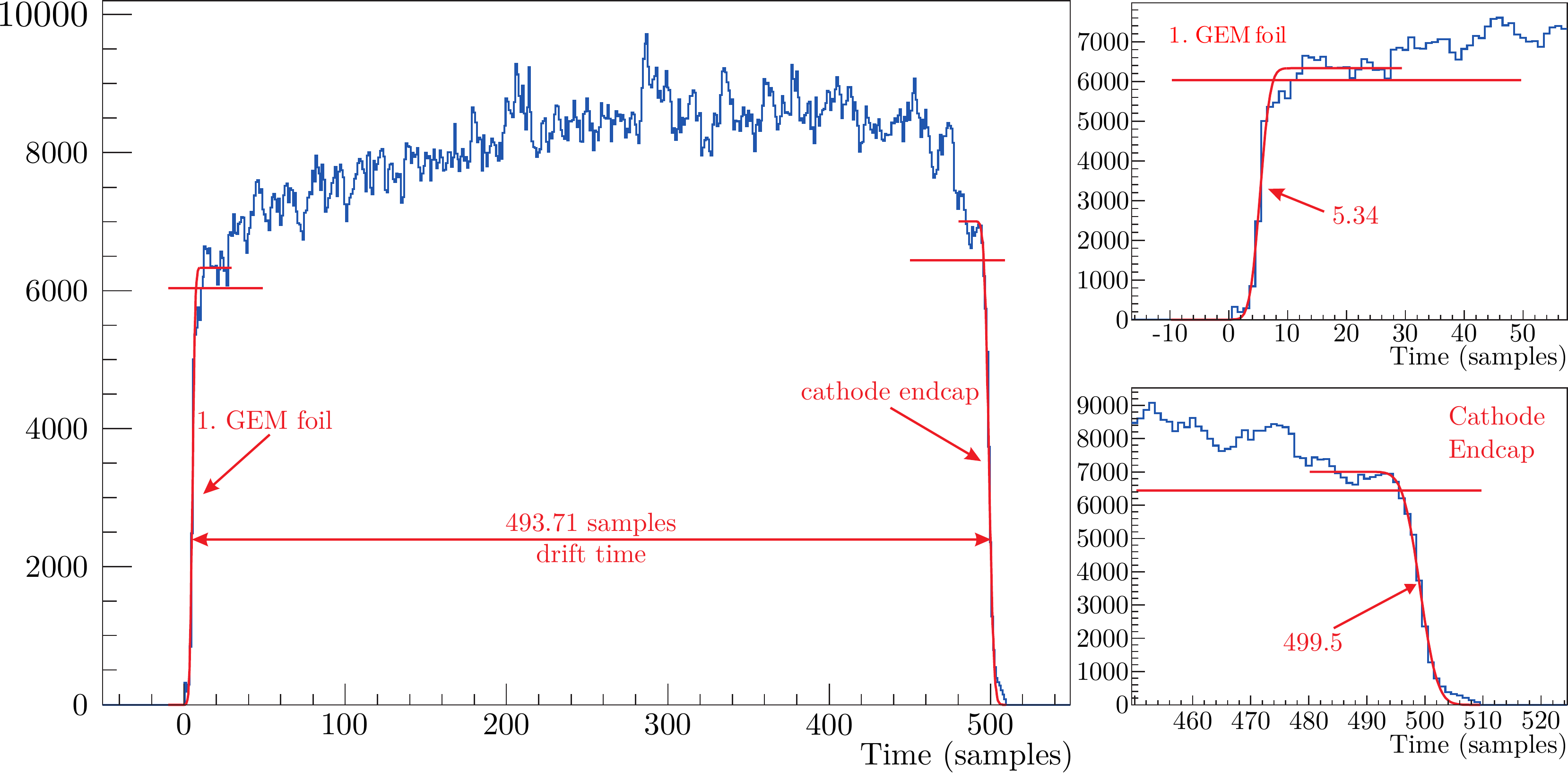}
    \caption{Time spectrum of clusters from cosmics with fitted error
      functions. For the measurement shown, the drift field was set to
      \SI{309.6}{\volt\per\centi\meter} in Ar/CO$_2$ (90/10).}
    \label{fig:TSpec-fit}
  \end{center}
\end{figure}

\begin{figure}[!ht]
  \begin{center}
    \includegraphics[width=\textwidth]{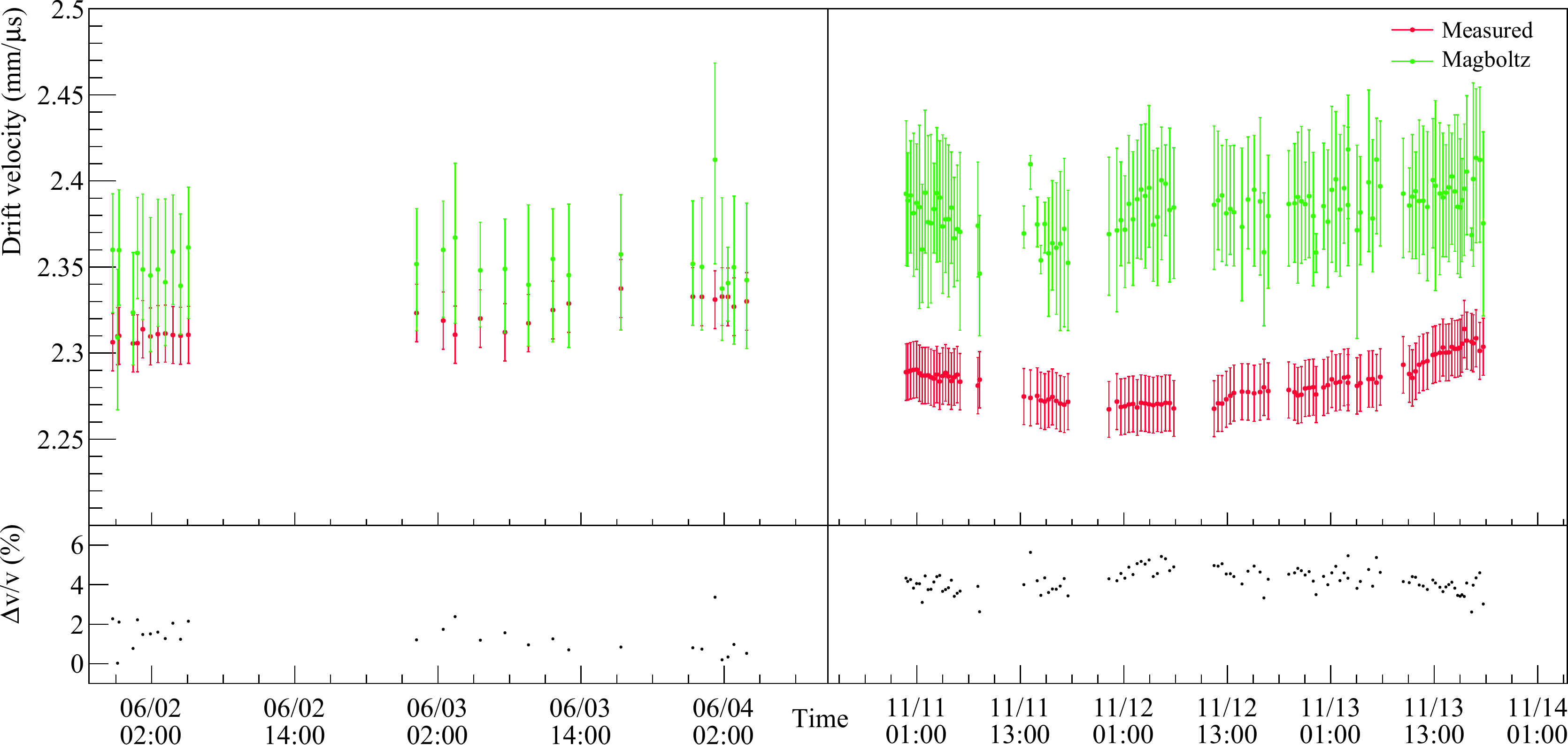}
    \caption[labelInTOC]{Comparison of (red) measured and (green)
      calculated drift 
      velocities as a function 
      of time during the (left) pion beam and (right) cosmic muon
      measurement 
      campaigns. The calculation with Magboltz takes into account the
      measured values of 
      temperature, pressure, drift field and their respective
      variations during one measurement. The uncertainties for the
      calculated drift velocities are mainly governed by the rather big
      variations of the temperature and pressure
      measurements. 
      The lower part of the image shows the relative difference
      between the drift velocity calculated by Magboltz and the
      measured drift velocity.}
    \label{fig:Drift-vel}
  \end{center}
\end{figure}

\noindent
The resolution of a TPC relies critically on the knowledge of the
drift velocity of electrons. The drift velocity depends on many
parameters such as the exact gas mixture, potential impurities like
water and oxygen, the temperature and pressure of the gas, and the
exact values of the electric and magnetic fields. Since not all of
these parameters might be known with the required accuracy, it is
important to be able to
determine the drift velocity in situ from the data.\\
%
If the maximum drift time of electrons in the TPC is smaller than the
readout time window given by the product of the maximal number of
samples (511) and the time length per sample ($64.3\,\ns$ for a
sampling frequency of $15.55\,\MHz$), the full drift window of the TPC
is read out.
The average drift velocity can then be determined using the drop of
the cluster occupancy at the geometric boundaries imposed by the
cathode and the GEM foil closest to the drift volume.  The measured
cluster occupancy distribution as a function of the time obtained from
cosmic muon data at a drift field of \SI{309.6}{\volt\per\cm} is shown
in \figref{fig:TSpec-fit}.
The clusters used for this procedure are constructed as described in
\ref{sec:ClusterReco}. \Tabref{tab:cluster-select} shows the cuts
applied to select good clusters and to reduce background
contributions.
\begin{table}
  \centering
  \begin{tabular}{ l r }
    \hline\hline
    selection criteria 		& Value 				\\
    \hline
    cluster size 				& $>$ 2 pad hit 		\\
    cluster amplitude 		& $>$ 50 ADC channels 	\\
    radial position (min.) 	& $>$ \SI{57}{\mm} 		\\ 
    radial position (max.) 	& $<$ \SI{148}{\mm}		\\
    \hline \hline
  \end{tabular}
  \caption{Cuts applied to select clusters for the measurement of the
    drift velocity.}
  \label{tab:cluster-select}
\end{table} 
Error functions are fitted to the occupancy distribution in order to
determine the position of the cathode and the first GEM foil. The
inflection points of these functions are used to determine the edges
of the drift volume.
%
With the difference between the two edges in units of ADC samples,
$N_\mathrm{cathode}-N_\mathrm{GEM1}$, the sampling rate $S\!R$ of the
front-end chip and the known distance $l$ between the chamber edges,
the drift velocity averaged over the full volume can be calculated by
\begin{equation}
  v_\mathrm{drift} = \frac{ l \cdot S\!R}{N_\mathrm{cathode} 
    - N_\mathrm{GEM1}}.
\end{equation}
  %
%
%
\begin{figure}[ht]
  \begin{center} \centering
    \includegraphics[width=0.9\textwidth]{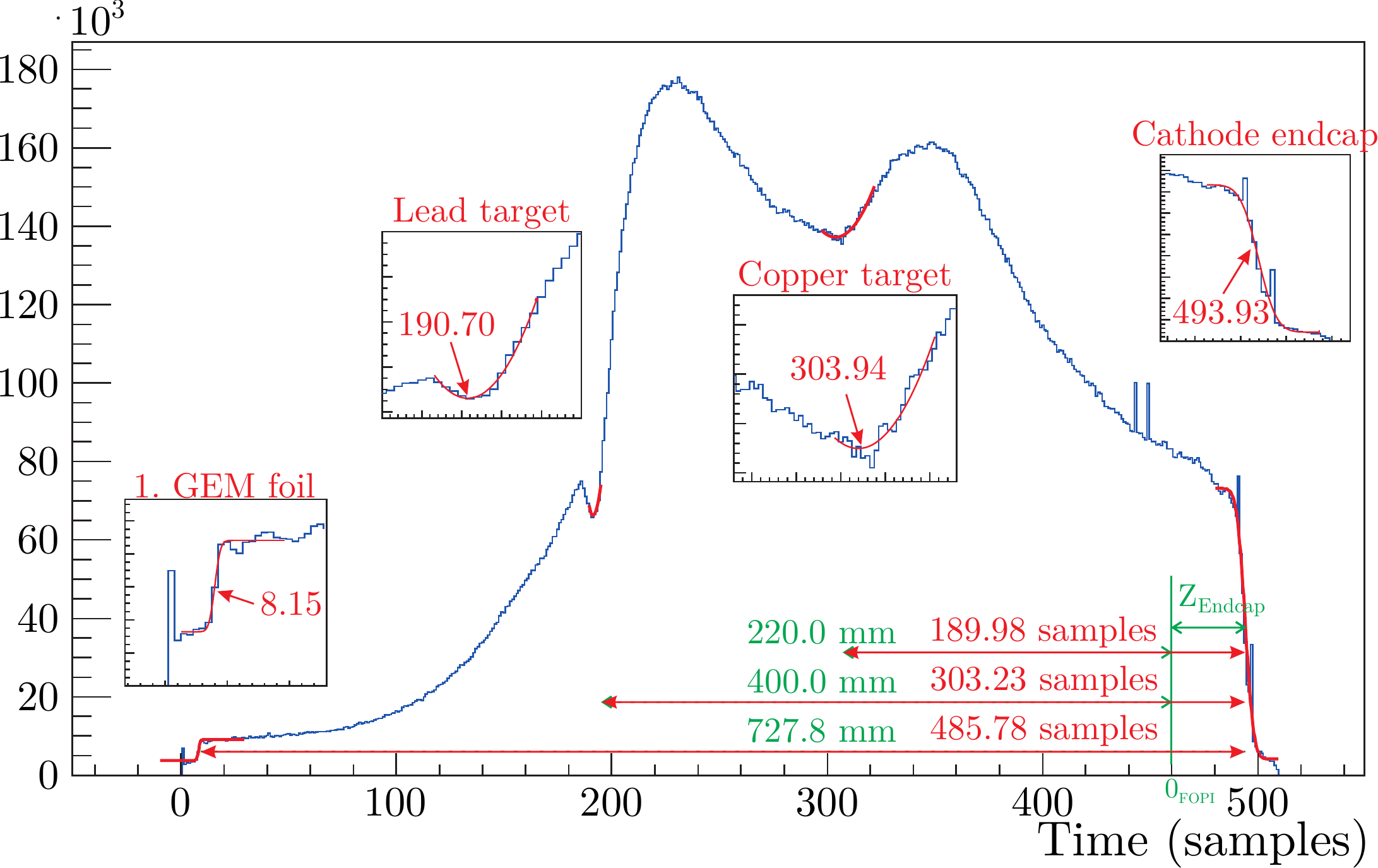}
    \caption{Time spectrum of clusters from a measurement with beam
      pions hitting a Pb and a Cu target inserted into the inner bore
      of the TPC. The spectrum has been taken during the $\pi^{-}$
      beam time 2011 at the FOPI experiment in Ar/CO$_2$ (90/10) gas
      with a drift field of $E_\mathrm{drift} =
      \SI{302.4}{\volt\per\cm}$.}
    \label{fig:TSpec-target-1}
  \end{center}
\end{figure}
The same method can also be used for measurements with particle beams,
provided the drift velocity is large enough so that the full drift
length is visible.  In \figref{fig:TSpec-target-1} one can see a
cluster time spectrum obtained with a $1.7\,\GeV/c$ $\pi^{-}$ beam at
a drift field of \SI{302.4}{\volt\per\cm} in Ar/CO$_2$ (90/10)
gas. Two targets were installed in the central bore of the TPC, a Pb
and a Cu disk \SI{180}{\mm} further downstream, which are visible as
pronounced drops in the occupancy in \figref{fig:TSpec-target-1}.  In
case the physical boundaries of the drift volume are not visible,
e.g.\ because of a smaller drift field, the known target positions may
be used to perform the drift velocity calibration instead
\cite{DKaiser:2014}.\\
%
%
%
Figure \ref{fig:Drift-vel} shows the time evolution of drift
velocities determined by fitting the end-cap positions for different
data taking periods (left: pion beam, right: cosmics), compared
to drift velocities calculated with Magboltz
\cite{Biagi1999234}. These take into account the gas temperatures and
pressures measured at the inlet to the TPC located in the media
flange.  The uncertainties shown for the Magboltz calculation take
into account the variations of the pressure and temperature
measurements around their average values. The uncertainties for the
measured values are the statistical uncertainties of the error
function fits.\\
One can see that the relative differences between measurement and
Magboltz calculations are below \SI{4}{\percent} for the pion beam
measurement and in between \SI{2}{\percent} and \SI{6}{\percent} for
the cosmic muon measurement.
%
The systematic shift between measured and calculated drift velocity is
attributed to a possible bias in the temperature and pressure
measurements, which may not reflect the actual gas conditions and to a
systematic uncertainty of the actual drift length.
For the future, it is planned to extract the drift velocity from the
alignment procedure using particle tracks which are measured both in
the GEM-TPC and the CDC of the FOPI experiment. In addition, a more
precise measurement of the actual gas pressure and temperature is
envisaged.

%% file: conclusions/conclusions.tex
\section{Conclusions}
\label{Con}

\noindent A large TPC with GEM amplification and without gating grid 
was built and successfully tested to
demonstrate its applicability in future high-rate experiments such as
ALICE at CERN. 
The detector features a number of innovative elements such as an
extremely light-weight and gas-tight vessel, which serves as field
cage and gas container at the same
time, a stack of large disk-shaped sectorized GEM foils integrated in
a media flange which forms the detector end cap, or a readout plane
with $10,254$ hexagonal pads in order to minimize the pad angular effect. 
The signals induced on the pad-plane are read out using AFTER/T2K
ASICs 
mounted on fully custom-made front-end cards and ADCs. 

First commissioning results of the system have been shown, which prove
the feasibility of this novel detector type. 
The chamber itself as well as all custum-made supporting
infrastructure such as the cooling circuit, the gas supply, and
the slow-control system have shown excellent
performance during data taking with cosmic rays and various particle 
beams as part of the FOPI spectrometer. 
Calibration procedures have been established and put into operation,
such as pedestal and noise determination, gain calibration and equalization
based on the injection of
metastable $^{83}$Kr into the detector gas, and 
drift velocity determination.  

These results
raise the expectation that (i) the \gt will significantly
improve the performance of the FOPI experiment by enhancing the vertex
resolution, the secondary vertex reconstruction and the global event
reconstruction, and that (ii) an ungated TPC with GEM amplification
will allow us to overcome the large deadtime associated with
classical TPCs, and thus open many new areas of application of this
powerful detector type.   

%% file: ack/ack.tex
\section*{Acknowledgements}

We express our
gratitude to the engineers and technicians from 
GSI Darmstadt, TU M\"unchen, Universit\"at Bonn, and SMI Vienna,
who have contributed to the
construction and commissioning of the GEM-TPC. 

We acknowledge support from 
Bundesministerium f\"ur Bildung und For\-schung, 
(Grant agreement no. 06MT245I, 06MT9165I, 05P12WOCAA, 05P12WOGHH), 
DFG cluster of  excellence ``Origin and Structure of the Universe'',
DFG SFB/TR16 ``Elektromagnetische Anregung subnuklearer Systeme'',  
GSI For\-schungs- und Entwicklungsauftrag (Grant agreement no.  
TMPAUL, TMPAUL1012, TMFABI1012, TMLFRG1316F),
Helmholtz Association (Young Investigator Group VH-NG-330),
European Union FP7 (HadronPhysics2, Grant agreement no. 227431;
HadronPhysics3, Grant agreement no. 283286).